\documentclass[prx,reprint,floatfix,superscriptaddress,footinbib]{revtex4-1}

\usepackage{xcolor}
\usepackage{microtype}
\usepackage[bookmarks=true,
bookmarksnumbered=false,
bookmarksopen=false,
colorlinks=true,
linkcolor=blue]{hyperref}
\definecolor{webblue}{rgb}{0, 0, 0.5} % less intense blue
\hypersetup{colorlinks=true,urlcolor=blue,citecolor=blue}

\usepackage{graphicx}

\usepackage{scalerel}

\usepackage{comment}
\usepackage{pifont}
\usepackage{mathrsfs}
\usepackage{subfigure}
\usepackage{url}
\usepackage{hyperref}
\usepackage{inconsolata}
\usepackage{amsmath}
\usepackage[capitalize]{cleveref}
\usepackage{braket}
\usepackage{outlines}
\usepackage{enumitem}
\usepackage{soul}
\usepackage{color}
\usepackage{bm} 

\usepackage[utf8]{inputenc}
\usepackage{siunitx,booktabs}
\usepackage{multirow}
\usepackage{amssymb}
 \usepackage{array,multirow}
 \renewcommand{\vec}[1]{\boldsymbol{#1}}
 \newcommand{\pdagger}{{\phantom{\dagger}}}

 \usepackage{relsize}

\usepackage{soul}

\usepackage{pgfplots}
\usepackage{ifthen}

\usepackage{comment}
\usepackage{pifont}

\usepackage{tabularx}

\begin{document}

\title{Quadratic Dirac fermions and the competition of ordered states \\ in twisted bilayer graphene}

%Revival of quadratic Dirac fermions and the competition of ordered states in twisted bilayer graphene

%\title{Dirac revivals and the competition of ordered states in twisted bilayer graphene}
  
%Cascade of phase transitions from correlated Dirac revivals in twisted bilayer graphene

%Dirac revivals and competing orders in twisted bilayer graphene

\author{Julian Ingham}
\email{jingham@bu.edu}
\affiliation{Department of Physics, Columbia University, New York, NY, 10027, USA}
\affiliation{Physics Department, Boston University, Commonwealth Avenue, Boston, MA 02215, USA}

\author{Tommy Li}
\affiliation{Dahlem Center for Complex Quantum Systems and Fachbereich Physik, Freie Universit\"{a}t Berlin, Arnimallee 14, 14195 Berlin, Germany}

\author{Mathias S.~Scheurer}
\affiliation{Institute for Theoretical Physics III, University of Stuttgart, 70550 Stuttgart, Germany}

\author{Harley D.~Scammell}
\affiliation{School of Mathematical and Physical Sciences, University of Technology Sydney, Ultimo, NSW 2007, Australia}

\date{\today}

\begin{abstract}

Magic-angle twisted bilayer graphene (TBG) exhibits a captivating phase diagram as a function of doping, featuring superconductivity and a variety of insulating and magnetic states. The bands host Dirac fermions with a reduced Fermi velocity; experiments have shown that the Dirac dispersion reappears near integer fillings of the moir\' e unit cell --- referred to as the \textit{Dirac revival} phenomenon. The reduced velocity of these Dirac states leads us to propose a scenario in which the Dirac fermions possess an approximately quadratic dispersion. The quadratic momentum dependence and particle-hole degeneracy at the Dirac point results in a logarithmic enhancement of interaction effects, which does not appear for a linear dispersion. The resulting non-trivial renormalisation group (RG) flow naturally produces the qualitative phase diagram as a function of doping -- with nematic and insulating states near integer fillings, which give way to superconducting states past a critical relative doping.  The RG method further produces different results to strong-coupling Hartree-Fock treatments: producing T-IVC insulating states for repulsive interactions, explaining the results of very recent STM experiments, alongside nodal $A_2$ superconductivity near half-filling, whose properties explain puzzles in tunnelling studies of the superconducting state. The model explains a diverse range of additional experimental observations, unifying many aspects of the phase diagram of TBG.  
 
\end{abstract}

\maketitle

\section{Introduction}
\label{intro}
Twisted bilayer graphene (TBG) has become a central focus of theoretical and experimental condensed matter physics \cite{Cao2018a,Cao2018b,Yankowitz2019,Lu2019,Cao2020,Polshyn2019,Xie2019,Jiang2019,Choi2019,Kerelsky2019,Tomarken2019, Saito2021,Das2022, Cao2021b, Paul2022,Wong2020,Morissette2023, Oh2021,Kim2022b,Park2021c,Nuckolls2023,Kim2023,Pierce2021,Stepanov2021,Bhowmik2022,Bhowmik2022,Choi2021,Nuckolls2020,Park2021,Xu2020,Xie2021,Das2021b,Serlin2019,Polski2022,Saito2021b,Sharpe2019,OrbMagReview,Sharpe2021b,Lin2022,Grover2022,Uri2020,Tschirhart2021,Kim2021,Po2018,Po2019,Xu2018, Tarnopolsky2019,Bernevig2021a,Bernevig2021b, Bernevig2021c, Bernevig2021f,Song2019, Song2022,Isobe2018,You2019,Liu2018,Gonzalez2019,Classen2019,Lin2019,Chichinadze2020,Chichinadze2020b,Guinea2018,Bultinck2020,Bernevig2021d,Bernevig2021e,Kwan2021b,Liu2021c,Liu2021b,PhononsTIVC,Wagner2022,Christos2022,BernevigTrilayer,TBorNotTB,Yu2023,Peltonen2018, Xu2018kek,Lian2019, Wu2018, Zhang2019, Yuan2018, Kang2019,Ahn2019,Julku2020,Padhi2020,Fernandes2021,Kwan2021,Padhi2018,Thomson2021,Yu2021,Khalaf2022,Mathias2020,Lake2022,Zhang2019b,Lee2019,Christos2023,Roy2019,Islam2023,Brillaux2022,Parthenios2023}. Since the original discovery of unconventional superconductivity and correlated insulating states \cite{Cao2018a,Cao2018b}, intense experimental scrutiny has uncovered a rich phase diagram as a function of temperature, electron density, and magnetic field -- featuring orbital magnetism, nematic ordering, and Kekul\'e textures \cite{Cao2018a,Cao2018b,Yankowitz2019,Lu2019,Cao2020,Polshyn2019,Xie2019,Jiang2019,Choi2019,Kerelsky2019, Saito2021,Das2022, Cao2021b, Paul2022,Wong2020,Morissette2023, Tomarken2019,Oh2021,Kim2022b,Park2021c,Nuckolls2023,Kim2023,Pierce2021,Stepanov2021,Bhowmik2022,Bhowmik2022,Choi2021,Nuckolls2020,Park2021,Xu2020,Xie2021,Das2021b,Serlin2019,Polski2022,Saito2021b,Sharpe2019,OrbMagReview,Sharpe2021b,Lin2022,Grover2022,Uri2020,Tschirhart2021,Kim2021}.

Stacking two sheets of graphene and twisting by a relative angle $\theta$, the composite system is no longer periodic with the lattice constant of monolayer graphene, but is periodic at a larger moir\' e scale $\sim a/(2\sin(\theta/2))$, folding the monolayer graphene dispersion into a mini Brillouin zone \cite{Santos2012,Bistritzer2011,Koshino2018,Po2018obstr,Khalaf2019idk,Yoo2019,Tritsaris2020,Carr2019b,Carr2019,Kang2018,Zang2022}. The coupling between the two layers hybridises the monolayer Dirac points of graphene, forming mini bands, and when the twist angle is reduced to the so-called magic angle, the fourfold degenerate bands near charge neutrality flatten and the velocity of the Dirac points grows small, enhancing interaction effects and giving rise to a diverse collection of interesting material properties.

Compressibility measurements observe an interesting property of TBG referred to as the \textit{Dirac revival} -- when the density of electrons is increased to an integer number of electrons per unit cell, additional electrons `reset' to the charge neutrality point, and are described by the Dirac dispersion but with a reduced degeneracy \cite{Zondinger2020}. For instance, at one electron per moir\'e unit cell $\nu=1$ (Fig. \ref{f:revival}), one of the four-fold degenerate bands spontaneously becomes fully occupied, and as the density is increased the remaining three-fold degenerate bands refill starting from the charge neutrality point \cite{endnote1}. These revived Dirac fermions appear at much higher temperatures ($\approx 20$ K) than the insulating and superconducting states ($\approx 1$ K), and constitute the parent state of the correlated physics.

\begin{figure*}[t]
\hspace{0.3cm}
\includegraphics[width=17.5cm]{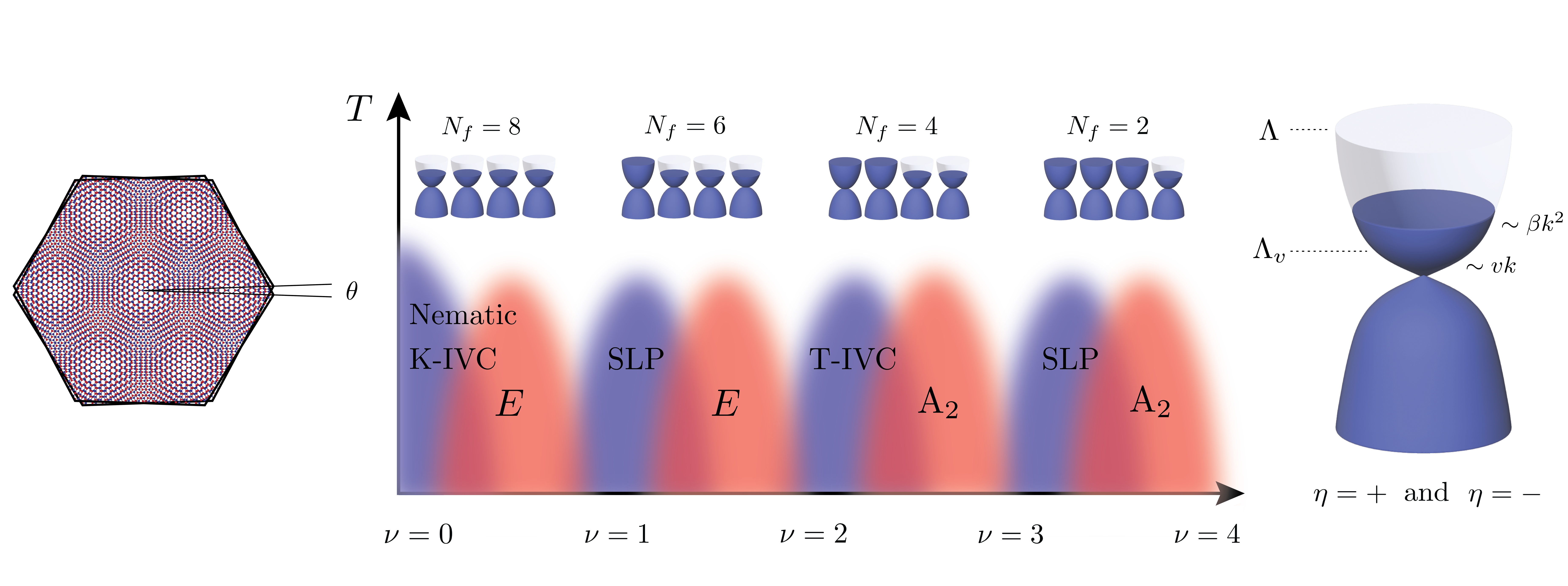}
\vspace{-0.55cm}
\caption{\textbf{Theoretical model.} Left: Two stacks of graphene with a relative twist of $\theta$, Right: Our model for the Dirac fermions near integer filling. The dispersion is linear beneath an infrared cutoff $\Lambda_v$, and quadratic between $\Lambda_v$ and an ultraviolet cutoff $\Lambda$. In the temperature window $\Lambda_v<T<\Lambda$, interactions are logarithmically enhanced. Middle: Our proposed phase diagram features interlaced superconductors (red) and correlated insulators (blue), including nematic/K-IVC order at $\nu=0$ and T-IVC order with proximate $E$/$A_2$ superconductivity near $\nu=2$.}
\label{f:Model}
\end{figure*}

While the magic angle tunes the Fermi velocity to be small, there is nonetheless a nonzero bandwidth; we suggest that the smallness of the linear term in the Dirac dispersion makes the dispersion approximately quadratic for an energy window near the Dirac points (Fig.~\ref{f:Model}c). Due to the Dirac revival, the quadratic Dirac dispersion near charge neutrality then characterises the physics of TBG near all integer fillings. A quadratic dispersion plus particle-hole degeneracy at the Fermi level in two dimensions results in a logarithmic enhancement of interaction effects, and a non-trivial renormalisation group (RG) flow. In this work we analyse the RG flow of these quadratic Dirac fermions and derive a number of interesting results:

\begin{enumerate}

\item Near integer fillings, the particle-hole degeneracy of the Dirac point causes insulating and nematic states, driven by particle-hole fluctuations, to compete strongly against superconductivity. Doping above the Dirac point lifts the particle-hole degeneracy, leaving superconductivity as the dominant order.

\item The result is a phase diagram with nematic or insulating states at integer fillings and superconductors in between (see Fig.~\ref{f:Model}b), consistent with the experimentally obtained phase diagram of TBG.

\item The quadratic Dirac theory  predicts different ground states to pre-existing mean-field analyses; we list the resulting order parameters in Table \ref{symmetries}. For instance, Hartree-Fock studies of the Coulomb repulsion favour K-IVC over T-IVC \cite{Guinea2018,Bultinck2020,Bernevig2021d,Bernevig2021e,Kwan2021b,Liu2021c,Liu2021b,PhononsTIVC,Wagner2022,Christos2022}, requiring phonon coupling for T-IVC to contend, whereas the RG flow can favour T-IVC for purely repulsive interactions.  Very recent STM experiments have found evidence for T-IVC rather than K-IVC order near $\nu=2$ \cite{Nuckolls2023,Kim2023}; our mechanism therefore provides a natural explanation of this puzzle.

\item Intervalley $A_2$ and $E$ superconductors appear as instabilities alongside T-IVC order near $\nu\gtrsim 2$; the properties of the $A_2$ state can explain the transition from U- to V- shaped tunnelling profiles seen in studies of the superconducting state \cite{Oh2021,Kim2022b,Park2021c}.

%$\item {\color{red} Wess-Zumino-Witten terms appearing in the Landau-Ginzburg free energy  predict a continuous transition between T-IVC and $A_2$ superconductivity, which may explain the pseudogap region and appearance of a kekul\'e pattern in the superconducting state near half filling}

\end{enumerate}

Incorporating a small finite Dirac velocity does not dramatically modify the RG behaviour of the Dirac theory, simply introducing an IR cutoff $\Lambda_v$ on the RG flow. The enhancement of the interactions occurs within a window of temperatures associated to the energy window in which the dispersion appears quadratic (c.f. Fig. \ref{f:Model}c).

We contrast our theory with previous theoretical models. Firstly, interacting theories with linearly-dispersing Dirac fermions (e.g. Refs. \cite{Roy2019,Islam2023,Brillaux2022,Parthenios2023}) do not naturally feature insulating and nematic states as weak coupling instabilities. As we explain in Sec \ref{erg}, the quadratic scaling of the dispersion is essential to the presence of insulating and nematic states near integer fillings. Secondly, the starting point of our analysis is to treat the bands as dispersive rather than approximately flat (c.f. Refs \cite{Kang2019,Bultinck2020,Bernevig2021d,Christos2022,TBorNotTB}), motivated by the experimental observation of dispersive Dirac states \cite{Zondinger2020}, and a bandwidth $\approx 40$ meV, much larger than that predicted by bandstructure \cite{Tomarken2019}. Thirdly, we stress that van Hove singularities (vHS) and Fermi surface nesting \cite{Isobe2018,You2019,Liu2018,Gonzalez2019,Classen2019,Lin2019,Chichinadze2020,Chichinadze2020b} do not feature in our model. Since the Dirac revival resets the dispersion to that of the Dirac point near integer fillings, in this regime the Fermi surface is neither nested nor located near a vHS. Experiments have found that superconductivity is seen near resets, and consistently suppressed at twist angles where revivals disappear and vHS are observed \cite{endnote2}.

By contrast, our model argues that the superconducting and insulating states arise from the RG flow from a quadratic dispersion, with superconductivity dominant when the insulating states are suppressed via doping. Superconductivity therefore does not originate from an insulating parent state, but appears as a competing phase. The competing order scenario is supported by the presence of superconductivity in the absence of insulating states at smaller twist angles or when TBG is strongly screened by external gates \cite{Liu2021,Stepanov2020,Saito2020b} (though we comment that interpretation of these experiments is complicated by the presence of disorder), the appearance of the insulating state under the superconducting dome when superconductivity is suppressed by a magnetic field, along with the comparable magnitudes of the superconducting and insulating $T_c$. Given this last point, it is particularly notable that our framework allows a simultaneous treatment of the insulating and superconducting states on equal footing, unlike Hartree-Fock studies which are well-suited to describing the insulating states.

We lastly note that signatures of the Dirac revival phenomena are also seen in twisted trilayer graphene (tTLG) \cite{Park2021a,Hao2021,Cao2021a,Siriviboon2021,Zhang2021b}, and so we anticipate that our analysis is likely relevant to a range of moir\'e systems. 

\section{Model and symmetry constraints}
\label{model}
\subsection{Quantum numbers and Symmetries}
\begin{figure}[t]
\includegraphics[width=8.6cm]{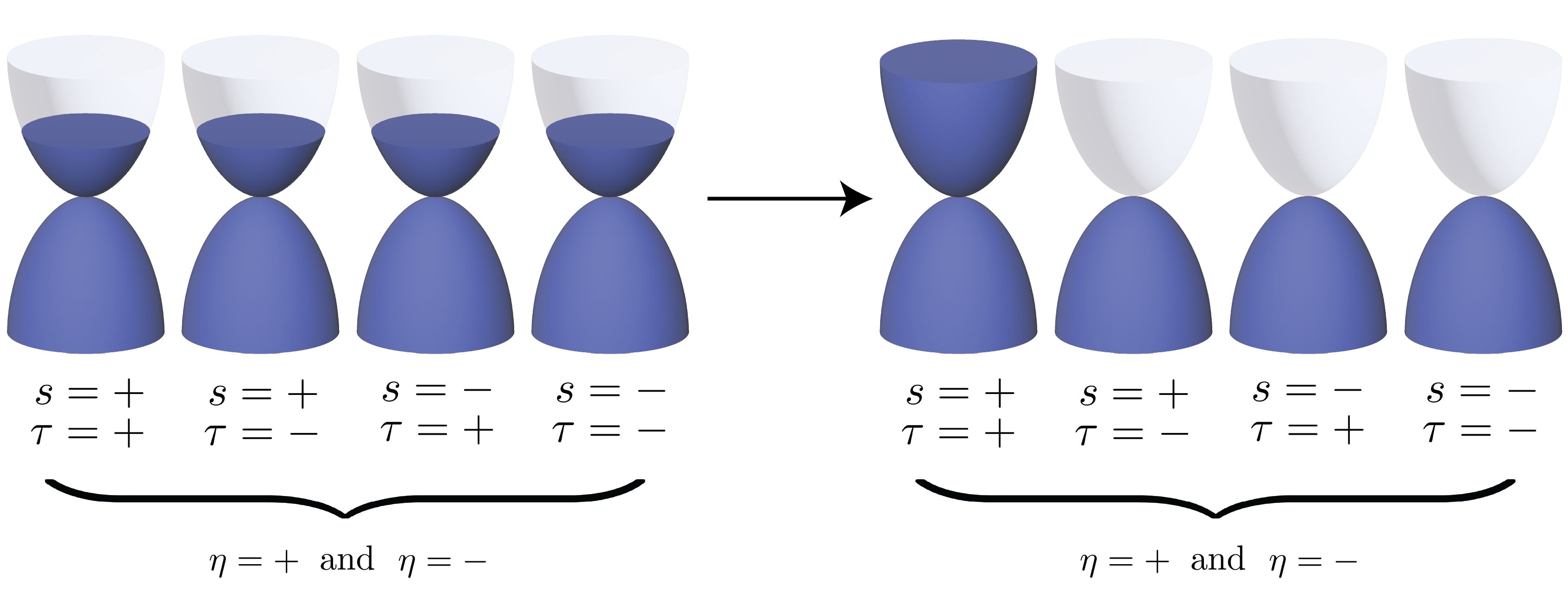}
\caption{\textbf{Dirac revival.} At each moir\'e valley $\eta=\pm$ there are four flavours corresponding to the spin $s=\pm$ and monolayer valley $\tau=\pm$ degeneracy, which for $|\nu| \gtrsim 0$ populate equally as the density increases (Left). When the bands are quarter filled at $|\nu|=1$, all the density is spontaneously transferred to one flavour, and any additional electrons are added near the band touching point of the remaining Dirac states (Right).} 
\label{f:revival}
\end{figure}

The bands near charge neutrality are four-fold degenerate, originating from the spin and monolayer valley degeneracy. The conduction and valence bands exhibit Dirac points at the moir\'e $K$-points; we index these band touching points by sublattice $\sigma$, monolayer valley $\tau$, and moir\'e valley $\eta=\pm$, corresponding to the Bloch states near quasimomenta $\tau R_{\pm \theta/2} \bm{K}$ where $\bm{K}$ is the monolayer valley momentum. Counting the number of Dirac cones gives $N_f=8$ species of Dirac fermions (see Fig. \ref{f:revival} left).

We describe the valley and spin quantum numbers as ``flavours''; after each Dirac revival, a flavour is projected out reducing the degeneracy by two, as shown in Fig. \ref{f:revival}, i.e. $N_f = 2(4-\lfloor \nu \rfloor)$ for $\nu>0$. In our analysis, we will not attempt to explain the origins of the revivals, but take the polarised Dirac theory as an input parent state. It is observed (e.g. Ref. \cite{Polski2022}) that this parent state appears at different angles for electron $\nu>0$ and hole $\nu<0$ doping. Our argument that the correlated phases arise from the revived Dirac parent state therefore naturally explains the observed electron-hole asymmetry of the superconducting phase diagram. In what follows we will take $\nu>0$ with the understanding that our results apply to all integer $-4<\nu<4$ at which revivals occur.

TBG possesses threefold rotational symmetry in the plane $C_{3z}$, twofold rotational symmetry about the $x$ axis $C_{2x}$, twofold rotational symmetry in the plane $C_{2z}$, %with centre of rotation $(L/(2\sqrt{3}),0)$, 
i.e. the $D_6$ point group, along with time-reversal symmetry $\Theta$ (TRS). The system maintains SU(2) spin rotational symmetry due to absence of spin-orbit coupling. In addition, TBG has approximate symmetries, which we shall take to be exact in our model: independent spin rotations in the two monolayer valleys results in an enlarged SU(2)$ \times$ SU(2) spin symmetry. This symmetry is broken in experiment by the small yet finite Hund's coupling $J_H$. In the small twist angle limit, TBG also possesses a particle-hole symmetry ${\cal P}$ \cite{endnote3}; combining particle-hole and TRS gives an anti-commuting chiral symmetry represented by ${\cal S}=\mathcal{P}\Theta=\sigma_x\tau_x\eta_y$, with action ${\cal S}{\cal H}_0(\bm k){\cal S}^\dag = -{\cal H}_0(\bm k)$ on the single-particle Hamiltonian $\mathcal{H}_{0}$.

\subsection{Single-particle Hamiltonian}
The above symmetries allow us to construct the most general single-particle Hamiltonian describing the Dirac states near the moir\'{e} $K$-points, which to quadratic order we find to be
\begin{align}
\label{singleHeq}
\mathcal{H}_0 = v \tau_{z} (k_{+} \alpha_{-}+k_{-} \alpha_{+}) +i \beta \eta_z \left(k_{+}^{2} \alpha_{+}-k_{-}^{2} \alpha_{-}\right) 
\end{align}
where $(\alpha_x,\alpha_y) = (\sigma_x,\tau_z\sigma_y)$, $\alpha_\pm = \alpha_x \pm i \alpha_y$, and $k_\pm = k_x \pm i k_y$. Strikingly, we find that restricting to quadratic order in the momentum expansion results in an emergent commuting chiral symmetry $[{\cal C},\mathcal{H}_0]=0$ with ${\cal C}=-i\sigma_z {\cal S}=\sigma_y \tau_x \eta_y$; terms which break this symmetry may only appear at cubic and higher order in momentum. The symmetry ${\cal C}$ has been studied in previous works, where a so-called `chiral limit' \cite{Tarnopolsky2019} results in ${\cal C}$ as an exact symmetry \cite{endnote4}. Here we do not impose the chiral limit, yet we find that this symmetry appears as an emergent low-energy symmetry of the Dirac effective theory.

Our approach shall be to assume $\Lambda_v = v^2/\beta$ is small compared to the UV cutoff $\Lambda \gg \Lambda_v$, so that there is a range of energies in which the dispersion can be treated as quadratic, allowing us to neglect the linear term, Fig. \ref{f:Model}c. In TBG, there are natural reasons to expect $\Lambda \gg \Lambda_v$ -- in the limit where $\mathcal{P}$ is taken to be an exact symmetry, it has been shown \cite{Raquel} that the velocity can be made to vanish by tuning only a single parameter. Motivated by these results, in the Supplementary Material we show that for a wide range of tunnelling couplings, the $\mathcal{P}$-symmetric Bistritzer-MacDonald model \cite{Bistritzer2011} possesses a twist angle at which the Dirac points exhibit a quadratic dispersion \cite{endnote5}. However, our results are not reliant on the exact values of $v$ and $\beta$ -- we shall leave them as phenomenological constants, which may feasibly be investigated experimentally through compressibility measurements \cite{shahal}.

\subsection{Interactions}
Projecting the Coulomb interaction onto the basis of states near the Dirac points $|\sigma,\tau,\eta\rangle$ gives
\begin{align}
V = \tfrac{1}{2}\sum_{\mu\nu} V_{13;24} \psi^\dag_{1,\bm k} \psi_{3,\bm k-\bm q}\psi^\dag_{2,\bm p }\psi_{4,\bm p+ \bm q}
\label{V1234}
\end{align}
where $1$,$2$,$3$,$4$ are shorthand for the indices $\{\sigma, \tau, \eta\}$. A powerful approach is to write the interactions in the adjoint representation:
\begin{align}
V_{13;24} = V_{\mu\nu} \, (\Omega^\mu)_{13} (\Omega^\nu)_{24}
\end{align}
where $\Omega^\mu \in \{\sigma_a\tau_b\eta_c\}$, representing the Coulomb potential as a sum of tensor products in  $\sigma\tau\eta$ space \cite{Li2020,Scammell2021,Li2020b}. The potential is constrained by the requirement that only symmetry-invariant tensor products appear; in the Supplementary Material, we list the full set of symmetry-allowed products of bilinears. 
%We additionally prove that, 
Under the assumption of a real Coulomb potential, only $\Omega^\mu$ which commute with $\mathcal{S}$ and $\cal C$ may appear. These constraints result in only three possible bilinears: $\Omega^\mu\in\{\sigma _0\tau _0\eta _0,\sigma_z\tau_0\eta _z, \sigma_z\tau_0\tilde{\eta} _{\pm}\}$, where $\tilde\eta_\pm = \eta_x\pm i\tau_z\eta_y$. Renormalisation of the interactions, which we discuss further in the next section, generates the additional vertices $\eta_z \alpha_{x}\Omega^\mu,\eta_z \alpha_{y}\Omega^\mu, \alpha_{z}\Omega^\mu$, which commute with $\cal C$ but not $\cal S$. This results in a set of nine coupling constants,
\begin{widetext}
\begin{gather}
V= g_o (\sigma _0\tau _0\eta _0\otimes \sigma _0\tau _0\eta _0)+g_{x\tau}(\alpha_\pm\tau _z\eta _0\otimes \alpha_\mp\tau _z\eta _0)+g_z(\alpha_z\tau_0\eta _0\otimes \alpha_z\tau _0\eta _0) \nonumber \\
+\,v_{o\tau}(\sigma _0  \tau_z \tilde\eta _\pm\otimes \sigma _0  \tau_z \tilde\eta _\mp )+v_{x}(\alpha_\pm \tau_0 \tilde\eta_\pm \otimes \alpha_\mp \tau_0 \tilde\eta_\mp+\alpha_\pm \tau_0 \tilde\eta_\mp \otimes \alpha_\mp \tau_0 \tilde\eta_\pm)+v_z (\sigma _z\tau _0\tilde\eta _\pm\otimes \sigma _z\tau _0\tilde\eta _\mp )  \nonumber \\
 +u_{o\tau}(\sigma _0  \tau_z \eta _z\otimes \sigma _0  \tau_z \eta _z) +u_{x}(\alpha_\pm  \tau_0 \eta _z\otimes \alpha_\mp  \tau_0 \eta _z) +u_z(\sigma_z\tau_0\eta _z\otimes \sigma_z\tau _0\eta _z).
\label{interactions}
\end{gather} 
\end{widetext}
Based on the above arguments, our expectation is that $g_o$ and $v_z$ are likely the largest couplings near $\nu=0$, but after each Dirac revival the bare values of these couplings likely change. Our theory for TBG near integer filling comprises the single-particle Hamiltonian Eq. \eqref{singleHeq} along with the four-fermion interactions of Eq. \eqref{interactions}, $\mathcal{H}=\mathcal{H}_0 + V$. We argue this describes the normal state at each integer filling out of which the insulating and superconducting phases develop. Prior studies of Dirac theories \cite{Roy2019,Islam2023,Brillaux2022,Parthenios2023} have not explored the combination of quadratic band touching, $\eta$--dependent scattering, and filling factor-dependent degeneracy $N_f = 2(4- \lfloor \nu \rfloor)$, which we now elucidate.

\begin{figure*}
\includegraphics[width=17.5cm]{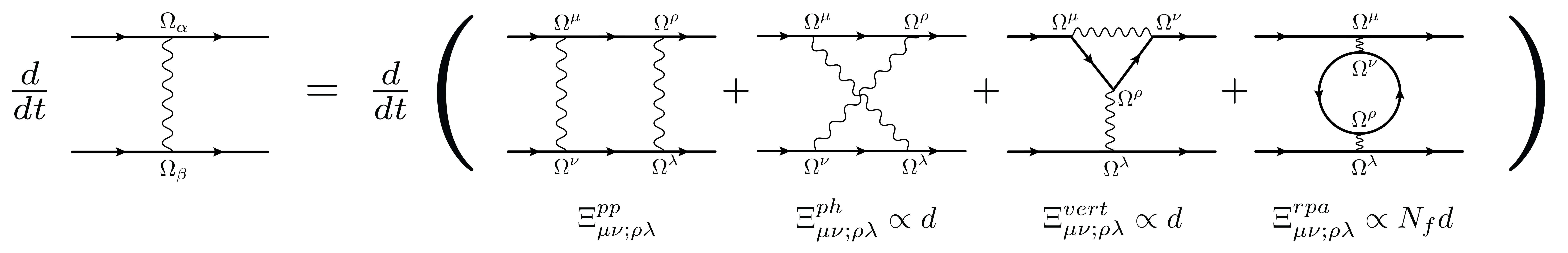}
\vspace{-0.2cm}
\caption{\textbf{Flow equation.} The approximately quadratic dispersion and particle-hole symmetry result in logarithmic divergences in each diagram. Removing particle-hole degeneracy by doping above the band-touching point weakens the latter three diagrams; we encode this effect approximately through a prefactor $d<1$. The RPA diagram $\propto N_f$ decreases after each Dirac revival.} 
\label{f:erg}
\end{figure*}
\section{Renormalisation flow equations}
\label{erg}
The field theory we derived has a number of interesting properties. The combination of the band-touching Dirac states and momentum scaling of the energy $\propto k^2$ in two dimensions results in a logarithmic enhancement of interaction effects \cite{Yang2010,Sun2009,Wang2017,Uebelacker2011,PavelCuprates}, analogous to how a linear dispersion in one dimension results in strongly interacting physics in the theory of Luttinger liquids \cite{Giamarchi2004}. 

Corrections to the interaction constants are proportional to the so-called particle-particle and particle-hole susceptibilities,
\begin{gather}
\label{susceptibilities1}
    \chi_{pp}(T) = \scaleobj{.95}{\sum_n} \scaleobj{.85}{\int} \mathcal{G}(i\omega_n,\bm{k})\mathcal{G}(-i\omega_n,-\bm{k}) \, d^2k\\
\label{susceptibilities2}
    \chi_{ph}(T) = \scaleobj{.95}{\sum_n} \scaleobj{.85}{\int} \mathcal{G}(i\omega_n,\bm{k})\mathcal{G}(i\omega_n,\bm{k})\, d^2k
\end{gather}
where $\mathcal{G}(i\omega_n,\bm{k})$ is the Matsubara Green's function,
\begin{align}
    \mathcal{G}(i\omega_n,\bm{k}) = \frac{1}{i\omega_n + \mu - i \beta \eta_z \left(k_{+}^{2} \alpha_{+}-k_{-}^{2} \alpha_{-}\right)},
\end{align}
and $\omega_n=(2n+1)\pi T$ are fermionic Matsubara frequencies. When the chemical potential is placed near the band-touching point, i.e. $\mu=0$, one finds that the scaling of the numerator $\sim d^2k$, and denominator $\sim \beta k^2$, results in $\chi_{pp}(T),\chi_{pp}(T)\rightarrow \log (\Lambda/T)/(4\pi\beta)$ as $T\rightarrow 0$ where $\Lambda$ is the UV cutoff, i.e. the corrections to the couplings diverge logarithmically. Doping away from the band-touching point via $\mu\neq 0$ weakens the divergence in $\chi_{ph}$ by removing the degeneracy of particle and hole excitations, while $\chi_{pp}$ remains logarithmically divergent.

By comparison, a linear dispersion would result in $\chi_{ph} \sim T$ as $T\rightarrow 0$, i.e. the associated corrections to the couplings would scale towards zero. In experiment, a small but finite velocity is observed; the effects of a finite velocity can be roughly incorporated as an IR cutoff on the RG flow $\Lambda_v \sim v^2/\beta$ -- as the temperature is lowered from $\Lambda$ to $\Lambda_v$, the quadratic dispersion results in a logarithmic enhancement of interactions, and for temperatures lower than $\Lambda_v$ the enhancement ceases. Hence, there exists a window of temperatures in which the RG flow is controlled by the quadratic dispersion, c.f. Fig. \ref{f:Model}.

To track the evolution of the effective couplings with temperature, we use the functional renormalisation group (fRG) method \cite{Platt2013,Metzner2012,Salmhofer2001,Polchinski1984,Wetterich1993,Kennes2019,Klebl2020}; we derive the method from a path integral treatment in the Supplementary Material. The couplings become functions of the dimensionless RG time $t = \Lambda/T$, where the values at $t=1$ are the unrenormalised values, and $t\rightarrow\infty$ describes the low temperature behaviour of the theory. We find the RG equations can be written in the simple form reflected in Fig. \ref{f:erg}; we obtain the analytic expression,
\begin{gather}
\tfrac{d}{dt} V_{\alpha\beta}\, \Omega_{\alpha} \otimes \Omega_{\beta} \nonumber \\
= V_{\mu\nu}V_{\rho\lambda} (\dot\Xi_{\mu\nu;\rho\lambda}^{pp} + \dot\Xi_{\mu\nu;\rho\lambda}^{ph} + \dot\Xi_{\mu\nu;\rho\lambda}^{rpa} + \dot\Xi_{\mu\nu;\rho\lambda}^{vert}  ) 
\label{ERG_flow}
\end{gather}
where Einstein summation is implied for indices, and the matrix-valued RG kernels $\Xi_{\mu\nu;\rho\lambda}$ correspond to the Feynman diagrams in Fig. \ref{f:erg}. The RG procedure is to take the bare interactions and evolve them according to \eqref{ERG_flow} until they grow large, resulting in a diverging susceptibility for some order parameters and a concomitant phase transition to an ordered state (see Sec. \ref{instab}). At weak coupling, this occurs as $t\rightarrow \infty$, and in this limit the fRG equations reduce to the well-known parquet equations \cite{Maiti2013,Classen2020,Schulz1987,Furukawa1998,Nandkishore2012,Wu2022,Li2022nest,Scammell2023}. The RG flow predicts a divergence of the renormalised couplings as the  flow proceeds into the deep IR, resulting in a re-emergence of strong coupling and a possible instability towards an ordered state.

The diagram $\Xi_{\mu\nu;\rho\lambda}^{pp}$ is the Cooper channel diagram $\propto \chi_{pp}$ familiar from Fermi liquid theory -- the internal lines have opposite momenta, and the diagram is proportional to the ``Cooper logarithm'' which drives the superconducting instability. The other diagrams $\Xi_{\mu\nu;\rho\lambda}^{ph}$, $\Xi_{\mu\nu;\rho\lambda}^{rpa}$, $\Xi_{\mu\nu;\rho\lambda}^{vert}$ $\propto \chi_{ph}$ are the so-called ``particle-hole'' diagrams, which diverge as a result of particle-hole degeneracy and the quadratic dispersion. As one dopes away from the band touching point, the contribution of these diagrams is weakened via a cut-off on the logarithmic divergence. We encode this effect of doping by multiplying the particle-hole diagrams by a constant $d=d(\mu)\leq 1$ which equals $1$ at the band touching point and grows smaller with increased doping away from the band touching point i.e. increasing deviation from particle-hole degeneracy -- a standard approximation in parquet RG \cite{endnote6}. Secondly, the RPA bubble diagram $\Xi_{\mu\nu;\rho\lambda}^{rpa}$ contains a fermionic trace which produces a factor $N_f$. After each Dirac revival, $N_f$ reduces by $2$, changing the renormalisation flow by weakening the RPA diagram, and altering the preferred ordered states near each integer filling.

%In summary, varying the filling $\nu$ has two key effects: first, at each successive integer filling the degeneracy $N_f$ is halved, which changes the interaction corrections. Second, doping away from integer fillings weakens the particle-hole diagrams relative to the Cooper channel via the parameter $d$, resulting in superconductivity dominating the interplay of possible ordered states. 

%while doping near a given integer filling factor determines the degeneracy $N_f$ as a result of the Dirac revival and therefore the precise competition of ordered states.

\begin{figure*}
\includegraphics[width=16cm]{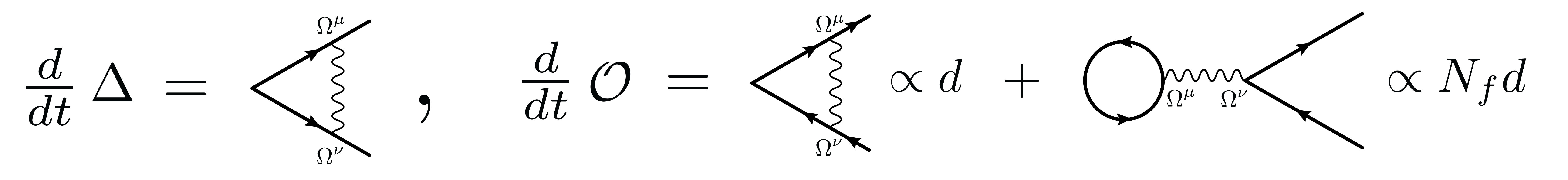}
\vspace{-0.0cm}
\caption{\textbf{Gap equations.} Diagrams contributing to the superconducting and particle-hole order parameters. The particle-hole diagrams weaken upon doping away from integer filling, i.e. decreasing $d$, while the bubble diagram $\propto N_f = 2(4-\lfloor \nu \rfloor)$. } 
\label{f:ops}
\end{figure*}

\section{Ordering instabilities of the Dirac theory}
\label{instab}

The ground state becomes unstable to an ordered phase when the associated order parameter develops a diverging susceptibility. The critical temperature for the ordered phase is given by $T_c=\Lambda/t_c$ where $t_c$ is the RG time at which the susceptibility diverges. The onset of an ordered state can be described by RG flow equations for the order parameter vertices, corresponding to the diagrams in Fig. \ref{f:ops}, and take the form
\begin{align}
\partial_t \mathcal{O}_i &= \lambda_{i}(t)  \,
\mathcal{O}_i \\
\partial_t \Delta_i &= \tilde{\lambda}_{i}(t)  \, \Delta_i
\end{align}
where $\lambda_{i}(t)$ and $\tilde{\lambda}_{i}(t)$ are henceforth referred to as `order parameter eigenvalues', and are expressions involving the renormalised couplings, as well as $\chi_{pp}$, and $\chi_{ph}$ (see the Supplementary Material). The susceptibilities for superconducting orders $\Delta_i \in \langle\psi\sigma_\mu\tau_\nu\eta_\rho \psi\rangle  $ are driven to diverge by the particle-particle diagram $\propto \chi_{pp}$ in Fig. \ref{f:ops}a, while susceptibilities for particle-hole orders $\mathcal{O}_i \in \langle \psi^\dag \sigma_\mu\tau_\nu\eta_\rho \psi \rangle$ are driven to diverge by the particle-hole diagrams $\propto \chi_{ph}$ in Fig. \ref{f:ops}b; the logarithmic divergences in these diagrams mean the $\Delta_i$ and $\mathcal{O}_i$ compete as weak coupling instabilities. The $\chi_{ph}$ in Fig. \ref{f:ops}b are proportional to $d$ and get weaker away from integer filling; decreasing $d$ suppresses the tendency towards $\mathcal{O}_i$. 

Considering now the role of Dirac revivals: First, the particle-hole diagram $\propto N_f$ changes after each Dirac revival, which has a non-trivial influence on the order parameters. Second, doping away from the band touchings at integer fillings decreases $\chi_{ph}$ via the factor $d$,  suppresses the tendency towards $\mathcal{O}_i$; in other words, near the band-touching point, fluctuations of the degenerate particle and hole states promote insulating states $\mathcal{O}_i$, while doping away from the band-touching point weakens these fluctuations allowing superconductivity $\Delta_i$ to dominate. Lastly, after each Dirac revival, the order parameters and couplings are projected onto flavour polarised bands. Denoting the projection operator onto the remaining flavours as $\mathcal{P}_f$, the order parameters transform as $\mathcal{O} \rightarrow \mathcal{P}_f\mathcal{O} \mathcal{P}_f^\dag$, $\Delta \rightarrow \mathcal{P}_f \Delta \mathcal{P}_f^T$. Since the operators $\mathcal{P}_f$ commute with the Hamiltonian, we can solve the RG equations in the unpolarised basis, then project the resulting order parameters onto the flavour polarised bands at a given filling.

To determine the leading instabilities, we employ two approaches. Firstly, at long RG times $t \rightarrow \infty$ the diverging couplings tend towards fixed constant ratios of each other referred to as {\it fixed rays} of the RG flow. All possible choices of initial coupling values flow to one of these possible sets of ratios in the deep infrared, which therefore represent universal properties of the model. At a fixed ray, the eigenvalues $\lambda_i$ ($\tilde{\lambda}_i$) which diverge sufficiently fast (see the Supplementary Material) at a given filling produce a corresponding ordered state. 

However, fixed rays are only approached at long RG times, and stronger initial couplings and/or a larger IR cutoff set by the Dirac velocity may mean that the flow is terminated by an instability before fixed ray behaviour is attained. Hence, in addition to describing the full set of fixed rays in our interacting model, a second approach is to explicitly integrate the RG equations and identify the leading diverging order parameter vertices, given some initial values of the couplings.

In the next section we will present the full set of fixed rays, and also analyse explicit solutions of the RG equations at specific filling factors.

\section{Properties of the ordered states}
\label{probes}

\subsection{Order parameters}

\renewcommand{\arraystretch}{1.2}
\begin{table*}
    \begin{center}
\caption{\textbf{Parent order parameters.} Names and transformation properties of the order parameters which appear as fixed ray solutions of the RG equations. After each Dirac revival, the parent order parameters are multiplied by the projector onto the flavour-polarised subspace $\mathcal{P}_f$, i.e. $\mathcal{O} \rightarrow \mathcal{P}_f\mathcal{O} \mathcal{P}_f^\dag$, $\Delta \rightarrow \mathcal{P}_f \Delta \mathcal{P}_f^T$. We list the irreducible representations (IRs) of $D_6$, as well as of $D_3$ for the states that break moir\'e-translational symmetry. The names of the particle-hole orders indicate their behaviour under spinful time-reversal ($\Theta$). The order parameters are grouped into intervalley coherent (IVC), flavour-polarised (polarised), density wave (DW), and nematic (N) particle-hole orders, as well as zero- ($Q=0$) and finite- momentum ($Q\neq 0$) superconductors. The right column associates the order parameter with a fixed ray eigenvalue from Table \ref{f:rays}, indicating the filling regions in which the order parameter appears.}
\label{symmetries}
\vspace{0.1cm}
\begin{ruledtabular} %p{0.25\linewidth}
 \begin{tabular} {clllllllllllllllllllll} \\[-3.5mm]
& Label & Name & Abbreviation & Order Parameter & IR of $D_6$ & IR of $D_3$ & $\lambda_j$  \\[1mm] \hline \vspace{-0.2cm} \\
\multirow{4}{*}{\rotatebox[origin=c]{90}{IVC}} 
& $O_7$ & $\Theta$-odd intervalley coherent & K-IVC & $\sigma_x\eta_y (\tau_x,\tau_y) $  & $A_2,B_2$ & $A_2$ & $\lambda_7$ \\
& $O_{7s}$  & $\Theta$-even spin-polarised intervalley coherent  & S-K-IVC & $\sigma_x\eta_y (\tau_x,\tau_y)  \bm s $  &  $A_2,B_2$ & $A_2$  & $\lambda_7$ \\ %\vspace{0.1cm}
& $O_8$ & $\Theta$-even intervalley coherent & T-IVC & $\sigma_x\eta_x (\tau_x,\tau_y)  $ & $A_1,B_1$ & $A_1$  & $\lambda_8$  \\ 
& $O_{8s}$ & $\Theta$-odd spin-polarised intervalley coherent & S-T-IVC & $\sigma_x\eta_x (\tau_x,\tau_y) \bm s $  & $A_1,B_1$ & $A_1$ & $\lambda_8$ \vspace{0.2cm} \\[1mm] \hline \vspace{-0.2cm} \\
\multirow{4}{*}{\rotatebox[origin=c]{90}{polarised}} & $O_{11}$ & $\Theta$-even/odd moir\'e-valley, sublattice polarised & MSLP$_\pm$ & $\sigma_z\eta_z (\tau_0,\tau_z)$ & $B_1,A_1$ & $A_1$ & $\lambda_{11},\lambda_8$ \\
& $O_{11s}$ & $\Theta$-odd/even spin, moir\'e-valley, sublattice polarised & S-MSLP$_\mp$ & $\sigma_z\eta_z  (\tau_0,\tau_z)\bm s$ & $B_1,A_1$ & $A_1$ & $\lambda_{8}$ \\
& $O_{12}$  & $\Theta$-even/odd sublattice polarised & SLP$_\pm$ & $\sigma_z(\tau_0 ,\tau_z)$ & $B_2,A_2$ & $A_2$ & $\lambda_7,\lambda_{12}$ \\
& $O_{12s}$ & $\Theta$-odd/even spin, sublattice polarised & S-SLP$_\mp$ & $\sigma_z (\tau_0,\tau_z)\bm s $ & $B_2,A_2$ & $A_2$ & $\lambda_{7}$ \vspace{0.2cm} \\[1mm] \hline \vspace{-0.2cm} \\
\multirow{1}{*}{\rotatebox[origin=c]{90}{DW}} 
& $O_9$ & $\Theta$-odd moir\'e density wave & MDW$_{-}$ & $(\tau_z\eta_x,\eta_y)$ & $-$ & $E$ & $\lambda_{9}$ \vspace{0.2cm} \\[1mm] \hline \vspace{-0.2cm} \\
\multirow{2}{*}{\rotatebox[origin=c]{90}{N}} 
& $O_1$ & $\Theta$-odd graphene nematic & N$_-$ & $(\tau_z\sigma_x, \sigma_y)$ & $E_1$ & $E$ & $\lambda_{1}$ \\
& $O_6$ & $\Theta$-even moir\'e-polarised graphene nematic & MPN$_+$ & $\eta_z(\sigma_x, \sigma_y\tau_z)$ & $E_2$ & $E$ & $\lambda_6$  \vspace{0.2cm} \\[1mm] \hline \vspace{-0.2cm} \\  
\multirow{4}{*}{\rotatebox[origin=c]{90}{$Q=0$\,}} 
& $\Delta_{5\tau\bar\tau}$ & $A_2$ intervalley spin-singlet & $A_2$-SSC & $\eta_z \tau_x is_y$ & $A_2$ & $A_2$ & $\tilde{\lambda}_5$ \\ %\vspace{0.1cm}
& $\Delta_{5\tau\bar\tau}^t$ & $B_2$ intervalley spin-triplet & $B_2$-TSC & $\eta_z \tau_y \boldsymbol{s} is_y$ & $B_2$ & $B_2$ & $\tilde{\lambda}_5$ \\
& $\Delta_{6\tau\bar\tau}$ & $A_1$ intervalley spin-singlet & $A_1$-SSC & $\tau_x is_y$ & $A_1$ & $A_1$ & $\tilde{\lambda}_6$\\
& $\Delta_{6\tau\bar\tau}^t$ & $B_1$ intervalley spin-triplet & $B_1$-TSC & $\tau_y \bm s is_y$ & $B_1$ & $B_1$ & $\tilde{\lambda}_6$ \vspace{0.2cm} \\[1mm] \hline \vspace{-0.2cm} \\
\multirow{8}{*}{\rotatebox[origin=c]{90}{$Q\neq0$}}
& $\Delta_{4\tau\bar\tau}$ & $E$ inter-moiré-valley spin-singlet & $E$-Q$_M$-SSC & $\sigma_z(\eta_x, \eta_y \tau_z)\tau_x is_y$ & $-$ & $E$ & $\tilde{\lambda}_4$ \\ 
& $\Delta_{4\tau\bar\tau}^t$ & $E$ inter-moiré-valley spin-triplet & $E$-Q$_M$-TSC & $\sigma_z(\eta_x \tau_z, \eta_y)\tau_x \bm s is_y$ & $-$ & $E$ & $\tilde{\lambda}_4$ \\ 
 & $\Delta_{4\tau\tau}^0 $ & $E_2$ intravalley spin-singlet & $E_2$-Q-SSC & $\sigma_x(\eta_0,\eta_z)is_y$ & $E_2$ & $E$ & $\tilde{\lambda}_4$ \\ %\vspace{0.1cm}
& $\Delta_{4\tau\tau}^z $ & $E_1$ intravalley spin-singlet & $E_1$-Q-SSC & $\sigma_x(\eta_0,\eta_z)\tau_z is_y$ & $E_1$ & $E$ & $\tilde{\lambda}_4$ \\ %\vspace{0.1cm}
& $\Delta_{5\tau\tau}^0 $ & $B_2$ intravalley spin-triplet & $B_2$-Q-TSC & $\sigma_y \eta_x \bm s is_y$ & $B_2$ & $A_2$ & $\tilde{\lambda}_5$ \\ %\vspace{0.1cm}
& $\Delta_{5\tau\tau}^z $ & $A_2$ intravalley spin-triplet & $A_2$-Q-TSC & $\sigma_y \eta_x \tau_z \bm s is_y$ & $A_2$ & $A_2$ & $\tilde{\lambda}_5$ \\ %\vspace{0.1cm}
& $\Delta_{6\tau\tau}^0 $ & $B_1$ intravalley spin-singlet & $B_1$-Q-SSC & $\sigma_y \eta_y is_y$ & $B_1$ & $A_1$ & $\tilde{\lambda}_6$ \\ %\vspace{0.1cm}
& $\Delta_{6\tau\tau}^z $ & $A_1$ intravalley spin-singlet & $A_1$-Q-TSC & $\sigma_y \eta_y \tau_z is_y$ & $A_1$ & $A_1$ & $\tilde{\lambda}_6$ \vspace{0.2cm} \\ 
 \end{tabular}
\end{ruledtabular}
\end{center}
\end{table*}

Table \ref{symmetries} contains the full set of order parameter structures which appear as a fixed ray, in a non-zero subrange of filling $0\leq \nu <4$. The order parameters are classified by the irreducible representations (irreps) of the spinless point group $D_6$, which is strictly only applicable in the unpolarised case of $0\leq\nu<1$, but straightforwardly modified at other filling ranges. 
    
The full set of parent ordered states include spin singlet and triplet T-IVC and K-IVC insulating states consisting of a gap which hybridises the two valleys -- phases which have been discussed in many prior works on TBG and multi-layer extensions \cite{Bultinck2020,Bernevig2021d,Christos2022,BernevigTrilayer,TBorNotTB,Kang2019}. The singlet K-IVC state breaks TRS -- consisting of a pattern of magnetisation currents which triple the graphene unit cell -- but preserves a modified ‘Kramers’-like TRS, consisting of TRS combined with a $U(1)$ $\tau$-rotation. By contrast, the T-IVC consists of a spatial modulation of charge which triples the graphene unit cell, but preserves TRS \cite{Calugaru2022,Hong2022}. Triplet, or `spin', order parameters also appear (S-K-IVC and S-T-IVC) with opposite behaviour under TRS. 
    
    In addition to the IVC states, RG-driven instabilities exist for moir\'e charge density waves (MDW$_-$), as well as polarised states (S-/MSLP$_\pm$ and SLP$_\pm$), which consist of Chern insulating, quantum spin Hall, and topologically trivial gaps. In experiment, multiple nearly degenerate Chern insulating states are seen near each filling factor, with the topologically non trivial states typically stabilised by a small applied magnetic field \cite{Pierce2021,Stepanov2021,Bhowmik2022,Choi2021,Nuckolls2020,Park2021,Xu2020,Xie2021,Das2021b,Serlin2019,Polski2022}. The SLP$_-$ state exhibits Chern numbers $C=4-\lfloor\nu\rfloor$, a sequence observed in experiment. A combination of IVC and moir\'e polarised order can account for the full set of observed Chern numbers; a more careful investigation of the signatures of these gaps is left for future studies. Lastly, we also find nematic states of the form $\propto \eta_{0,z}\tau_z \bm{v}\cdot \bm{\alpha}$, where $\bm{\alpha} = (\sigma_x, \tau_z\sigma_y)$, referred to as ``graphene nematicity'' in \cite{GrapheneNem1,GrapheneNem2}. These states do not open up a gap but instead split the quadratic band-touching into four Dirac points separated by the `nematic director' $\bm{v}$, spontaneously breaking threefold rotational symmetry.

    %our model provides a possible mechanism for moir\'e-translation-breaking states  

    The last column of Table \ref{symmetries} associates the order parameter with a fixed ray eigenvalue --  Table \ref{f:rays} provides the filling regions in which each eigenvalue, and therefore their associated order parameter(s), appear(s) as a leading instability. One sees from Table \ref{symmetries} that several distinct $D_6$ irreps have the same fixed ray eigenvalues, arising due to the additional symmetries of our chosen model. First, the SU(2)$_+\times$SU(2)$_-$ spin rotation symmetry results in degeneracies between `spin' and `charge' order, as has been discussed many times before \cite{You2019,Mathias2020,Kang2019,Bultinck2020,Bernevig2021d,Christos2022,BernevigTrilayer,TBorNotTB}. For instance, the T-IVC and S-T-IVC states are degenerate as they can be related by a spin-rotation in only one of the two valleys. In experiment, a small but finite inter-valley Hund's coupling \cite{Morissette2023} will split the degeneracy between these ``Hund's partners''.  Secondly, as discussed in Sec.~\ref{model}, our interacting model possesses a U(1) symmetry generated by ${\cal C}$, i.e. ${\cal U}_{\cal C}(\psi) = \exp(i\psi{\cal C}/2)$. For example, the T-IVC state and MSLP$_-$ are related by ${\cal O}_\text{T-IVC} = {\cal U}_{\cal C}(\pi/2) {\cal O}_\text{MSLP$_-$} {\cal U}_{\cal C}^\dag(\pi/2)$, and hence degenerate. %since ${\cal C}$ commutes with the interaction vertices in \eqref{interactions}. 
    This degeneracy is also lifted in a physical setting by finite subleading corrections which break particle-hole symmetry.

    Flavour polarisation is compactly treated in Table \ref{symmetries} by use of the projection operators. However, this compact notation obscures certain subtleties -- for instance, since the projection operator $\mathcal{P}_f$ can break $C_{2z}$ or TRS by imbalancing the two valleys, it is possible for the ordered states for $\nu\geq 1$, e.g. $\mathcal{P}_f \mathcal{O}_i \mathcal{P}_f^\dag$, to break time reversal or inversion symmetry even when $\mathcal{O}_i$ does not.

\subsection{$\nu=0$}
\begin{figure}[b]
\hspace{-0.55cm}
\includegraphics[width=7.80cm]{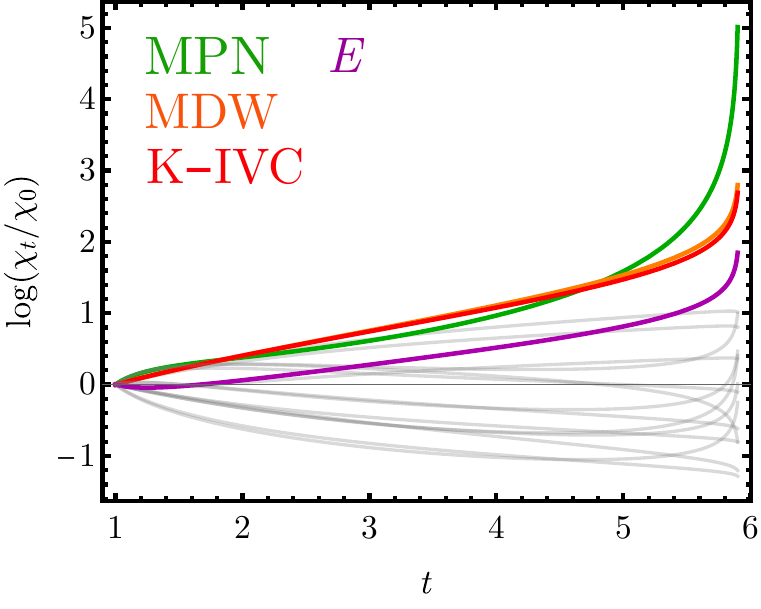}
\vspace{-0.40cm}
\caption{\textbf{RG flow near $\nu=0$.} RG flow of the quantities $\log(\chi(t)/\chi(0))$, where $\chi(t)$ is an order parameter susceptibility, for the 18 distinct order parameter structures with initial conditions $g_o = v_z = 0.25$, and $N_f=8$, $\mu=0$ as appropriate for the vicinity of $\nu=0$. Moir\'e-polarised nematic (MPN) order is the leading instability, with moir\'e density wave and K-IVC closely contending at short RG times.} 
\label{f:rg_flownu0}
\end{figure}

We begin by discussing the unpolarised case corresponding to the charge neutrality point $\nu=0$ (CNP).
Comparing with Table \ref{f:rays}, neither T- or K-IVC appear as fixed rays, i.e. weak coupling instabilities. Rather, nematic order, moir\'e density wave, and sublattice polarised order are the leading particle-hole orders. In Fig. \ref{f:rg_flownu0} we illustrate a characteristic example of an RG flow plot, demonstrating that for $g_o = v_z = 0.25$ the leading instability is moir\'e polarised nematic (MPN) order. However, at early RG times, moir\'e density wave (MDW), T- and K-IVC compete closely, so one may imagine that in the strong coupling regime -- where an instability is reached at shorter RG times -- %or for a larger Dirac velocity $v$, 
these may be candidate ground states as well.

The presence of nematic order as a candidate state explains the observation of nematic order near $\nu=0$ \cite{Jiang2019,Choi2019}, and the twofold reduction in the Landau fan degeneracy \cite{Zhang2019b}. Additionally, in recent STM studies, it was found that strained devices exhibit a gapless CNP, while very low strain devices feature a gap at the CNP \cite{Nuckolls2023}. This is quite natural in our description: a gap may be produced by a leading tendency towards K-IVC or MDW, while strain -- which couples to the nematic susceptibility -- should promote nematic order, leaving the CNP gapless.  

Interestingly, the only superconducting states which appear as fixed rays are exotic -- the finite-$Q$ pair density wave states $E$/$E_1$/$E_2$ in Table \ref{symmetries} \cite{endnote7}. Since pair density wave order is more susceptible to disorder, our prediction of this type of superconductor near $\nu=0$ is consistent with the fact that superconductivity is less commonly seen near this filling compared with the vicinity of $\nu=2$. %The observed superconducting state near $\nu=0$ was also observed to vanish in devices where screening gates were placed in proximity $\approx 7 nm$ to TBG \cite{Stepanov2020}, which generally exhibit greater disorder than those where the gates are placed further away.

\begin{table}[t!]
\begin{center}
\begin{ruledtabular}
 \begin{tabular} {lllllllllllllllllllll} \\[-3.5mm]
 Filling region & Fixed ray eigenvalues   \\[1mm] \hline \vspace{-0.2cm} \\ \vspace{0.2cm}
$0\leq \nu < 1$ & $\lambda_1, \lambda_6, \lambda_9,  \lambda_{11}, \lambda_{12}; \  \tilde{\lambda}_4$   \\   \vspace{0.2cm}
$1\leq \nu < 2$ & $\lambda_6,\lambda_7$, $\lambda_9, \lambda_{11}, \lambda_{12}; \  \tilde{\lambda}_4, \tilde{\lambda}_5,  \tilde{\lambda}_6$ \\   \vspace{0.2cm}
$2\leq \nu < 3$ & $\lambda_7,\lambda_8,\lambda_9, \lambda_{11}, \lambda_{12}; \  \tilde{\lambda}_4, \tilde{\lambda}_5, \tilde{\lambda}_6$ \\   \vspace{0.2cm}
$3\leq \nu < 4$ & $\lambda_7,\lambda_8,\lambda_9, \lambda_{11}, \lambda_{12}; \ \tilde{\lambda}_5$  \\
 \end{tabular}
\end{ruledtabular}
\end{center}
\vspace{-0.45cm}
\caption{\textbf{Fixed ray eigenvalues.} List of order parameter eigenvalues which dominate on a fixed ray, at and between each integer filling. The associated order parameters are presented in Table \ref{symmetries}, and the dependence on doping relative to integer filling is presented in the Supplementary Material.}
\label{f:rays}
\end{table}

\subsection{$\nu=2$}

%At $\nu=1,3$ the Dirac-revived bands necessarily break time-reversal symmetry, as the valleys are imbalanced at these fillings: at $\nu=3$ only a single species of valley remains, while at $\nu=1$ there is a spin sector in which a single species of valley remains. This also follows by noting that Kramers' degeneracy is incompatible with an odd number of active flavour degrees of freedom.

\begin{figure}[t!]
\includegraphics[width=7cm]{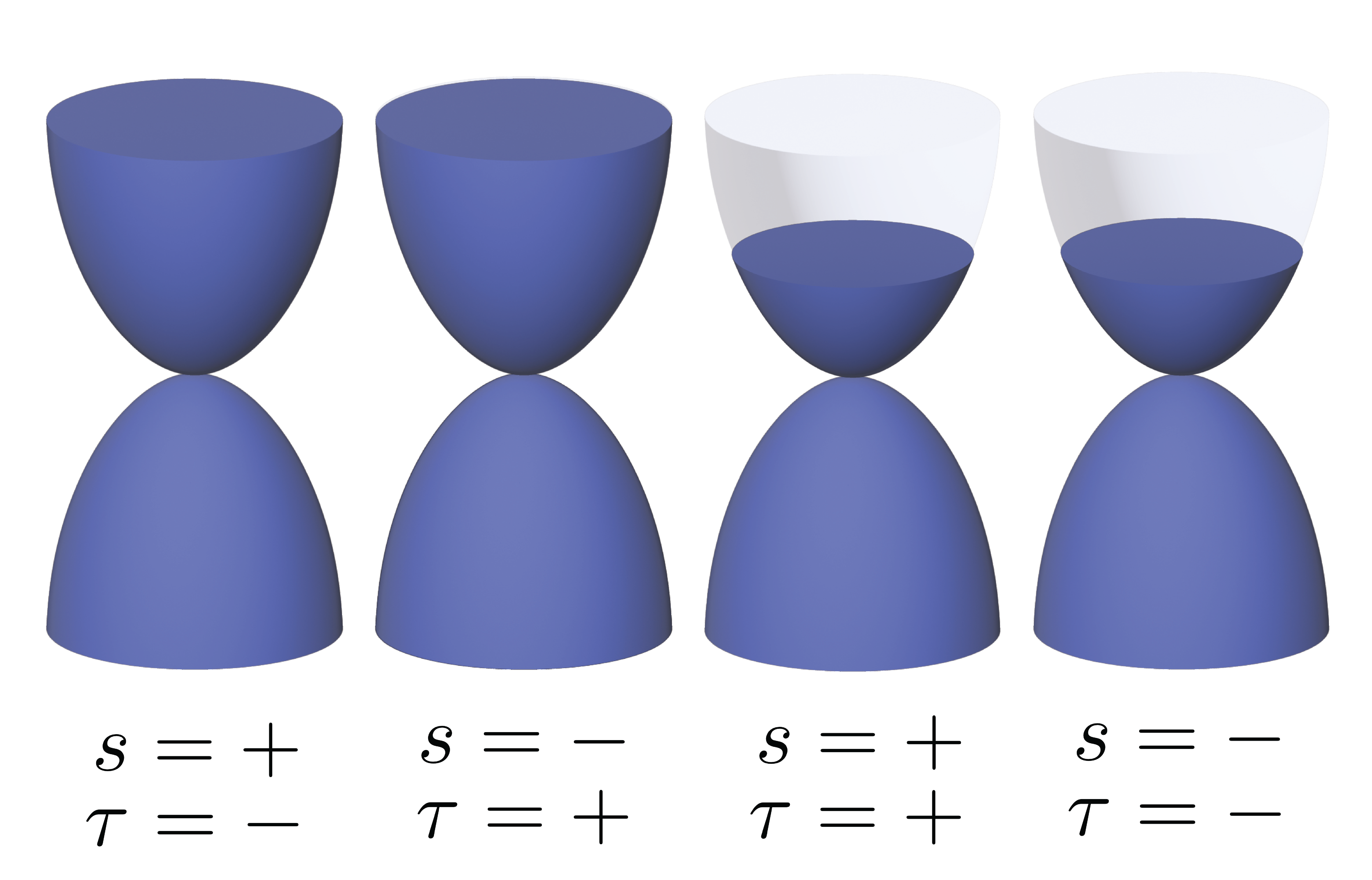}
\caption{\textbf{Flavour polarisation at $\bm{\nu=2}$.} Experimental evidence suggests the remaining flavours after the Dirac revival at $\nu=2$ consist of opposite spins at opposite valleys.} 
\label{f:nu2}
\end{figure}

At $\nu=2$, a flavour polarisation in the parent state which does not break time-reversal symmetry is possible -- namely, anti-alignment of the spins in opposite valleys, $s\tau=++,--$, as illustrated in Fig.~\ref{f:nu2}. This scenario is supported by (1) the observation of antiferromagnetic intervalley Hund's coupling in electron spin resonance \cite{Morissette2023}, and (2) the lack of hysteresis seen in unaligned TBG at $\nu=2$ \cite{endnote8}.

Assuming this spin-valley locked polarisation, the projection operator for $2\leq \nu < 3$ reads
\begin{align}
    \mathcal{P}^{\nu=2}_f = \tfrac{1}{2}(\tau_0 s_0 + \tau_z s_z). \label{ProjectorForDiracRevival}
\end{align}
We show the projected fixed-ray order parameters for $2\leq \nu < 3$ in Table \ref{nu2}, using a convenient choice of notation in which we define an `isovalley' quantum number $|\gamma=\pm\rangle \equiv |\tau=s=\pm\rangle$. The order parameters can no longer be categorised as `spin' and `charge', as spin triplet and singlet mix after projection -- which we may interpret as a consequence of broken $C_{2z}$. However, the system retains a spinful $D_6^s$; we classify the fixed ray orders by their $D_6^s$ irreps in Table \ref{nu2}.
 %(as well as $D_3^s$, which is especially important for the orders that break moir\'e translational symmetry)

\begin{figure}[b!]
\hspace{-0.55cm}
\includegraphics[width=7.80cm]{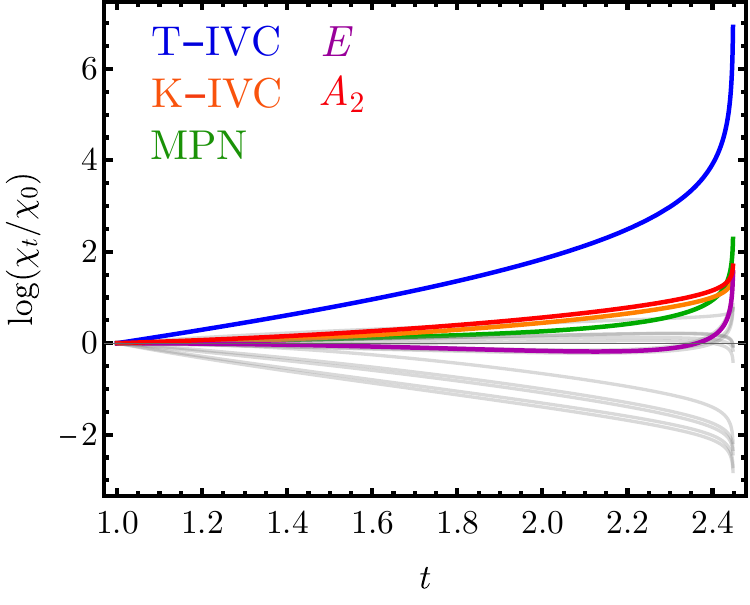}
\vspace{-0.40cm}
\caption{\textbf{RG flow near $\nu=2$.} RG flow of the quantities $\log(\chi(t)/\chi(0))$, where $\chi(t)$ is an order parameter susceptibility, for the 18 distinct order parameter structures with initial conditions $g_o=u_z=0.2$, $v_z=0.05$, $v_{x\tau}=0.12$, and $N_f=4$, $\mu=0$ as appropriate for the vicinity of $\nu=2$. T-IVC order dominates with a subleading $E_2$ superconducting instability; moir\'e-polarised nematic order and $A_2$ superconductivity appear as subleading competitors.} 
\label{f:rg_flownu2}
\end{figure}

The fixed ray analysis is `unbiased' -- it makes no assumption on the bare couplings. To complement this, we calculate an explicit RG flow which allows us to establish which order parameters are dominant given a physically motivated set of bare interaction couplings.    
Hence, in addition to the fixed rays of Table \ref{nu2}, we show an RG plot in Fig. \ref{f:rg_flownu2} for $\nu=2$, i.e. we set $N_f=4$. We demonstrate the emergence of T-IVC order for purely repulsive interactions: taking $g_o=u_z=0.2$, $v_z=0.05$, $v_{x\tau}=0.12$, i.e. positive couplings, but with $v_{x\tau}>v_z$. We find that the competition between K-IVC and T-IVC is determined by the relative magnitude of $v_{x\tau}$ and $v_z$: when these couplings are approximately equal, the IVC orders are nearly degenerate, while increasing the bare value of $v_{x\tau}$ ($v_z$) tends to promote T-IVC (K-IVC). 

As mentioned earlier, the polarised state MSLP$_-$ is degenerate with T-IVC, as a result of $\cal C$-symmetry. Subleading to T-IVC and MSLP$_-$ order are $E$ superconductivity, along with K-IVC, MPN$_+$ and $A_2$ superconductivity; these conclusions are true for a range of coupling values varied around the choice shown in Fig.~\ref{f:rg_flownu2}.

\renewcommand{\arraystretch}{1.2}
\begin{table*}
    \begin{center}
\caption{\textbf{Dominant flavour-polarised order parameters at $\bm {\nu=2}$.} We assume the time-reversal-symmetric flavour polarisation $\{\tau s\} = \{++, --\}$, with corresponding projection operator ${\cal P}=(\tau_0s_0 + \tau_zs_z)/2$. We define an {\it isovalley} quantum number $\gamma=\tau=s$ as described in the main text; working in this basis automatically implements the flavour-projection. Due to projection, for $\nu=2$ the point group and TRS have a different representation than at $\nu=0$.}
\label{nu2}
\vspace{0.1cm}
\begin{ruledtabular} %p{0.25\linewidth}
 \begin{tabular} {clllllllllllllllllllll} \\[-3.5mm]
& Name & Abbreviation & Order Parameter & IR of $D^s_6$ & IR of $D^s_3$ & $\lambda_j$  \\[1mm] \hline \vspace{-0.2cm} \\
\multirow{1}{*}{\rotatebox[origin=c]{90}{IVC}} 
& $\Theta$-odd intervalley coherent & K-IVC & $\sigma_x\eta_y (\gamma_x,\gamma_y)  $ & $B_2,A_2$ & $A_2$  & $\lambda_7$\\
& $\Theta$-even intervalley coherent & T-IVC & $\sigma_x\eta_x (\gamma_x,\gamma_y)  $ & $B_1,A_1$ & $A_1$  & $\lambda_8$ \vspace{0.2cm} \\[1mm] \hline \vspace{-0.2cm} \\
\multirow{3}{*}{\rotatebox[origin=c]{90}{polarised}} 
& $\Theta$-odd moir\'e-valley, sublattice polarised & MSLP$_-$ & $\sigma_z\eta_z  \gamma_z$ & $B_1$ & $A_1$ & $\lambda_{8}$ \\
& $\Theta$-even moir\'e-valley, sublattice polarised & MSLP$_+$ & $\sigma_z\eta_z$ & $A_1$ & $A_1$ & $\lambda_{11}$ \\
& $\Theta$-even sublattice polarised & SLP$_+$ & $\sigma_z$ & $A_2$ & $A_2$ & $\lambda_{7}$\\
& $\Theta$-odd sublattice polarised & SLP$_-$ & $\sigma_z\gamma_z$ & $B_2$ & $A_2$ & $\lambda_{12}$  \vspace{0.2cm} \\[1mm] \hline \vspace{-0.2cm} \\
\multirow{1}{*}{\rotatebox[origin=c]{90}{DW}} 
& $\Theta$-odd moir\'e density wave & MDW$_{-}$ & $(\gamma_z\eta_x,\eta_y)$ & $-$ & $E$ & $\lambda_{9}$ \vspace{0.2cm} \\[1mm] \hline \vspace{-0.2cm} \\
\multirow{2}{*}{\rotatebox[origin=c]{90}{$Q=0$\,}} 
 & $A_2$ intervalley & $A_2$-SC & $\eta_z \gamma_y$ &  $A_2$ &  $A_2$ & $\tilde{\lambda}_5$ \\ %\vspace{0.1cm}
 & $A_1$ intervalley & $A_1$-SC & $\gamma_y$ & $A_1$ & $A_1$ &  $\tilde{\lambda}_6$
%& $\Delta_{6\tau\bar\tau}^t$ & $B_1$ intervalley spin-triplet & $B_1$-TSC & ${\cal P}\tau_y s_z is_y$ & $\tilde{\lambda}_6$ \vspace{0.2cm} 
\vspace{0.2cm} \\[1mm] \hline \vspace{-0.2cm} \\
\multirow{3}{*}{\rotatebox[origin=c]{90}{$Q\neq0$}}
& $E$ \ intervalley & $E$-Q$_M$-SC & $\sigma_z(\eta_x, \eta_y \gamma_z)\gamma_y$ & $-$ & $E$ & $\tilde{\lambda}_4$ \\
& $A_2$ intravalley & $A_2$-Q-SC & $\sigma_y \eta_x$ & $A_2$ & $A_2$ & $\tilde{\lambda}_5$ \\ %\vspace{0.1cm}
& $B_2$ intravalley & $B_2$-Q-SC & $\sigma_y \eta_x \gamma_z$ & $B_2$ & $A_2$ & $\tilde{\lambda}_5$\vspace{0.2cm} \\ 
 \end{tabular}
\end{ruledtabular}
\end{center}
\end{table*}

Previous studies based on Hartree-Fock treatments and perturbation theory around the flat-band limit in twisted bilayer \cite{Bultinck2020,Bernevig2021d} and trilayer graphene \cite{Christos2022,BernevigTrilayer,TBorNotTB} have led to the conventional wisdom that Coulomb interactions should favour K-IVC over T-IVC order  -- with the T-IVC state estimated to be significantly higher in energy ($\sim 5$ meV) than K-IVC order \cite{Bultinck2020,Christos2022}. It has been proposed that phonon-mediated attractive interactions combined with spin-valley polarisation can promote T-IVC \cite{PhononsTIVC}, though T- and K-IVC remain close contenders. The RG flow enhances T-IVC, demonstrating that repulsive electronic interactions, without phonons, can allow T-IVC to dominate K-IVC order. We reiterate that recent STM studies of the spatial texture of the insulating state near $\nu=2$ see T-IVC order in low-strain devices \cite{Nuckolls2023,Kim2023} -- our results provide a natural resolution of this puzzle. 
%Since the K-IVC state does not produce a kekul\'e distortion, we also note the possibility that K-IVC could coexist with the observed T-IVC state, a scenario which would arise in our model for small $v_{x\tau}$, $v_z$ 

%It is therefore plausible that more disordered samples may produce $A_2$ superconductivity, although we note that STM has seen evidence of a kekul\'e distortion in the superconducting state \cite{Nuckolls2023}, which implies either coexisting T-IVC order or a pair density wave state with the properties of our $E_1$ and $E_2$. \change{[If we want to be really careful about this, we will have to check that there is a combination associated with a minimum of a Ginzburg-Landau expansion of the free energy where translation cannot be compensated by a U(1) gauge transformation; otherwise, you would not see any modulation in STM. My hunch is that this is indeed the case, but I haven't had the time to check it. We can also keep it vague (as is) here and study this in the follow-up, including the density modulation itself.]}

Upon doping away from the band touching point, so that $\nu\gtrsim 2$, the leading superconducting instabilities are the $E$ states, with $A_2$ appearing as a close competitor. While $E$/$E_{1,2}$ are the leading instabilities, as already noted they are finite momentum states and so may be more susceptible to disorder than $A_2$. The presence of $A_2$ superconductivity can resolve a second experimental puzzle, related to tunnelling conductance measurements \cite{Oh2021,Kim2022b,Park2021c} of the superconducting state near $\nu=2$ which show a `U-shaped' density of states attributed to a full superconducting gap, that can become `V-shaped' upon doping  -- typically attributed to gap nodes (aside from thermal fluctuations as a possible origin \cite{Fluctuations}). Since the $A_2$ state is odd under $C_{2x}$, the gap function has opposite signs at the two moir\'e valleys and vanishes at the $C_{2x}$-invariant $\Gamma M$ line in the Brillouin zone. While our theory becomes increasingly unjustified as doping is increased away from the Dirac point, if we assume that the superconducting order does not undergo a phase transition to a different irrep then the gap closes as the nodal lines of the $A_2$ state approach the Fermi level, which may explain the observed tunnelling conductance. 
Furthermore, we expect that the symmetry-imposed sign change of the $A_2$ intervalley state can also lead to subgap peaks as seen in tunnelling experiments in the strong-coupling limit \cite{Kim2022b,CruzStronCoupl}. However, we leave a quantitative analysis of this aspect to future work. 

% \change{[I would be careful about the $E$ state: since it breaks translational symmetry, its spectrum might be a bit non-trivial -- one has to compute it to be sure under which conditions it is nodal or not. Simplest solution: let's focus on $A_2$ here, where we are safe.]}

%\change{[Carefully check that this coupling is allowed by moiré translational symmetry and time-reversal]} Some experiments have observed a twofold anisotropy in the in-plane critical field for the superconducting state near $\nu=2$; another interesting property of our $E$ state predicted to be the leading instability, is that it couples nematically to an applied field, producing just such an anisotropy. %{\color{blue} The possible fragility of $E$ superconductivity is a natural explanation of the fact that some experiments observe an absence of nematicity in the superconducting state near $\nu=2$.}

\subsection{$\nu=1,3$}

We will now describe the differences between the leading order parameters observed at $\nu=1$ and $3$. 
%The symmetries of the polarised order parameters have important physical consequences -- as stated earlier, the Dirac-revived Fermi normal state at $\nu=1,3$ necessarily breaks time-reversal-symmetry due to an imbalance in the valleys.
The simplest case is the region $3\leq\nu<4$, in which the Dirac revival has necessarily polarised all but one flavour, i.e. only a single species of spin and a single species of valley remain. Letting the remaining flavour be $\{s\tau\}=++$, the associated projection operator is $\mathcal{P}_f^{\nu=3} = (s_0+s_z)(\tau_0+\tau_z)/4$. After projection there are only two insulating states which appear, $\sigma_z \mathcal{P}^{\nu=3}_f$ and $\sigma_z \eta_z\mathcal{P}^{\nu=3}_f$, which have Chern numbers $\pm 1$ and 0 respectively -- both of which have been seen in experiment near $\nu=3$ \cite{Pierce2021}. Any possible superconducting state is necessarily intra-flavour; the possible fixed-ray superconducting states significantly reduce to $\Delta_{5,\tau\tau}=(\sigma_y\eta_x)\mathcal{P}^{\nu=3}_f$ , i.e. an $A_2$ $Q \neq 0$ superconductor. To the best of our knowledge, superconductivity has been observed near $\nu=0,1,2$ but not $\nu=3$ -- consistent with the theoretical fragility of this state, though we speculate it may be observable in low-disorder devices \cite{endnote9}.

Near $\nu=1$, the projection operator may, without loss of generality, be written as $\mathcal{P}_f^{\nu=1} = s_0\tau_0 - (s_0-s_z)(\tau_0+\tau_z)/4$, i.e. we project out the flavour $\{s\tau\}=+-$. The formation of an IVC state at $\nu\gtrsim 1$ leaves one flavour ungapped, which means a fully gapped state near either $\nu=1,3$ requires an order parameter $\propto \sigma_z$. We speculate that the states $\propto \sigma_z$ are more fragile to disorder, as experimental studies have seen a gap at $\nu=1,3$ mainly in low-disorder devices; further, many studies have found these insulating states do not appear in transport, but do appear in local compressibility measurements -- suggesting the formation of local regions in which the insulating state forms, but which are shorted due to disorder-induced conductive channels \cite{Pierce2021}.

%Scanning SOT measurements near $\nu=1$ have explicitly imaged the formation of insulating states in local regions of opposite Chern numbers \cite{Grover2022}, broken up by disorder -- referred to as a `Chern mosaic' -- and an applied field may align these local regions akin to hystersis in a ferromagnet. 

%imagining orbital ferromag paper: To date, quantum anomalous Hall effects have been observed at band fillings \nu = 1 and \nu = 3 in various heterostructures (4–6)

\section{Wess-Zumino-Witten terms}
\label{wzw}

Having described the ordering instabilities of the Dirac theory, we now consider the interplay of the insulating and superconducting states. A direct second-order transition between certain insulating and superconducting states is possible when the Landau-Ginzburg free energy possesses a so-called Wess-Zumino-Witten (WZW) term -- a scenario which has been recently discussed in several studies on graphene-based systems \cite{GroverSenthil,Shubhayu,Khalaf2020sciadv,Christos2020,Shubhayu2,Yveskyrm}.

The presence of this term results in the skyrmion defects of a three-component particle-hole order parameter $\vec{m}$ carrying charge $2 \mathcal{N} e$ with $\mathcal{N} \in \mathbb{N}$, so that the proliferation of vortices -- and the associated destruction of the particle-hole order -- leads to superconductivity \cite{GroverSenthil}. Conversely, superconducting vortices carry the quantum numbers of the associated particle-hole order. We emphasise that our mechanism for superconductivity is a Fermi liquid instability via the RG-enhanced Coulomb interaction, arising from the quadratic Dirac dispersion -- our analysis of skyrmion defects in the insulating phases will serve to demonstrate that continuous transitions between our superconducting and insulating phases are possible.

The WZW term can be written explicitly by defining $\vec{n} = (m_1,m_2,m_3, \text{Re} \Delta, \text{Im} \Delta)$, and adding an auxiliary dimension $u$ to the spacetime dimensions $(\tau,x,y,z)$:
\begin{eqnarray}
\mathcal{S}_{\text{WZW}} &=& i \frac{2 \pi \mathcal{N}}{\Omega_4} \int_0^1 du \int d \tau dx dy 
\sum_{abcde=1}^5 \epsilon_{abcde} \nonumber \\
&~&~~~~~~~~~~\times n_a \partial_u n_b \partial_\tau n_c \partial_x n_d \partial_y n_e\,, \label{SWZW}
\end{eqnarray}
where $\Omega_4 = 8 \pi^2/3$.  The general criterion for the emergence of a WZW term in TBG as well as all possible compatible choices of $\vec{m}$ and zero-momentum superconducting order parameters have been worked out in Ref.~\cite{Christos2020}, which we will next apply to our results (we present further details in the Supplemental Material).

%In a complementary view, $\mathcal{S}_{\text{WZW}}$ also attaches quanta of the particle-hole order to the vortices in the superconductor \cite{GroverSenthil}; as such the proliferation of vortices induces the corresponding particle-hole partner order.

We first note that there is no single set of particle-hole and superconducting orders among the fixed rays for the region $0\leq \nu < 1$ around CNP consistent with a WZW term. This results from the fact that there is only one possible fixed-ray eigenvalue ($\tilde{\lambda}_4$) associated with superconductivity in this filling range, see Table~\ref{f:rays}, and the associated superconducting order parameters in Table~\ref{symmetries} are all inconsistent with a WZW term. This observation might provide another reason for why superconductivity is less commonly observed for $|\nu| < 1$ in experiment. 

%the form of the possible WZW terms crucially depends on the nature of the Dirac revival \cite{Christos2020}. For concreteness, we here focus on spin-valley locking as in Table~\ref{nu2}. 
%, since the $A_2$ order parameter is the prime candidate for superconductivity for $2\leq \nu < 3$ within our theory, it can explain the nodal-to-gapped transition \cite{Oh2021,Kim2022b}, the T-IVC state can dominate the RG flow close to $\nu=2$ and has been observed in experiment \cite{Nuckolls2023}

Moving on to $\nu=2$, our RG-dominant $E$ intervalley superconductor does not allow for a skyrmion-mediated transition, however the closely competing and less fragile $A_2$ state does. Most importantly, a WZW term between the $A_2$ intervalley superconductor, the two components of T-IVC order, and SLP$_+$ is possible. This is the most plausible scenario for a skyrmion-mediated critical point within our analysis, though we note that the SLP$_+$ is not close in energy to the T-IVC state within the RG without fine-tuning of parameters. Apart from this, the only other WZW term we find is the one between the $A_1$ intervalley superconductor, K-IVC, and SLP$_+$ order -- the scenario of Ref.~\cite{Khalaf2020sciadv}.
Our conclusions are that superconducting vortices in our primary candidate superconductor, the $A_2$ intervalley state, can carry quanta of the T-IVC \cite{GroverSenthil}; they could therefore exhibit a Kekul\'e pattern similar to that seen in STM analysis of superconducting state near $\nu=2$ \cite{Nuckolls2023}. The analysis further shows that the transition from T-IVC to $A_2$ superconductivity may be second order.

\section{Discussion}
\label{disc}

We have argued that TBG near integer fillings is described by a Dirac theory as a result of the observed revivals, with the assumption that the Fermi velocity is small enough that the fermions have a quadratic dispersion in some range of momenta near the Dirac points. %; so long as $v^2/(\beta \Lambda) \ll 1$, where $\Lambda$ is an upper cutoff on the quadratic dispersion, there is a range of energy scales in which this assumption is valid. 
The result is a non trivial renormalisation group flow across an associated window of temperatures: the particle-hole fluctuations near the band touching points result in nematic and insulating states, while doping away from the band-touching results in superconductivity dominating. Our theory is able to simultaneously describe both the insulating and superconducting states, a major advantage over alternative methods such as Hartree-Fock \cite{Guinea2018,Bultinck2020,Bernevig2021d,Bernevig2021e,Kwan2021b,Liu2021c,Liu2021b,PhononsTIVC,Wagner2022,Christos2022}. We motivated our assumption of a quadratic energy regime by appealing to an approximate particle-hole symmetry TBG possesses, however a direct verification of the physical values of $v$ and $\beta$ is experimentally feasible, via measurements of the electronic compressibility \cite{shahal}.

The theory can explain a great deal of the observed phenomena in TBG. Firstly, the theory provides a unified explanation of the phase diagram of TBG throughout the entire region $-4<\nu<4$, which consists of interlaced insulating/nematic and superconducting states. The theory explains why insulating/nematic states appear near each integer filling, and why superconducting states have been observed in the absence of insulating states -- the phases have a common origin, rather than a `parent-child' relationship. Furthermore, the theory naturally accounts for a gapped CNP in low-strain devices and a gapless nematic CNP in the presence of strain, consistent with experiment, and can account for the sequence of Chern numbers associated to the insulating states near integer filling. 

Secondly, recent STM studies of the insulating states near $\nu=2$ have found evidence of T-IVC order in low-strain devices \cite{Nuckolls2023,Kim2023};. Prior mean-field treatments and strong-coupling calculations with Coulomb interactions have favoured the K-IVC state over T-IVC (see, e.g., \cite{Bultinck2020,Bernevig2021d,Christos2022,BernevigTrilayer,TBorNotTB}), but here we find that RG provides a mechanism for the appearance of T-IVC order, relying on repulsive interactions rather than a resort to phonons. The T-IVC state exhibits a spatial pattern known as a Kekul\'e distortion, and in the presence of strain can result in a spatial texture known as an `incommensurate Kekul\'e spiral' (IKS) -- a state introduced in Ref. \cite{Kwan2021}, which \cite{Nuckolls2023,Kim2023} also observed in strained devices.

Thirdly, our results suggest a resolution of another recent experimental puzzle. Tunnelling conductance measurements of the superconducting state near $|\nu|=2$ show a transition from a V-shaped density of states to a U-shaped density of states as a function of doping \cite{Oh2021,Kim2022b,Park2021c}, indicating a transition between nodal and fully-gapped superconductivity. Our prediction of $A_2$ superconductivity in the Dirac theory near $|\nu|=2$ provides a possible microscopic mechanism which naturally accounts for these features. Note that the U-shaped regime has only been reported in tTLG, however it is generally believed that this system shares the same pairing symmetry as TBG.

Fourthly, our proposed link between revivals and superconductivity can explain the asymmetry of the phase diagram between electron- $\nu>0$ and hole- $\nu<0$ doping. Experiments have observed that that the Dirac revivals appear in a different window of angles for $\nu>0$ and $\nu<0$ -- e.g.  Ref. \cite{Polski2022} observed revivals for $\nu>0$ in the range $\theta\approx 0.88^\circ-1.04^\circ$, but observed revivals for $\nu<0$ in the range $\theta\approx 0.97^\circ-1.23^\circ$ \cite{endnote10}. In our theory, the Dirac revivals create the parent state from which superconductivity emerges at low temperatures -- i.e. the quadratic momentum regime near the Dirac point -- consistent with the observed asymmetry in the superconducting phase diagram.

Fifthly, the Dirac revival picture also offers two possible explanations of why the superconducting states at $\nu=0,1,3$ appear to be less robust -- firstly that the leading superconducting orders which appear are finite momentum states fragile to disorder, and secondly that the Dirac revival does not always appear at $|\nu|=1,3$.

Finally, the theory explains why superconductivity is generally absent when TBG is aligned with an hBN substrate, which breaks $C_{2z}$ symmetry and gaps out the Dirac points, obviating the interaction physics of the band-touching point.

%Two new predictions of the theory are the appearance of pair density wave order associated to the $E$ states, and the near-degeneracy of MSLP$_-$ order with T-IVC near $|\nu|=2$; we leave the further analysis of these experimental consequences to future work. 

%Comments on Chern insulators:  The stabilising of the Chern insulating states by applied field is likely related to their susceptibility to disorder as a result of broken TRS. As has been explicitly seen in experiments, the insulator may form in local regions of opposite Chern numbers broken up by disorder -- referred to as a `Chern mosaic' -- and an applied field may align these local regions akin to hystersis in a ferromagnet. 

A host of other moir\'e systems -- including twisted multi-layer graphene and twisted transition metal dichalcogenides -- are characterised by Dirac particles near charge neutrality with flattened dispersions. Our RG results open up a possible approach to studying the interaction physics of these systems -- in fact, experiments on twisted trilayer graphene also indicate signatures of Dirac revivals at integer fillings \cite{Park2021a,Hao2021,Cao2021a,Siriviboon2021}, as well as multi-layer graphene proximitised with WSe$_2$ \cite{Zhang2021b}, so we anticipate the physics of quadratic Dirac fermions is directly relevant in these systems as well. Moreover, Bernal-stacked bilayer graphene proximitised with WSe$_2$ -- recently found to exhibit superconductivity \cite{Zhang2023} -- also possesses a flavour-polarised Fermi surface characterised by a spin-valley locking equivalent to our scenario for TBG near $|\nu|=2$. Our analysis suggests that flavour polarisation and band-touching Dirac states are the essential ingredient in the emergence of insulating and proximate superconducting states in these systems.

 \section*{Acknowledgements}
The authors thank Eva Andrei, Maine Christos, Shahal Illani, Eslam Khalaf, Yves Kwan, Ryan Lee, Kevin Nuckolls, Raquel Queiroz, and Senthil Todadri for discussions and comments on the manuscript.
M.S.S.~acknowledges funding by the European Union (ERC-2021-STG, Project 101040651---SuperCorr). Views and opinions expressed are however those of the authors only and do not necessarily reflect those of the European Union or the European Research Council Executive Agency. Neither the European Union nor the granting authority can be held responsible for them.

\let\oldaddcontentsline\addcontentsline
\renewcommand{\addcontentsline}[3]{}

\widetext
%\end{document}

\newpage  \newpage

\begin{center}
\textbf{\large Supplementary Material}
\end{center}

\setcounter{equation}{0}
\setcounter{table}{0}
\setcounter{section}{0}
\setcounter{figure}{0}
\makeatletter
\renewcommand{\theequation}{S\arabic{equation}}
\renewcommand{\thetable}{S\arabic{table}}
\renewcommand{\thefigure}{S\arabic{figure}}
\renewcommand{\thesection}{S\arabic{section}}
%\renewcommand{\citenumfont}[1]{S#1}
%\begin{widetext}

\section{Symmetry analysis}

\subsection{Continuum model}
The twisted bilayer may be described via an effective continuum Hamiltonian, valid at long-wavelengths compared to the lattice spacing of graphene. Two monolayers, each with linear dispersion near the $K$-points, are coupled due to interlayer tunnelling; the Bistritzer-MacDonald (BM) Hamiltonian is given by $H=H_0+V$ where
\begin{align}
\label{contmodel}
H_0 &= \int d^2 \bm{r} \sum_\ell   \psi_\ell^\dag (\bm{r}) e^{-i\tau_z {\sigma_z \ell} \theta/2} v(\tau_z \sigma_x ( -i\partial_x) + \sigma_y (-i\partial_y))\psi_{\ell} (\bm{r}) +  \sum_{j,{\ell< \ell'}} e^{i{\tau_z}\bm{G}_j \cdot \bm{r}} \psi^\dag_\ell (\bm{r}) T^{j} \psi_{\ell'}(\bm{r}) + \text{h.c.}\\
&\equiv \int d^2 \bm{r} \sum_{\ell,\ell'} \psi_\ell^\dag \left[ h_0(\bm r) \ell_0+ h_T(\bm r)\ell_- + h_T^\dag(\bm r)\ell_+\right]_{\ell,\ell'} \psi_{\ell'} \nonumber
\end{align}
with the monolayer and tunnelling Hamiltonians,
\begin{align}
h_0(\bm r) &= e^{-i\tau_z {\sigma_z \ell_z} \theta/2} v(\tau_z \sigma_x ( -i\partial_x) + \sigma_y (-i\partial_y))\\
h_T(\bm r) &=  e^{i{\tau_z}\bm{G}_j \cdot \bm{r}} T^{j} 
\end{align}
where $(\psi^\dag_{\ell}(\bm{r}))_{\tau,\sigma,s} = \psi^\dag_{\ell,\tau,\sigma,s}$ for $\ell = \pm$ corresponding to separate layers, is an eight-component spinor with monolayer valley ($\tau$), sublattice pseudospin ($\sigma$) and spin ($s$) degrees of freedom. We have also used Pauli matrices $\ell_0$, $\ell_\pm = (\ell_x \pm i \ell_y)/2$ and $\ell_z$. The tunnelling matrices coupling the two layers $T^{j}$ are given by
\begin{align}
\label{Tmat}
T^{j}_{\sigma \sigma'} = \left(\begin{array}{cc} w_0 e^{-\frac{2\pi i}{3} j \tau_z} & w_1 e^{\frac{2\pi i}{3} j \tau_z}  \\  w_1 & w_0 e^{-\frac{2\pi i}{3} j \tau_z}\end{array}\right)_{\sigma \sigma'}
\end{align}
with tunnelling constants $w_0$ and $w_1$, index $j=\{0,\pm1\}$
and moir\'e wavevectors $\bm{G}_j=(8\pi/3)\sin(\theta/2)R_{\phi_j}(0,-1)$, with $R_{\phi_j}$ is an in-plane rotation through an angle $\phi_j=0,2\pi/3,-2\pi/3$. The Coulomb interaction is 
\begin{align}
V =\tfrac{1}{2}\int d^2 \bm{r} d^2 \bm{r}'  \sum_{\ell,\ell'} V(\bm{r}-\bm{r}')
 \psi^\dag_\ell(\bm{r}) \psi_\ell(\bm{r})  \psi^\dag_{\ell'}(\bm{r}') \psi_{\ell'}(\bm{r}')  
\end{align}
where the precise form of the screened Coulomb potential $V(\bm{r}-\bm{r}')$ shall be discussed later in this Supplement. The spatial modulation of the interlayer coupling results in the formation of a mini Brillouin zone corresponding to a triangular superlattice with spacing $L = a/(2\sin(\theta/2))$. For small $\theta$, $L\gg a$ which justifies the continuum approximation.

\subsection{Symmetries of the continuum model}

The bilayer satisfies threefold rotational symmetry in the plane, $C_{3z} \psi^\dag_{\ell,\tau,\sigma,s}(\bm{r}) C^{-1}_{3z} = e^{\frac{\pi i}{3} \tau (\sigma -\ell)} \psi^\dag_{\ell,\tau,\sigma,s}(C_{3z}^{-1}\bm{r})$, twofold rotational symmetry about the $x$ axis, $C_{2x} \psi^\dag_{\ell,\tau,\sigma,s}(x,y) C_{2x}^{-1} = \psi^\dag_{\bar\ell,\tau,\bar\sigma,s}(x,-y)$, as well as a twofold rotational symmetry in the plane $C_{2z} \psi^\dag_{\ell,\tau,\sigma,s}(\bm{r})C_{2z}^{-1} = \psi^\dag_{\ell,\bar\tau,\bar\sigma,s}(\bm{D-r})$ with $\bm{D} = (L/(\sqrt{3}),0)$. The system maintains an SU(2) spin rotational symmetry due to absence of spin-orbit coupling, which we may combine with physical time-reversal symmetry to introduce a spin-independent time-reversal operation $\Theta$ satisfying $\Theta \psi^\dag_{\ell,\tau,\sigma,s}(\bm{r})\Theta^{-1} = \psi^\dag_{\ell,\bar\tau,\sigma,s}(\bm{r})$.   In addition to the point group symmetries and time-reversal, the continuum model possesses an SU(2) $\times$ SU(2) symmetry corresponding to independent spin rotations in the two valleys. This symmetry is broken by interactions -- specifically, the weak but finite intervalley Hund's coupling $J_H$ -- but in our analysis we neglect such terms, and so the SU(2) $\times$ SU(2) symmetry is exact.

In addition to the above symmetries, which hold for the BM model as printed in \eqref{contmodel}, for the analysis in the main text we work solely in the PH symmetric limit, i.e. taking $e^{-i\tau_z {\sigma_z \ell} \theta/2}\approx 1$. For convenience we explicitly state the PH-symmetric BM model,
\begin{gather}
h(\bm r) =h_0(\bm r) \ell_0+  \sum_{j}  h_T(\bm r)\ell_- + \text{h.c.}\\
h_0(\bm r) = v(\tau_z \sigma_x ( -i\partial_x) + \sigma_y (-i\partial_y))\\
h_T(\bm r)=  e^{i{\tau_z}\bm{G}_j \cdot \bm{r}} T^{j} 
\end{gather}
Then it is straightforwardly checked that ${\cal P} \tilde{h}(\bm k){\cal P}^\dag = \tilde{h}(-\bm k)$ for PH anti-unitary operator ${\cal P}= \sigma_x \ell_y K$, where the Fourier transform of $h(\bm r)$ is denoted $\tilde{h}(\bm k)$

\subsection{Relation to the Dirac theory}
As discussed in the main text, we can index the states at the Dirac points using a basis of eight orbital states $|\sigma,\tau,\eta\rangle = |\sigma\rangle \otimes |\tau\rangle \otimes |\eta\rangle$ with $\sigma=\pm$ indexing the twofold degenerate states at the same minivalley. 
The $\eta$-index is chosen such that in the limit of decoupled layers (zero tunnelling) $e^{i\bm k\cdot \bm r}|\sigma,\tau,\eta\rangle \to e^{i\bm k\cdot \bm r}|\sigma,\tau,\ell\rangle$, where $\bm k$ is measured from the valley momentum $\bm K_{\ell}$ of the layer $\ell$. At this high symmetry point $|\sigma,\tau,\eta\rangle$ and $|\sigma,\tau,\ell\rangle$ have identical transformation properties under $D_6$ and $\Theta$.  Turning on tunnelling between layers changes the relationship between $|\sigma,\tau,\eta\rangle$ and $|\sigma,\tau,\ell\rangle$, but since $D_6$ and $\Theta$ remain symmetries of the model, it does not change the transformation properties of $|\sigma,\tau,\eta\rangle$. 

\subsection{Symmetry representations in the Dirac theory}
\subsubsection{$\nu=0$}
Considering only the spinless transformations, the action of the $D_6$ group is represented by the unitary operators
\begin{align}
C_{2z}:& \quad U_{C_{2z}} =  \tau_x \sigma_x\\
C_{2x}:& \quad U_{C_{2x}} =  \eta_x \sigma_x\\
C_{3z}:& \quad U_{C_{3z}} =  e^{i\pi \tau_z (\sigma_z-\eta_z)/3}
\end{align}
here $\tau_\mu, \eta_\mu, \sigma_\mu$ are Pauli matrices acting on graphene valley, moir\'e valley and sublattice indices. In addition, there are non-spatial symmetries
\begin{align}
\text{(TRS)} \quad \Theta &=  \tau_x K,\\
\text{(PHS)} \quad {\cal P} &=  \sigma_x \eta_y K, \quad ({\cal P}^2=-1)\\
\text{(Chiral)} \quad{\cal S} &=  {\cal P}\Theta = \sigma_x \eta_y \tau_x\\
 \quad{\cal C} &=  -i\sigma_z {\cal S} = \sigma_y \eta_y \tau_x.
\end{align}
where $K$ denotes complex conjugation. Note that ${\cal C}$ is an emergent, low-energy symmetry, see the discussion in Sec \ref{kphamiltonian}. For convenience, we present the transformation laws of bilinears in Table \ref{sf:sym_transf}. 

\subsubsection{$\nu=2$}
Considering transformations of the spin-valley locked system near $\nu=2$, we construct the following representation of spinful $D_6^s$ and TRS, 
\begin{align}
C_{2z}':& \quad U_{C_{2z}} =  i\gamma_x \sigma_x\\
C_{2x}':& \quad U_{C_{2x}} =  \eta_x \sigma_x\\
C_{3z}':& \quad U_{C_{3z}} =  e^{i\pi \gamma_z (\sigma_z-\eta_z)/3}\\
\Theta':& \quad \Theta =  i\gamma_y K.
\end{align}
Here $\gamma_\mu$ acts on the spin-valley locked index $\gamma=s=\tau$. We use this representation in Table \ref{nu2}.

\begin{table}[t]
\begin{center}
\caption{Transformation laws for relevant bilinears.} 
\label{sf:sym_transf}
\vspace{0.1cm}
\begin{ruledtabular}
 \begin{tabular} {lllllllllllllllllllll} \\[-3.5mm]
     & $C_{3z}$ & $C_{2x}$ & $C_{2z}$ & $\Theta$ & ${\cal P}$ & ${\cal C}$   \\[1mm] \hline \vspace{-0.2cm} \\ \vspace{0.2cm}
$q_\pm$ & \  $q_{\pm} e^{\pm \frac{2 \pi i}{3}}$ & $q_{\mp}$ & $-q_{\pm}$ & $-q_{\mp}$ & $-q_{\mp}$ & $q_{\mp}$  \\   \vspace{0.2cm}
$q_\pm^2$  & \  $q_{\pm}^{2}$ $e^{\pm \frac{4 \pi i}{3}}$ & $q_{\mp}^{2}$ & $q_{\pm}^{2}$ & $q_{\pm}^{2}$ & $q_{\pm}^{2}$ &  $q_{\pm}^{2}$  \\   \vspace{0.2cm}
$q_{\pm}^{3}$ & \ $q_{\pm}^{3}$ & $q_{\mp}^{3}$ & $-q_{\pm}^{3}$ & $-q_{\mp}^{3}$ & $-q_{\mp}^{3}$ & $q_{\mp}^{3}$ \\ \vspace{0.2cm}
$\alpha_{\pm}$ & $e^{\pm \frac{2 \pi i}{3}}$ $\alpha_{\pm}$ & $\alpha_{\mp}$ & $\alpha_{\pm}$ & $\alpha_{\mp}$ & $\alpha_{\pm}$ & $-\alpha_{\pm}$ \\ \vspace{0.2cm}
 $\tau_z$ & \ $\tau_{z}$ & $\tau_{z}$ & $-\tau_{z}$ & $-\tau_{z}$ & $\tau_{z}$ & $-\tau_{z}$  \\ \vspace{0.2cm}
$\eta_z$ & \ $\eta_{z}$ & $-\eta_{z}$ & $\eta_{z}$ & $\eta_{z}$ & $-\eta_{z}$ & $-\eta_{z}$ \\ \vspace{0.2cm}
$\sigma_z$ & $\sigma_{z}$ & $-\sigma_{z}$ & $-\sigma{z}$ & $\sigma_{z}$ & $-\sigma_{z}$ & $-\sigma_{z}$
 \end{tabular}
\end{ruledtabular}
\end{center}
\end{table}

\subsection{Effective Dirac Hamiltonian}
\label{kphamiltonian}

From the transformation properties of the $\sigma$ matrices and products of $\tau_{z}, \eta_{z}, \sigma_{z}$ in Table \ref{sf:sym_transf} we can construct the single particle Hamiltonian order by order in $q$; the result is the most general single-particle Hamiltonian permitted by symmetry for the Dirac states. Since no $q$-independent operators are invariant under all symmetries, the Hamiltonian must vanish at $q=0$. Using the constraints of $D_6$ and $\Theta$, one finds
\begin{gather}
    \mathcal{H}_0 = v\tau_z(e^{i\phi_1\eta_z}k_-\alpha_+ +e^{-i\phi_1\eta_z}k_+\alpha_- )+ \beta(e^{i\phi_2\eta_z}k_+^2\alpha_+ +e^{-i\phi_2\eta_z}k_-^2\alpha_- ) + \beta' k^2\alpha_0   \nonumber \\
+\, \gamma_1\left(e^{i\phi_3 \eta_z}k_{+}^{3}+e^{-i\phi_3 \eta_z}k_{-}^{3}\right) +\gamma_2\tau_{z}k^2\left(e^{i\phi_4 \eta_z}k_{+} \alpha_{-}+e^{-i\phi_4 \eta_z}k_{-} \alpha_{+}\right) + ...
\end{gather}
Further, the $\eta$-dependent phases can be fixed using the constraints of ${\cal S} = {\cal P}{\cal T}$; the effective Dirac Hamiltonian, up to quadratic-order in $\bm k$, is 
\begin{align}
\mathcal{H}_0 = v \tau_{z} (k_{+} \alpha_{-}+k_{-} \alpha_{+}) +i \beta \eta_z \left(k_{+}^{2} \alpha_{+}-k_{-}^{2} \alpha_{-}\right).
\end{align}
As explained in the main text, this low-energy model hosts an emergent symmetry $[{\cal C},\mathcal{H}_0]=0$, with ${\cal C}=-i\sigma_z {\cal S}$. This emergent symmetry only holds in the infrared -- the first term which breaks this ${\cal C}$-symmetry occur at cubic-order in $\bm k$, i.e.
\begin{align}
\delta \mathcal{H}_0 = \gamma \tau_{z} (k_{+}^{3} + k_{-}^{3}).
\end{align}
Hence, the effective theory close to the Dirac points acquires the approximate emergent symmetry ${\cal C}$ in addition to ${\cal S}$, $D_6$ and ${\cal T}$.

\subsection{Interacting Hamiltonian}

We now consider the Coulomb interaction. In proximity to metallic gates, the $2d$ Coulomb potential reads
\begin{align}
V(\bm{q}) = \tfrac{2\pi e^2}{\varepsilon_r q} \tanh(qD)
\end{align}
where $D$ is the distance to the adjacent gates symmetrically spaced above and below the device. The interaction at large $q$ is $\propto 1/q$ and is unscreened by gates, and saturates to $\sim 2\pi e^2 D/\varepsilon_r$ for $q<1/D$. In our analysis, we will neglect the long-range momentum dependence $\propto 1/q$ of the Coulomb interactions

We will also restrict our attention to interactions involving momenta at the Dirac points which conserve $\tau$ -- as we have discussed, monolayer valley is a good quantum number in TBG due to the large momentum transfer, relative to the moir\'e momentum scale $Q_m=(8\pi/3)\sin(\theta/2)$, required to scatter between valleys. We also neglect the momentum dependence of the interaction constants near the Dirac points; an extension of the RG analysis to include the long-range part of the Coulomb potential may be left for future work.

Using the symmetry representations of the previous sections, we may derive the symmetry-allowed form of the interactions in terms of symmetry-invariant products of fermion bilinears. For a real-valued potential which only depends on the spatial coordinates, only time-reversal-invariant bilinears are allowed \cite{Li2020} -- a restriction widely unappreciated in the literature. This condition is lifted by the RG flow. A variation of the proof in \cite{Li2020} in the presence of the approximate particle-hole symmetry shows that similarly only $\mathcal{P}\Theta$-invariant bilinears are allowed to appear -- and this criterion is not lifted by the RG flow. Since this symmetry is approximate, we shall begin by stating the full set of interactions allowed by $D_6$ and TRS, before indicating the $\mathcal{P}\Theta$-invariant subset.

The full set of interactions allowed by $D_6$ and $\Theta$ are
\begin{gather}
\label{bare_total}
V= g_o \sigma _0\tau _0\eta _0\otimes \sigma _0\tau _0\eta _0+g_x(\sigma _x\tau _0\eta _0\otimes \sigma _x\tau _0\eta _0 +\sigma _y\tau _z\eta _0\otimes \sigma _y\tau _z\eta _0 )+g_{z\tau} \sigma_z \tau_0\eta _0\otimes \sigma _z\tau_0\eta _0 \nonumber \\
 +v_o(\sigma _0\tau _0\eta _x\otimes \sigma _0\tau _0\eta _x+\sigma _0\tau _z\eta _y\otimes \sigma _0\tau _z\eta _y ) +v_x(\sigma _x\tau _0\eta _x\otimes \sigma _x\tau _0\eta _x +\sigma _x\tau _z\eta _y\otimes \sigma _x\tau _z\eta _y +\sigma _y\tau _0\eta _y\otimes \sigma _y\tau _0\eta _y +\sigma _y\tau _z\eta _x\otimes \sigma _y\tau _z\eta _x ) \nonumber \\
 +v_z(\sigma _z\tau _0\eta _x\otimes \sigma _z\tau _0\eta _x+\sigma _z\tau _z\eta _y\otimes \sigma _z\tau _z\eta _y) +iw( \sigma _0\tau _0\eta _x\otimes \sigma _z\tau _z\eta _y - \sigma _0\tau _z\eta _y\otimes \sigma _z\tau _0\eta _x+\sigma _z\tau _z\eta _y\otimes \sigma _0\tau _0\eta _x- \sigma _z\tau _0\eta _x\otimes \sigma _0\tau _z\eta _y ) \nonumber \\
 +u_o\sigma _0\tau _0\eta _z\otimes \sigma _0\tau _0\eta _z +u_x(\sigma _y\tau _z\eta _z\otimes \sigma _y\tau _z\eta _z +\sigma _x\tau _0\eta _z\otimes \sigma _x\tau _0\eta _z)+u_z\sigma _z\tau _0\eta _z\otimes \sigma _z\tau _0\eta _z
\end{gather}
along with the set of interactions generated by $\Omega^\mu\in\{\sigma _0\tau _0\eta _0,\sigma_z\tau_0\eta _z, \sigma_z\tau_0\tilde{\eta} _{\pm}\}$, where $\Omega^\mu$ is a bilinear taken from the above set. This large set of interactions is the interacting Hamiltonian one must work with when particle-hole symmetry is relaxed, and we shall analyse it further in a future publication. As stated in the main text, if we insist on only including $\mathcal{P}\Theta$-invariant bilinears, along with bilinears that commute with the emergent ${\cal C}$ which applies near the Dirac point, this large set of interactions becomes remarkably constrained, including only three couplings
\begin{align}
    V= g_o (\sigma _0\tau _0\eta _0\otimes \sigma _0\tau _0\eta _0)+v_z (\sigma _z\tau _0\tilde\eta _\pm\otimes \sigma _z\tau _0\tilde\eta _\mp )  +u_z(\sigma_z\tau_0\eta _z\otimes \sigma_z\tau _0\eta _z).
\end{align}
where $\tilde\eta_\pm = \eta_x\pm i\tau_z\eta_y$, along with the set of interactions generated by $\Omega^\mu\in\{\sigma _0\tau _0\eta _0,\sigma_z\tau_0\eta _z, \sigma_z\tau_0\tilde{\eta} _{\pm}\}$, i.e. the interacting Hamiltonian presented in the main text.

% Consistent notation would have denote $g_z$ as $g_{z\tau}$...maybe change this in a future version

%In order to control a perturbative analysis of the interactions, one may consider the limit where the Coulomb potential is strongly screened, ie $Q_m D >1$ where $Q_m = (8\pi/3)\sin(\theta/2)$ is the  moir\'e momentum scale -- in this strongly screened limit, the couplings of Eq. \eqref{V1234}, $V_{13;24} \propto D$. %This regime has been realised in experiment, where it was found that gates placed very close to the device enhanced superconductivity and suppressed the observed insulating states. 

\section{Quadratic Dirac fermions in the $\mathcal{P}$-symmetric Bistrizer-MacDondald Model}

\begin{figure}[b]
\hspace{-0.3cm}
\includegraphics[height=5.80cm]{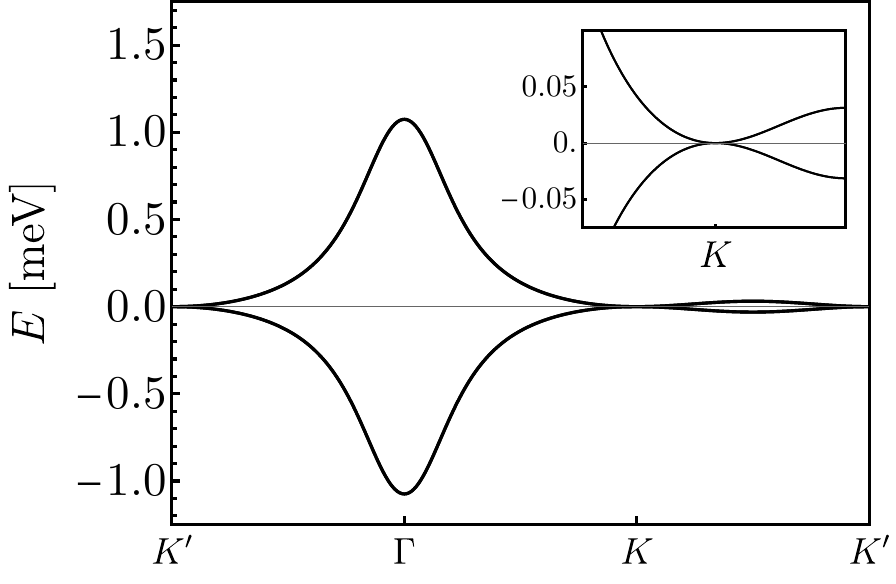} 
\raisebox{-0.7cm}{
\includegraphics[height=6.50cm]{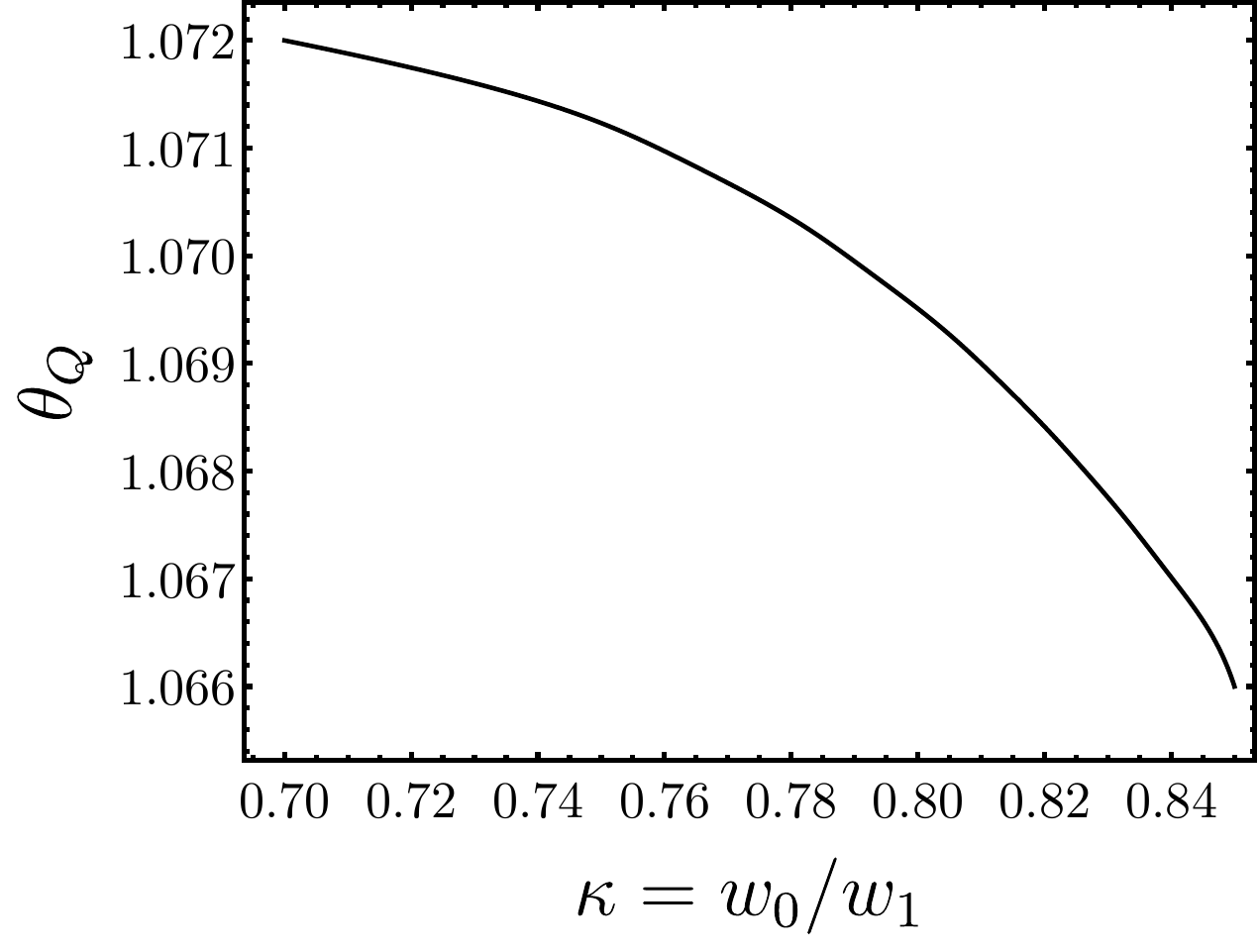}}
\caption{\textbf{Quadratic band-touching in the $\mathbf{\mathcal{P}}$-symmetric continuum model.} Left: bandstructure of the $\mathbf{\mathcal{P}}$-symmetric continuum model, with inset magnifying the dispersion near the moir\'e $K$-points, for $\theta=1.072^\circ$ and $w_0/w_1=0.74$. Right: a plot of the critical angle at which the dispersion is quadratic $\theta_Q$ against the ratio of tunnel couplings $w_0/w_1$. In both plots we have fixed $w_1=0.097$ eV. } 
\label{f:PsymBM}
\end{figure}

With the inclusion of lattice relaxation $w_0\neq w_1$, the continuum model Eq. \eqref{contmodel} does not generally possess a twist angle at which the Dirac velocity vanishes. However, it has been shown \cite{Raquel} that the presence of particle-hole symmetry $\mathcal{P}$ in a Dirac system allows the velocity to be tuned to zero with a single parameter. Motivated by this observation, in this section we impose $\mathcal{P}$ as an exact symmetry, as described earlier, and compute the bandstructure as a function of twist angle and the ratio of tunnelling amplitudes $w_0/w_1$, while setting $w_1=0.097$ eV (following Ref. \cite{Koshino2018}). We find for a range of $w_0$ and $w_1$, there always exists a twist angle $\theta_Q$ for which the dispersion is quadratic, i.e. the velocity vanishes and the second derivative of the dispersion is nonzero, see Fig. \ref{f:PsymBM}. See Ref. \cite{Raquel} for the dependence of $v$ on twist angle in the $\mathbf{\mathcal{P}}$-symmetric continuum model.

%These results motivate the naturalness of a quadratic dispersion for the Dirac fermions in TBG, as near the angle $\theta_Q$, any finite Dirac velocity arises due to small particle-hole symmetry-breaking terms.

\newpage

 \section{Analytic exact renormalisation group}

\subsection{Wetterich equation}
This appendix will describe the exact analytic RG formalism. We begin by deriving the Wetterich equation for a fermionic system \cite{Polchinski1984,Wetterich1993}. The derivation will closely follow Ref. \cite{Platt2013}. We have the generating functional
\begin{align}
W[\bar\eta,\eta] = \int D\psi D\bar\psi e^{ -S[ \bar\psi,\psi ] + \int \bar\eta\psi + \bar\psi \eta }
\end{align}
and its logarithm,
\begin{align}
\mathcal{G}[\bar\eta, \eta] = -\log \left( W[\bar\eta,\eta] \right)
\end{align}
whose derivatives evaluated at $\bar{\eta}=\eta=0$ give the connected $n$-particle Green's functions,
\begin{align}
G_{1,...,n;1',...,n'} = -\frac{1}{Z}\int D\psi D\bar\psi \, e^{ -S[ \bar\psi,\psi ]  } \ \psi_{1}...\psi_{n}\bar{\psi}_{1'}...\bar{\psi}_{n'}
\end{align}
We employ DeWitt notation, in which each of the indices $i\in \{1,...,n;1',...,n'\}$ is shorthand for all quantum numbers: frequencies $\omega_i$, momentum $\bm{k}_i$, $\sigma_i$, $\tau_i$, $\eta_i$ and spin $s_i$. Integration over momentum and frequencies, as well as summation over pseudospin etc., all appear as summation over DeWitt indices. We shall also employ Einstein summation notation throughout. We seek to derive an exact flow equation for the generating functional, allowing us to circumvent performing the path integral. We write the action,
\begin{align}
S[ \bar\psi,\psi ] = -\bar{\psi}_a Q_{ab}\psi_b +  U_{1234} \bar{\psi}_{1}\bar{\psi}_{2}\psi_{3}\psi_{4}
\end{align}
and we make the replacement $Q_{ab}\rightarrow Q_{ab}^\Lambda$ where $Q_{ab}^\Lambda\rightarrow \infty$ as $\Lambda\rightarrow \infty$ and $Q_{ab}^\Lambda\rightarrow  Q_{ab} = G^{-1}$ the inverse Green's function as $\Lambda \rightarrow 0$. Accordingly, we obtain a cutoff-dependent generating functional $\mathcal{G}^\Lambda$ with the boundary conditions 
\begin{align}
\mathcal{G}^\Lambda= \begin{cases}
  \mathcal{G}  & \text{for} \ \Lambda \rightarrow 0 \\
  0 & \text{for} \ \Lambda \rightarrow \infty
\end{cases}
\end{align}
Note that in $S[\bar\psi,\psi]$, only the Green's function $Q^\Lambda$ contains any $\Lambda$ dependence, but that is adequate to regularise the theory. We follow the trajectory $\mathcal{G}^\Lambda$ from the known starting point at large $\Lambda$ to the full functional at $\Lambda=0$. We determine the derivative $\partial_\Lambda \mathcal{G}^\Lambda$ and consider the extrapolation to $\Lambda \rightarrow 0$ as an initial value problem. To do this we replace the functional integration with a formally exact differential equation: 
\begin{align}
\partial_\Lambda \mathcal{G}^\Lambda[\bar\eta, \eta] &= - e^{\mathcal{G}^\Lambda[\bar\eta, \eta] } \partial_\Lambda  e^{-\mathcal{G}^\Lambda[\bar\eta, \eta] } \nonumber \\
&= - e^{\mathcal{G}^\Lambda[\bar\eta, \eta] } \int D\psi D\bar\psi \, \bar{\psi}_a \partial_\Lambda Q_{ab}\psi_b\, e^{ -S[ \bar\psi,\psi ] + \bar\eta\psi + \bar\psi \eta } \nonumber \\
&= \partial_\Lambda Q_{ab} \, ( e^{\mathcal{G}^\Lambda[\bar\eta, \eta] }  \partial_{\eta_{a}}\partial_{\bar{\eta}_b} e^{-\mathcal{G}^\Lambda[\bar\eta, \eta] } ) \nonumber \\&=   \partial_\Lambda Q_{ab}(\partial_{{\eta}_a}\mathcal{G}^\Lambda[\bar\eta, \eta]\partial_{\bar{\eta}_b}\mathcal{G}^\Lambda[\bar\eta, \eta] - \partial_{{\eta}_a}\partial_{\bar{\eta}_b}\mathcal{G}^\Lambda[\bar\eta, \eta] )
\end{align}
Upon Taylor expanding the functional $\mathcal{G}^\Lambda[\bar\eta, \eta] $, the last line gives us an infinite hierarchy of differential equations for the Taylor coefficients, namely the connected Green's functions. The result is the so-called \textit{Polchinski equation}. In practice, it is more convenient to write this equation in terms of the Legendre transform of the generating functional, namely the effective action $\Gamma^\Lambda[\bar\zeta, \zeta]$,
\begin{align}
\label{legendre}
\Gamma^\Lambda[\bar\zeta, \zeta] = \mathcal{G}^\Lambda[\bar\eta, \eta]  + \bar\eta_a \zeta_a +  \bar\zeta_b \eta_b
\end{align}
where the conjugate fields are defined as 
\begin{align}
\label{conj}
\zeta_a = -\partial_{\bar\eta_a} \mathcal{G}^\Lambda,  \ \ \ \ \ \ \ \ \ \
\bar\zeta_a = \partial_{\eta_a} \mathcal{G}^\Lambda
\end{align}
The 1PI vertex functions are the $\zeta$ derivatives of the effective action, ie 
\begin{align}
V^\Lambda_{1,...,n;1',...,n'} = \frac{\partial^{(2n)}\Gamma^\Lambda[\bar\zeta, \zeta]}{\partial\bar\zeta_{1}...\partial\bar\zeta_{n}\partial\zeta_{1'}...\partial\zeta_{n'}} \Big\rvert_{\zeta_i=\bar\zeta_i=0}
\end{align}
With these definitions we construct the flow equation for $\Gamma$. The Legendre transform \eqref{legendre} implies the Grassmann Hessians of $\Gamma$ and $\mathcal{G}$ are inverses of one another:
\begin{align}
\bm{\partial }^2\Gamma^\Lambda[\bar\zeta,\zeta]=\begin{pmatrix}
\tfrac{\partial^2 \Gamma^\Lambda}{\partial\bar\zeta_{1}\partial\zeta_{1'} } & \tfrac{\partial^2 \Gamma^\Lambda}{\partial\bar\zeta_{1}\partial\bar\zeta_{1'}}\\ 
\tfrac{\partial^2 \Gamma^\Lambda}{\partial\zeta_{1}\partial\zeta_{1'} } & \tfrac{\partial^2 \Gamma^\Lambda}{\partial\zeta_{1}\partial\bar\zeta_{1'} }
\end{pmatrix} = \begin{pmatrix}
-\tfrac{\partial^2 \mathcal{G}^\Lambda}{\partial\bar\eta_{1}\partial\eta_{1'} } & \tfrac{\partial^2 \mathcal{G}^\Lambda}{\partial\bar\eta_{1}\partial\bar\eta_{1'}}\\ 
\tfrac{\partial^2 \mathcal{G}^\Lambda}{\partial\eta_{1}\partial\eta_{1'} } & -\tfrac{\partial^2 \mathcal{G}^\Lambda}{\partial\eta_{1}\partial\bar\eta_{1'} }
\end{pmatrix}^{-1}
\end{align}
Using this, as well as \eqref{conj} the flow equation follows fairly straightforwardly: 
\begin{align}
\partial_\Lambda \Gamma^\Lambda[\bar\zeta,\zeta] &= \partial_\Lambda \mathcal{G}^\Lambda[\bar\eta, \eta] \nonumber \\
&= \partial_\Lambda Q_{ab}(-\bar{\zeta}_a \zeta_{b} - [\bm{\partial }^2\Gamma^\Lambda[\bar\zeta,\zeta]^{-1}]_{11})
\end{align}
where $[]_{11}$ means the $11$ matrix element. In the literature, this is typically written in a more elegant way, as a trace over auxiliary fermionic indices,
\begin{align}
\partial_\Lambda \Gamma^\Lambda[\bar\zeta,\zeta] = -\zeta_a\partial_\Lambda Q_{ab}\bar{\zeta}_{b} -\tfrac{1}{2}\text{Tr} (\partial_\Lambda \bm{Q} \,[\bm{\partial }^2\Gamma^\Lambda[\bar\zeta,\zeta]]^{-1})
\end{align}
where $\bm{Q} = \text{diag}(Q_{ab},-Q_{ba})$.

To solve this flow equation, one introduces an ansatz by expanding $\Gamma^\Lambda$ as a series in $\zeta$ and then equating coefficients on either side. The result is an infinite set of coupled differential equations for the $(2n)$-point vertices. We start by writing $\Gamma^\Lambda[\bar\zeta,\zeta] $ as a sum over powers of the fermion fields,
\begin{align}
\Gamma^\Lambda[\bar\zeta,\zeta] = \sum_{m=0}^\infty  \mathcal{A}^{(2m)\Lambda}[\bar\zeta,\zeta]
\end{align}
where
\begin{align}
\mathcal{A}^{(2m)\Lambda}[\bar\zeta,\zeta] = \tfrac{(-1)^m}{(m!)^2}\,  V^{(2m)\Lambda}_{1',...,m';1,...,m}\, \bar{\zeta}_{1'} ...\bar{\zeta}_{m'}{\zeta}_{1} ...{\zeta}_{m}
\end{align}
We then split $\Gamma^\Lambda$ up into the two-point term and all the higher point functions represented by $\mathcal{U}^{(2m)\Lambda}[\bar\zeta,\zeta]$
\begin{align}
\mathcal{U}^{(2m)\Lambda}[\bar\zeta,\zeta] &= \bm{\partial }^2\Gamma^\Lambda[\bar\zeta,\zeta]\bigr\rvert_{\zeta=\bar{\zeta}=0} - \bm{\partial }^2\Gamma^\Lambda[\bar\zeta,\zeta] \nonumber \\
&= -\sum_{m=2}^\infty  \bm{\partial }^2\mathcal{A}^{(2m)\Lambda}[\bar\zeta,\zeta]
\end{align}
Then using 
\begin{align}
(\bm{\partial }^2\Gamma^\Lambda[\bar\zeta,\zeta]\bigr\rvert_{\zeta=\bar{\zeta}=0})_{ab} = G_{ab}
\end{align}
and the matrix identity
\begin{align}
    (A-B)^{-1} = (1 + A^{-1}B + (A^{-1}B)^2 + ...) A^{-1}
\end{align}
substitution into the Wetterich equation and equating terms on either side with the same number of fields gives the coupled set of equations
 \begin{align}\label{eq:flowgrandpot}
  \partial_\Lambda\mathcal{A}^{(0)\Lambda} =& 
  - \text{tr}\left(\dot{Q}^{\Lambda}G^{\Lambda}\right)\\ 
  \partial_\Lambda\mathcal{A}^{(2)\Lambda} =& 
  - \tfrac{1}{2}\text{tr}\left(\bm{S}^{\Lambda}\bm{\partial}^2\mathcal{A}^{(4)\Lambda}\right) -\left(\overline{\zeta},\dot{Q}^{\Lambda}\zeta\right)\\ 
  \partial_\Lambda\mathcal{A}^{(4)\Lambda} =& - \tfrac{1}{2}\text{tr}\left(\bm{S}^{\Lambda}\bm{\partial}^2\mathcal{A}^{(6)\Lambda}\right) 
  \label{fourpoint}
+\tfrac{1}{2}\text{tr}\left(\bm{S}^{\Lambda}\bm{\partial}^2\mathcal{A}^{(4)\Lambda}\bm{G}^{\Lambda}\bm{\partial}^2\mathcal{A}^{(4)\Lambda}\right)\\ 
 \partial_\Lambda\mathcal{A}^{(6)\Lambda}  =& -\tfrac{1}{2}\text{tr}\left(\bm{S}^{\Lambda}\bm{\partial}^2\mathcal{A}^{(8)\Lambda}\right)+\ldots\quad
   \end{align}
 where we have defined the so-called single scale propagator \textbf{$S^\Lambda = - G^\Lambda \dot{Q} G^\Lambda = \dot{G}$}. The result is an infinite hierarchy of flow equations for each of $2m$-point functions; for e.g. the equation for $\mathcal{A}^{(2m)\Lambda}$ always contains a contribution from $\mathcal{A}^{(2m+2)\Lambda}$ (which may be interpreted as a tadpole diagram). We note that the method we shall use below -- of conveniently representing the irreducible vertices in terms of tensor products and evaluating the right hand side as matrix products -- is a fairly general scheme, and could in general allow an efficient evaluation of the flow equations for higher-point functions. However, for simplicity, we shall truncate this set of coupled equations and exclusively consider the RG evolution of the two-particle interaction $\mathcal{A}^{(4) \Lambda}$, ie neglect $\mathcal{A}^{(2m) \Lambda}$ for $m>2$. We also neglect self-energy feedback, ie we neglect the flow of $\mathcal{A}^{(2) \Lambda}$ -- an approximation which is justified in the limit where the interactions are momentum independent. In this approximation, the four-point vertex has exactly the same tensor form as the nine symmetry allowed interactions discussed previously. We proceed to obtain flow equations for these couplings.

\subsection{Flow equation for the four-point vertex}
Hence, we focus our attention on \eqref{fourpoint}, neglecting $\mathcal{A}^{(6) \Lambda}$,
\begin{align}
  \partial_\Lambda\mathcal{A}^{(4)\Lambda} = 
\tfrac{1}{2}\text{tr}\left(\bm{S}^{\Lambda}\bm{\partial}^2\mathcal{A}^{(4)\Lambda}\bm{G}^{\Lambda}\bm{\partial}^2\mathcal{A}^{(4)\Lambda}\right)
\end{align}
Performing the trace over the auxiliary fermionic indices, and denoting $V^{(4)\Lambda}_{12;34}$ simply as $V_{12;34}$,
\begin{gather}
\dot{V}_{12;34} = \tfrac{\partial}{\partial {\zeta}_4 }\tfrac{\partial}{\partial {\zeta}_3}\tfrac{\partial}{\partial \bar{\zeta}_2 }\tfrac{\partial}{\partial \bar{\zeta}_1 } \times\nonumber \\
\left\{\dot{G}_{\mu\nu} \left(\tfrac{\partial}{\partial \bar{\zeta}_\nu }\tfrac{\partial}{\partial \bar{\zeta}_\rho }V_{ab;cd} \bar{\zeta}_{a}\bar{\zeta}_{b}{\zeta}_{c}{\zeta}_{d}\right){G}_{\rho\lambda} \left(\tfrac{\partial}{\partial \bar{\zeta}_\lambda }\tfrac{\partial}{\partial \bar{\zeta}_\mu }V_{\alpha\beta;\gamma\delta} \bar{\zeta}_{\alpha}\bar{\zeta}_{\beta}{\zeta}_{\gamma}{\zeta}_{\delta} \right)        -         \dot{G}_{\mu\nu} \left(\tfrac{\partial}{\partial \bar{\zeta}_\nu }\tfrac{\partial}{\partial \bar{\zeta}_\rho }V_{ab;cd} \bar{\zeta}_{a}\bar{\zeta}_{b}{\zeta}_{c}{\zeta}_{d}\right){G}_{\rho\lambda} \left(\tfrac{\partial}{\partial \bar{\zeta}_\lambda }\tfrac{\partial}{\partial \bar{\zeta}_\mu }V_{\alpha\beta;\gamma\delta} \bar{\zeta}_{\alpha}\bar{\zeta}_{\beta}{\zeta}_{\gamma}{\zeta}_{\delta} \right)        \right.\nonumber \\
\left.-\dot{G}_{\mu\nu} \left(\tfrac{\partial}{\partial \bar{\zeta}_\nu }\tfrac{\partial}{\partial \bar{\zeta}_\rho }V_{ab;cd} \bar{\zeta}_{a}\bar{\zeta}_{b}{\zeta}_{c}{\zeta}_{d}\right){G}_{\rho\lambda} \left(\tfrac{\partial}{\partial \bar{\zeta}_\lambda }\tfrac{\partial}{\partial \bar{\zeta}_\mu }V_{\alpha\beta;\gamma\delta} \bar{\zeta}_{\alpha}\bar{\zeta}_{\beta}{\zeta}_{\gamma}{\zeta}_{\delta} \right)          +            \dot{G}_{\mu\nu} \left(\tfrac{\partial}{\partial \bar{\zeta}_\nu }\tfrac{\partial}{\partial \bar{\zeta}_\rho }V_{ab;cd} \bar{\zeta}_{a}\bar{\zeta}_{b}{\zeta}_{c}{\zeta}_{d}\right){G}_{\rho\lambda} \left(\tfrac{\partial}{\partial \bar{\zeta}_\lambda }\tfrac{\partial}{\partial \bar{\zeta}_\mu }V_{\alpha\beta;\gamma\delta} \bar{\zeta}_{\alpha}\bar{\zeta}_{\beta}{\zeta}_{\gamma}{\zeta}_{\delta} \right)\right\} \nonumber 
\end{gather}
Now performing the Grassmann algebra explicitly, we arrive at the analytic formula for the four-point vertex:
\begin{gather}
\dot{V}_{12;34} =\dot{G}_{\mu\nu}G_{\rho\lambda} V_{ab;cd} V_{\alpha\beta;\gamma\delta}\left(\left(\delta _{{a1}} \delta _{{b\nu }}-\delta
   _{{a\nu }} \delta _{{b1}}\right) \left(\delta _{\alpha
   \lambda } \delta _{{\beta 2}}-\delta _{{\alpha 2}}
   \delta _{\beta \lambda }\right)+\left(\delta _{\alpha \lambda }
   \delta _{{\beta 1}}-\delta _{{\alpha 1}} \delta _{\beta
   \lambda }\right) \left(\delta _{{a\nu }} \delta
   _{{b2}}-\delta _{{a2}} \delta _{{b\nu
   }}\right)\right) \nonumber \\
   \times \left(\left(\delta _{{c3}} \delta
   _{{d\rho }}-\delta _{{c\rho }} \delta
   _{{d3}}\right) \left(\delta _{\gamma \mu } \delta
   _{{\delta 4}}-\delta _{{\gamma 4}} \delta _{\delta \mu
   }\right)+\left(\delta _{\gamma \mu } \delta _{{\delta
   3}}-\delta _{{\gamma 3}} \delta _{\delta \mu }\right)
   \left(\delta _{{c\rho }} \delta _{{d4}}-\delta
   _{{c4}} \delta _{{d\rho }}\right)\right) \nonumber \\
   -\left( \mu \leftrightarrow \nu \right)-\left( \rho \leftrightarrow \lambda \right) + \left( \mu \leftrightarrow \nu, \  \rho \leftrightarrow \lambda \right)
\end{gather}
Expanding the product, the full expression can be separated into four contributions: the particle-particle and particle-hole ladders, the vertex corrections, and the RPA bubbles, $\dot V_{12;34}=\dot\Xi^{pp} + \dot\Xi^{ph} + \dot\Xi^{rpa} + \dot\Xi^{vert}$, represented diagrammatically in Fig. \ref{f:erg}. Explicitly, the RPA contribution is given by
\begin{gather}
\dot\Xi^{rpa}=\dot{G}_{\mu\nu}G_{\rho\lambda} V_{ab;cd} V_{\alpha\beta;\gamma\delta}\{\left(\delta _{{\alpha 1}} \delta _{{a2}}-\delta _{{a1}}
   \delta _{{\alpha 2}}\right) \left(\delta _{{b\nu }}
   \delta _{\delta \mu }-\delta _{{b\mu }} \delta _{\delta \nu
   }\right) \left(\delta _{{\gamma 3}} \delta _{{c4}}-\delta
   _{{c3}} \delta _{{\gamma 4}}\right) \left(\delta _{\beta
   \lambda } \delta _{{d\rho }}-\delta _{\beta \rho } \delta
   _{{d\lambda }}\right)\nonumber \\ +\left(\delta _{{a2}} \delta
   _{{\beta 1}}-\delta _{{a1}} \delta _{{\beta
   2}}\right) \left(\delta _{{b\nu }} \delta _{\gamma \mu
   }-\delta _{{b\mu }} \delta _{\gamma \nu }\right) \left(\delta
   _{{c4}} \delta _{{\delta 3}}-\delta _{{c3}} \delta
   _{{\delta 4}}\right) \left(\delta _{\alpha \lambda } \delta
   _{{d\rho }}-\delta _{\alpha \rho } \delta _{{d\lambda
   }}\right)\nonumber \\ +\left(\delta _{{a\nu }} \delta _{\delta \mu }-\delta
   _{{a\mu }} \delta _{\delta \nu }\right) \left(\delta
   _{{\alpha 2}} \delta _{{b1}}-\delta _{{\alpha 1}}
   \delta _{{b2}}\right) \left(\delta _{\beta \lambda } \delta
   _{{c\rho }}-\delta _{\beta \rho } \delta _{{c\lambda
   }}\right) \left(\delta _{{\gamma 4}} \delta
   _{{d3}}-\delta _{{\gamma 3}} \delta
   _{{d4}}\right)\nonumber \\ +\left(\delta _{{a\nu }} \delta _{\gamma \mu
   }-\delta _{{a\mu }} \delta _{\gamma \nu }\right) \left(\delta
   _{{\beta 1}} \delta _{{b2}}-\delta _{{b1}} \delta
   _{{\beta 2}}\right) \left(\delta _{\alpha \lambda } \delta
   _{{c\rho }}-\delta _{\alpha \rho } \delta _{{c\lambda
   }}\right) \left(\delta _{{\delta 3}} \delta
   _{{d4}}-\delta _{{\delta 4}} \delta _{{d3}}\right)\}
\end{gather}
The contribution of vertex corrections is
\begin{gather}
\dot\Xi^{vert}=\dot{G}_{\mu\nu}G_{\rho\lambda} V_{ab;cd} V_{\alpha\beta;\gamma\delta}\{\left(\delta _{{a1}} \delta _{{\alpha 2}}-\delta
   _{{\alpha 1}} \delta _{{a2}}\right) \left(\delta
   _{{b\nu }} \delta _{\gamma \mu }-\delta _{{b\mu }}
   \delta _{\gamma \nu }\right) \left(\delta _{{c4}} \delta
   _{{\delta 3}}-\delta _{{c3}} \delta _{{\delta
   4}}\right) \left(\delta _{\beta \lambda } \delta _{{d\rho
   }}-\delta _{\beta \rho } \delta _{{d\lambda
   }}\right) \nonumber \\ +\left(\delta _{{a1}} \delta _{{\beta 2}}-\delta
   _{{a2}} \delta _{{\beta 1}}\right) \left(\delta
   _{{b\nu }} \delta _{\delta \mu }-\delta _{{b\mu }}
   \delta _{\delta \nu }\right) \left(\delta _{{\gamma 3}} \delta
   _{{c4}}-\delta _{{c3}} \delta _{{\gamma 4}}\right)
   \left(\delta _{\alpha \lambda } \delta _{{d\rho }}-\delta
   _{\alpha \rho } \delta _{{d\lambda }}\right) \nonumber \\ +\left(\delta
   _{{\alpha 1}} \delta _{{a2}}-\delta _{{a1}} \delta
   _{{\alpha 2}}\right) \left(\delta _{{b\nu }} \delta
   _{\delta \mu }-\delta _{{b\mu }} \delta _{\delta \nu }\right)
   \left(\delta _{\beta \lambda } \delta _{{c\rho }}-\delta
   _{\beta \rho } \delta _{{c\lambda }}\right) \left(\delta
   _{{\gamma 4}} \delta _{{d3}}-\delta _{{\gamma 3}}
   \delta _{{d4}}\right) \nonumber \\ +\left(\delta _{{a1}} \delta
   _{{\beta 2}}-\delta _{{a2}} \delta _{{\beta
   1}}\right) \left(\delta _{{b\nu }} \delta _{\gamma \mu
   }-\delta _{{b\mu }} \delta _{\gamma \nu }\right) \left(\delta
   _{\alpha \lambda } \delta _{{c\rho }}-\delta _{\alpha \rho }
   \delta _{{c\lambda }}\right) \left(\delta _{{\delta 3}}
   \delta _{{d4}}-\delta _{{\delta 4}} \delta
   _{{d3}}\right) \nonumber \\ +\left(\delta _{{a\nu }} \delta _{\delta \mu
   }-\delta _{{a\mu }} \delta _{\delta \nu }\right) \left(\delta
   _{{\alpha 1}} \delta _{{b2}}-\delta _{{\alpha 2}}
   \delta _{{b1}}\right) \left(\delta _{{c3}} \delta
   _{{\gamma 4}}-\delta _{{\gamma 3}} \delta
   _{{c4}}\right) \left(\delta _{\beta \lambda } \delta
   _{{d\rho }}-\delta _{\beta \rho } \delta _{{d\lambda
   }}\right) \nonumber \\ +\left(\delta _{{a\nu }} \delta _{\gamma \mu }-\delta
   _{{a\mu }} \delta _{\gamma \nu }\right) \left(\delta
   _{{\beta 1}} \delta _{{b2}}-\delta _{{b1}} \delta
   _{{\beta 2}}\right) \left(\delta _{{c3}} \delta
   _{{\delta 4}}-\delta _{{c4}} \delta _{{\delta
   3}}\right) \left(\delta _{\alpha \lambda } \delta _{{d\rho
   }}-\delta _{\alpha \rho } \delta _{{d\lambda
   }}\right) \nonumber \\ +\left(\delta _{{a\nu }} \delta _{\gamma \mu }-\delta
   _{{a\mu }} \delta _{\gamma \nu }\right) \left(\delta
   _{{\alpha 1}} \delta _{{b2}}-\delta _{{\alpha 2}}
   \delta _{{b1}}\right) \left(\delta _{\beta \lambda } \delta
   _{{c\rho }}-\delta _{\beta \rho } \delta _{{c\lambda
   }}\right) \left(\delta _{{\delta 4}} \delta
   _{{d3}}-\delta _{{\delta 3}} \delta
   _{{d4}}\right) \nonumber \\ +\left(\delta _{{a\nu }} \delta _{\delta \mu
   }-\delta _{{a\mu }} \delta _{\delta \nu }\right) \left(\delta
   _{{b1}} \delta _{{\beta 2}}-\delta _{{\beta 1}}
   \delta _{{b2}}\right) \left(\delta _{\alpha \lambda } \delta
   _{{c\rho }}-\delta _{\alpha \rho } \delta _{{c\lambda
   }}\right) \left(\delta _{{\gamma 3}} \delta
   _{{d4}}-\delta _{{\gamma 4}} \delta _{{d3}}\right)\}
\end{gather}
Finally, the ladder contributions separate into the particle-hole channel:
\begin{align}
\dot\Xi^{ph}=\dot{G}_{\mu\nu}G_{\rho\lambda} V_{ab;cd} V_{\alpha\beta;\gamma\delta}\{\left(\delta _{{a\nu }} \delta _{\gamma \mu }-\delta _{{a\mu }} \delta
   _{\gamma \nu }\right) \left(\delta _{\beta \lambda } \delta _{{d\rho }}-\delta
   _{\beta \rho } \delta _{{d\lambda }}\right) \left(\delta _{{\alpha 2}}
   \delta _{{b1}} \delta _{{c3}} \delta _{{\delta 4}}+\delta
   _{{\alpha 1}} \delta _{{b2}} \delta _{{c4}} \delta _{{\delta
   3}}\right)\}
\end{align}
and the particle-particle channel:
\begin{align}
\dot\Xi^{pp}=-\dot{G}_{\mu\nu}G_{\rho\lambda} V_{ab;cd} V_{\alpha\beta;\gamma\delta}\{\left(\delta _{{a\nu }} \delta _{\gamma \mu }-\delta _{{a\mu }}
   \delta _{\gamma \nu }\right) \left(\delta _{\beta \lambda } \delta _{{d\rho
   }}-\delta _{\beta \rho } \delta _{{d\lambda }}\right) \left(\delta
   _{{\alpha 2}} \delta _{{b1}} \delta _{{c4}} \delta _{{\delta
   3}}+\delta _{{\alpha 1}} \delta _{{b2}} \delta _{{c3}} \delta
   _{{\delta 4}}\right)\}
\end{align}
Writing the interactions in the adjoint representation, $V_{ab;cd}= V_{ij} \Omega^i_{ab} \otimes \Omega^j_{cd}$, then the four contributions can be compactly represented as tensor combinations of matrix products in DeWitt space, as well as traces. %For instance, the first contribution of the RPA terms can be written as
%\begin{align}
%V_{ij}V_{kl}\left( (\Omega^i)_{24} (\Omega^k)_{13}-(\Omega^i)_{14} (\Omega^k)_{23}- (\Omega^i)_{23} (\Omega^k)_{14}+(\Omega^i)_{13} (\Omega^k)_{24}\right) \nonumber \\
%\times \left(\delta _{{B\nu }}
%   \delta _{\delta \mu }-\delta _{{B\mu }} \delta _{\delta \nu
%   }\right)  \left(\delta _{\beta
%   \lambda } \delta _{{D\rho }}-\delta _{\beta \rho } \delta
%   _{{D\lambda }}\right)\dot{G}_{\mu\nu}G_{\rho\lambda}(\Omega^j)_{BD} (\Omega^l)_{\beta \delta} \nonumber 
%\end{align}
%\begin{align}
%\left(\delta _{{B\nu }}
%   \delta _{\delta \mu }-\delta _{{B\mu }} \delta _{\delta \nu
%   }\right)  \left(\delta _{\beta
%   \lambda } \delta _{{D\rho }}-\delta _{\beta \rho } \delta
%   _{{D\lambda }}\right)\dot{G}_{\mu\nu}G_{\rho\lambda}%(\Omega^j)_{BD} (\Omega^l)_{\beta \delta} \nonumber \\
%   =\text{Tr}( \Omega^j  G  \, \Omega^l  \dot{G} ) + ...
%\end{align}
%\begin{align}
%\text{Tr}( \Omega^j  G  \, \Omega^l  \dot{G} ) + ...
%\end{align}
Suppressing DeWitt indices, the four contributions are given by
\begin{align}
\dot\Xi^{pp} &=-V_{ij}V_{kl} \ (\Omega^i G \Omega^l) \otimes (\Omega^j \dot{G} \Omega^l) + ...\\
\dot\Xi^{ph} &=V_{ij}V_{kl} \ (\Omega^i G \Omega^l) \otimes (\Omega^l \dot{G} \Omega^j)+ ...\\
\dot\Xi^{rpa} &=V_{ij}V_{kl} \ \text{Tr}[ \Omega^j  G  \, \Omega^l  \dot{G} ]\, \,\Omega^i\otimes \Omega^k+ ...\\
\dot\Xi^{vert} &=V_{ij}V_{kl} \ \{ (\Omega^i G \Omega^k \dot{G} \Omega^j) \otimes \Omega^l +\Omega^l\otimes(\Omega^i G \Omega^k \dot{G} \Omega^j)\} + ...
\end{align}
where $...$ denotes the appropriate symmetrisation/antisymmetrisations. The RG kernels defined in the main text are defined as the coefficients of the couplings appearing above, denoted $\Xi^{rpa} = V_{ij}V_{kl} \,\Xi^{rpa}_{ij;kl}$ etc.

\subsection{Evaluation of the RG kernels}
We now present explicit expressions for the four RG kernels presented above. We first choose a regulator scheme. As described in the main text, we implement a hard UV cutoff in momentum space $\Lambda$, and then implement an IR cutoff by working at finite temperature $T$; the RG flow then has the natural interpretation of lowering $T$ and tracking the evolution of the effective couplings/ground state. Starting at $T=\Lambda$ the couplings equal their bare values, and then $T\rightarrow 0$ corresponds to following the RG flow; defining a dimensionless RG time $t=\Lambda/T$, the deep IR corresponds to the long RG time limit of $t\rightarrow \infty$. We proceed by writing the Green's function in terms of the projection operator into the lower/upper band,
\begin{align}
\mathcal{G}(i\omega_n, \bm k) = \sum_{n;s=\pm} \frac{\mathcal{P}_{\bm{k},s} }{i\omega_n - s\varepsilon_k} \equiv \sum_{n;s=\pm} \mathcal{P}_{\bm{k},s}G_s(i\omega_n,  k)
\end{align}
where $\omega_n=(2n+1)\pi T$ are fermionic Matsubara frequencies, the dispersion $\varepsilon_k=\beta k^2$ and the projection operator is given by $\mathcal{P}_{\bm{k},s}= 1 + s\bm{d}(\bm{k})\cdot \bm{\alpha}$ with $d_x = \cos2\theta_{\bm k}$, $d_y=\sin2\theta_{\bm k}$, and $\bm{\alpha}= (\sigma_x\eta_z, \sigma_y\tau_z \eta_z)$. The matrix structure of the Green's function is entirely encoded in the projection operator, which only depends on the angle of $\bm k$ and not the magnitude. Hence the frequency and $k$ dependence can be extracted from each of these integrals, giving a matrix structure involving the projection operator times the frequency/momentum integral:
\begin{gather}
\label{XIPPSUPP}  
\Xi_{ij;kl}^{pp}=-\sum_{s,s'}\int d\theta  \ (\Omega^i \mathcal{P}_s \Omega^l) \otimes (\Omega^j \mathcal{P}_{s'} \Omega^l)  \ \sum_n\int dk\  G_s(i\omega_n,  k) G_{s'}(-i\omega_n,  k)+ ...\\
\label{XIPHSUPP}
\Xi_{ij;kl}^{ph}= \sum_{s,s'} \int d\theta\ (\Omega^i \mathcal{P}_s \Omega^l) \otimes (\Omega^l \mathcal{P}_{s'} \Omega^j) \sum_n\int dk\  G_s(i\omega_n,  k) G_{s'}(i\omega_n,  k)+ ...\\
\label{XIRPASUPP}
\Xi_{ij;kl}^{rpa}= \Omega^i\otimes \Omega^k   \sum_{s,s'}\int d\theta\ \text{Tr}( \Omega^j  \mathcal{P}_s  \, \Omega^l  \mathcal{P}_{s'} ) \,\sum_n\int dk\  G_s(i\omega_n,  k) G_{s'}(i\omega_n,  k)+ ...\\
\label{XIVERTSUPP}
\Xi_{ij;kl}^{vert}=  \sum_{s,s'}\int d\theta\ (\Omega^i \mathcal{P}_s \Omega^k \mathcal{P}_{s'} \Omega^j) \otimes \Omega^l \,\sum_n\int dk\  G_s(i\omega_n,  k) G_{s'}(i\omega_n,  k) + ...
\end{gather}
where ... denotes the symmetrisations/antisymmetrisations $-\left( \mu \leftrightarrow \nu \right)-\left( \rho \leftrightarrow \lambda \right) + \left( \mu \leftrightarrow \nu, \  \rho \leftrightarrow \lambda \right)$, as before. For both particle-particle and particle-hole diagrams, there are two distinct frequency/momentum integrals -- one pair of integrals is logarithmically divergent; at finite chemical potential, the logarithmic divergence in the particle-hole channel is weakened.  The other pair of integrals is not logarithmically divergent and correspond to additional contributions to the RG which come from our analytic fRG method. The additional contributions play an essential role at intermediate to strong coupling. Explicitly, these momentum/frequency integrals are given by
\begin{align}
\mathscr{L}_{pp} &\equiv \sum_{n,s} \int dk\  G_s(i\omega_n,  k) G_{s}(-i\omega_n,  k)\\
\mathscr{L}_{ph} &\equiv \sum_{n,s} \int dk\  G_s(i\omega_n,  k) G_{-s}(i\omega_n,  k)\\
\mathscr{N}_{pp} &\equiv \sum_{n,s} \int dk\  G_s(i\omega_n,  k) G_{-s}(-i\omega_n,  k)\\
\mathscr{N}_{ph}  &\equiv \sum_{n,s} \int dk\  G_s(i\omega_n,  k) G_{-s}(i\omega_n,  k)
\end{align}
Performing the Matsubara sum explicitly results in the expressions
\begin{gather}
\mathscr{L}_{pp} = \tfrac{1}{2\pi\beta}\sum_s \int^{s\Lambda}_0 \tfrac{s}{2(\varepsilon-\mu)}(n_T(\varepsilon-\mu)-n_T(\mu-\varepsilon) ) \, d\varepsilon\\
\mathscr{L}_{ph} = \tfrac{1}{2\pi\beta}\int^{\Lambda}_0 \tfrac{1}{\varepsilon}(n_T(\mu-\varepsilon)-n(\mu+\varepsilon)) \, d\varepsilon \\
\mathscr{N}_{pp} = \tfrac{1}{2\pi\beta}\sum_s\int^{s\Lambda}_0 \tfrac{s }{2\mu}(n_T(\mu-\varepsilon)-n_T(-\mu-\varepsilon) ) d\varepsilon  \\
\mathscr{N}_{ph} = \tfrac{1}{2\pi\beta}(n_T(\mu-\Lambda) - n_T(\mu+\Lambda))
\end{gather}
where $n_T(\varepsilon)$ is the Fermi-Dirac function at temperature $T$. These integrals are purely functions of the dimensionless ratio $\Lambda/T$ (and the fixed ratio between the chemical potential and cutoff $\mu/\Lambda$) and so we define a dimensionless RG time $t=\Lambda/T$ which we systematically \textit{increase}, equivalent to lowering the IR cutoff, when we differentiate with respect to the IR cutoff $T$. When $\mu=0$ note that as $T\rightarrow 0$, $\mathscr{N}\rightarrow -\nu_0$ where $\nu_0$ is the density of states at the Fermi level, while $\mathscr{L}\rightarrow \nu_0\log(\Lambda/T)$.  In this limit, the expressions are very simple, $\mathscr{L}_{pp}=\mathscr{L}_{ph}=\mathscr{L}$, $\mathscr{N}_{pp}=\mathscr{N}_{ph}=\mathscr{N}$, with
\begin{gather}
\mathscr{L}=-\tfrac{1}{2\pi\beta}\sum_{n,s} \int_0^\Lambda \tfrac{1}{i\omega_n -s\varepsilon}\tfrac{1}{-i\omega_n -s\varepsilon} \ d\varepsilon = \tfrac{1}{2\pi\beta}\int^\Lambda_0 \tfrac{1}{2\varepsilon}(n_T(\varepsilon)-n_T(-\varepsilon)) \ d\varepsilon \\
\mathscr{N}= \tfrac{1}{2\pi\beta}\sum_{n,s} \int_0^\Lambda \tfrac{1}{(i\omega_n -s\varepsilon)^2}\ d\varepsilon = \tfrac{1}{2\pi\beta}\sum_{n,s} \int_0^\Lambda\tfrac{\partial}{\partial \varepsilon} \tfrac{s}{i\omega_n -s\varepsilon} \ d\varepsilon = \tfrac{1}{2\pi\beta} \sum_{s}\int_0^\Lambda s\tfrac{\partial}{\partial \varepsilon} n_T(s\varepsilon) \ d\varepsilon = \tfrac{1}{2\pi\beta}(n_T(\Lambda)-n_T(-\Lambda))
\end{gather}
from which we obtain
\begin{align}
\dot{\mathscr{L}}(t) &= \tfrac{1}{2\pi\beta t }(n(t)-n(-t)) = -\tfrac{1}{2\pi\beta t } \tanh(\tfrac{t}{2})\\
\dot{\mathscr{N}}(t) &= \tfrac{1}{2\pi\beta}(n'(t)+n'(-t)) = -\tfrac{1}{2\pi\beta} \text{sech}^2(\tfrac{t}{2})
\end{align}
The problem remains to calculate the tensor structures in each RG kernel which arise due to performing the angular integral over the projection operators times interaction vertices. We make use of the fact that 
\begin{align}
\int d\theta  \ (\Omega^i \mathcal{P}_s \Omega^l) \otimes (\Omega^j \mathcal{P}_{s'} \Omega^l) =(\Omega^i \Omega^l) \otimes (\Omega^j \Omega^l) + s s' (\Omega^i \alpha_i \Omega^l) \otimes (\Omega^j \alpha^i\Omega^l)
\end{align}
and\begin{align}
\int d\theta\ (\Omega^i \mathcal{P}_s \Omega^k \mathcal{P}_{s'} \Omega^j) \otimes \Omega^l= (\Omega^i\Omega^k \Omega^j) \otimes \Omega^l + s s' (\Omega^i \alpha_i \Omega^k \alpha^i\Omega^j) \otimes \Omega^l
\end{align}
to evaluate the tensor structures which multiply the $\mathscr{L}$ and $\mathscr{N}$ functions. Comparing the coefficients of the tensor structures on either side of the flow equation gives us the flow equations for each coupling $V_{ij}$.

\newpage
\subsection{Flow equations for the couplings}
Explicitly, we find the  flow equations:
\begin{align}
\label{go} 
\dot{g}_o &= \dot{\mathscr{N}}_{ph} \left( -8 N_f g_o^2+8 g_o u_{o \tau }+16 g_o u_{\tau  x}+8 g_o u_z+16 g_o v_{z}+32 g_o
   v_{\tau  x}+16 g_o v_{o \tau }+20 g_o u_{x}+8 g_o g_z+10 g_o^2\right.\nonumber \\
&\left.\,\,\,\,\, +\,4 u_{x} g_z+4 u_{x}^2+2 g_z^2+2 u_{o
   \tau }^2+4 u_{o \tau } g_{x\tau}+4 v_{z}^2+8 v_{z} v_{\tau  x}+4
   g_{x\tau}^2+4 u_z g_{x\tau}+2 u_z^2+8 v_{\tau  x}^2+8 v_{o \tau } v_{\tau  x}+4 v_{o \tau }^2\right)   \nonumber \\
&+ \dot{\mathscr{N}}_{pp}  \left( 4 g_o u_{x}-2 g_o^2-4 u_{x} g_z-4 u_{x}^2-2 g_z^2-2 u_{o \tau }^2+4 u_{o \tau } g_{x\tau}-4
   v_{z}^2+8 v_{z} v_{\tau  x}-4 g_{x\tau}^2-4 u_z g_{x\tau}\right.\nonumber \\
&\left.\,\,\,\,\, -\, 2 u_z^2-8
   v_{\tau  x}^2-8 v_{o \tau } v_{\tau  x}-4 v_{o \tau }^2\right)  \nonumber
\\
&+ \dot{\mathscr{L}}_{ph} \left( -4 g_o u_{x}+2 g_o^2-4 u_{x} g_z+4 u_{x}^2+2 g_z^2+2 u_{o \tau }^2-4 u_{o \tau } g_{x\tau}+4
   v_{z}^2  -8 v_{z} v_{\tau  x}+4 g_{x\tau}^2 -4 u_z g_{x\tau}\right.\nonumber \\
&\left.\,\,\,\,\, +\,2 u_z^2+8
   v_{\tau  x}^2-8 v_{o \tau } v_{\tau  x}+4 v_{o \tau }^2 \right)   \nonumber
\\
&+ \dot{\mathscr{L}}_{pp}  \left( -4 g_o u_{x}-2 g_o^2+4 u_{x} g_z-4 u_{x}^2-2 g_z^2-2 u_{o \tau }^2-4 u_{o \tau } g_{x\tau}-4
   v_{z}^2-8 v_{z} v_{\tau  x} -4 g_{x\tau}^2 +4 u_z g_{x\tau}\right.\nonumber \\
&\left. \,\,\,\,\, -\,2 u_z^2-8
   v_{\tau  x}^2+8 v_{o \tau } v_{\tau  x}-4 v_{o \tau }^2 \right) 
\\
\label{ux}
\dot{u}_x &=  \dot{\mathscr{N}}_{ph} \left(4 u_{x} u_{o \tau }+8 u_{x} v_{z}+8 g_o u_{x}+2 g_o g_z+g_o^2-4 u_{x} u_z-8 u_{x} v_{o \tau }-4 N_f
   u_{x}^2+4 u_{x}^2+g_z^2+u_{o \tau }^2+4 u_{o \tau } g_{x\tau}+2 u_z u_{o \tau }\right.\nonumber \\
&\left.\,\,\,\,\, +\,2 v_{o
   \tau }^2+8 v_{z} v_{\tau  x}+4 v_{o \tau } v_{z}+4 g_{x\tau}^2+4 u_z u_{\tau 
   x}+u_z^2+8 v_{\tau  x}^2+8 v_{o \tau } v_{\tau  x}+2 v_{o \tau }^2\right)   \nonumber \\
&+ \dot{\mathscr{N}}_{pp}  \left( -4 g_o u_{x}-2 g_o g_z+g_o^2+4 u_{x} g_z+4 u_{x}^2+g_z^2+u_{o \tau }^2-4 u_{o \tau } u_{\tau 
   x}-2 u_z u_{o \tau }+2 v_{z}^2-8 v_{z} v_{\tau  x}-4 v_{o \tau } v_{z}+4
   g_{x\tau}^2\right.\nonumber \\
&\left.\,\,\,\,\, +\,4 u_z g_{x\tau}+u_z^2+8 v_{\tau  x}^2+8 v_{o \tau } v_{\tau  x}+2 v_{o \tau }^2\right)  \nonumber
\\
&+\dot{\mathscr{L}}_{ph} \left( 4 u_{x} u_{o \tau }+8 u_{x} v_{z}+8 g_o u_{x}-2 g_o g_z-g_o^2-4 u_{x} u_z-8 u_{x} v_{o \tau }-4 N_f
   u_{x}^2-4 u_{x}^2-g_z^2-u_{o \tau }^2+4 u_{o \tau } g_{x\tau}-2 u_z u_{o \tau }\right.\nonumber \\
&\left. \,\,\,\,\, -\,2 v_{o
   \tau }^2+8 v_{z} v_{\tau  x}-4 v_{o \tau } v_{z}-4 g_{x\tau}^2+4 u_z u_{\tau 
   x}-u_z^2-8 v_{\tau  x}^2+8 v_{o \tau } v_{\tau  x}-2 v_{o \tau }^2 \right)  \nonumber
\\
&  + \dot{\mathscr{L}}_{pp}  \left( -4 g_o u_{x}+2 g_o g_z-g_o^2+4 u_{x} g_z-4 u_{x}^2-g_z^2-u_{o \tau }^2-4 u_{o \tau } u_{\tau 
   x}+2 u_z u_{o \tau }-2 v_{z}^2-8 v_{z} v_{\tau  x}+4 v_{o \tau } v_{z}-4g_{x\tau}^2\right.\nonumber \\
&\left. \,\,\,\,\, +\,4 u_z g_{x\tau}-u_z^2-8 v_{\tau  x}^2+8 v_{o \tau } v_{\tau  x}-2 v_{o \tau }^2 \right)  
\\
\label{gz}
\dot{g}_z &=  \dot{\mathscr{N}}_{ph} \left( 4 g_o u_{x}+4 g_o g_z+4 u_{x} g_z+4 u_{x}^2+4 u_{o \tau } g_{x\tau}+4 u_z u_{o \tau }+8 v_{o
   \tau } v_{\tau  x}+8 v_{o \tau } v_{z}+4 g_{x\tau}^2+4 u_z g_{x\tau}+8 v_{\tau 
   x}^2+8 v_{o \tau } v_{\tau  x}\right)\nonumber \\
&\,\,\,\,\, +\, \dot{\mathscr{N}}_{pp}  \left( -4 g_o u_{x}-4 g_o g_z+4 u_{x} g_z+4 u_{x}^2-4 u_{o \tau } g_{x\tau}-4 u_z u_{o \tau }-8
   v_{z} v_{\tau  x}-8 v_{o \tau } v_{z}+4 g_{x\tau}^2+4 u_z g_{x\tau}+8 v_{\tau
    x}^2+8 v_{o \tau } v_{\tau  x} \right)\nonumber
\\
&+ \dot{\mathscr{L}}_{ph} \left( 8 g_z u_{o \tau }+16 g_z v_{z}-4 g_o u_{x}+12 g_o g_z-16 g_z g_{x\tau}+8 g_z
   u_z-32 g_z v_{\tau  x}+16 g_z v_{o \tau }-20 u_{x} g_z+4 u_{x}^2-8 N_f g_z^2+8 g_z^2 \right.\nonumber \\
&\left.\,\,\,\,\,  -\,4 u_{o \tau }g_{x\tau}+4 u_z u_{o \tau }-8 v_{z} v_{\tau  x}+8 v_{o \tau } v_{z}+4 u_{\tau 
   x}^2-4 u_z g_{x\tau}+8 v_{\tau  x}^2-8 v_{o \tau } v_{\tau  x} \right)  \nonumber
\\
&+ \dot{\mathscr{L}}_{pp}  \left( 4 g_o u_{x}-4 g_o g_z-4 u_{x} g_z+4 u_{x}^2+4 u_{o \tau } g_{x\tau}-4 u_z u_{o \tau }+8 v_{o
   \tau } v_{\tau  x} -8 v_{o \tau } v_{z}+4 g_{x\tau}^2 -4 u_z g_{x\tau} \right.\nonumber \\
&\left. \,\,\,\,\, +\,8 v_{\tau 
   x}^2-8 v_{o \tau } v_{\tau  x} \right)  
\\
\label{vz}
\dot{v}_{z} &= \dot{\mathscr{N}}_{ph} \left( 12 g_o v_{z}+20 u_{x} v_{z}+4 g_o v_{\tau  x}+8 g_z v_{z}+8 u_{x} v_{\tau 
   x}+4 g_z v_{\tau  x}+4 u_{x} v_{o \tau }+4 g_z v_{o \tau }-4 u_{o \tau } v_{z}-12 v_{z}
   g_{x\tau} \right.\nonumber \\
&\left.\,\,\,\,\, +\,4 u_{o \tau } v_{\tau  x}-8 u_z v_{z}-8 N_f v_{z}^2+8 u_{\tau 
   x} v_{\tau  x}+4 v_{o \tau } g_{x\tau}+4 u_z v_{\tau  x}+4 u_z v_{o \tau }\right)   \nonumber \\
&+ \dot{\mathscr{N}}_{pp}  \left( -4 g_o v_{z}+4 u_{x} v_{z}+4 g_o v_{\tau  x}-8 u_{x} v_{\tau  x}-4 g_z v_{\tau 
   x}-4 u_{x} v_{o \tau }-4 g_z v_{o \tau }+4 u_{o \tau } v_{z}-4 v_{z} g_{x\tau}\right.\nonumber \\
&\left.\,\,\,\,\, -\,4 u_{o\tau } v_{\tau  x}+8 g_{x\tau} v_{\tau  x}+4 v_{o \tau } g_{x\tau}+4 u_z v_{\tau  x}+4u_z v_{o \tau }\right)  \nonumber
\\
&+\dot{\mathscr{L}}_{ph} \left( 4 g_o v_{z}-4 u_{x} v_{z}-4 g_o v_{\tau  x}+8 u_{x} v_{\tau  x}-4 g_z v_{\tau 
   x}-4 u_{x} v_{o \tau }+4 g_z v_{o \tau }+4 u_{o \tau } v_{z}-4 v_{z} g_{x\tau} \right.\nonumber \\
&\left. \,\,\,\,\, -\,4 u_{o
   \tau } v_{\tau  x}+8 g_{x\tau} v_{\tau  x}-4 v_{o \tau } g_{x\tau}-4 u_z v_{\tau  x}+4
   u_z v_{o \tau } \right)  \nonumber
\\
&+ \dot{\mathscr{L}}_{pp}  \left( -4 g_o v_{z}-4 u_{x} v_{z}-4 g_o v_{\tau  x}-8 u_{x} v_{\tau  x}+4 g_z v_{\tau 
   x}+4 u_{x} v_{o \tau }-4 g_z v_{o \tau }+4 u_{o \tau } v_{z}+4 v_{z} g_{x\tau} \right.\nonumber \\
&\left. \,\,\,\,\, +\,4 u_{o
   \tau } v_{\tau  x}+8 g_{x\tau} v_{\tau  x}-4 v_{o \tau } g_{x\tau}-4 u_z v_{\tau  x}+4
   u_z v_{o \tau } \right)  
\\
\label{vx}
\dot{v}_{x\tau} &= \dot{\mathscr{N}}_{ph} \left(2 g_o v_{z}+8 g_o v_{\tau  x}+4 u_{x} v_{z}+2 g_z v_{z}+2 g_o v_{o \tau }+8 u_{x}
   v_{\tau  x}+4 u_{x} v_{o \tau }+2 g_z v_{o \tau }+2 u_{o \tau } v_{z}+4 v_{z} g_{x\tau}\right.\nonumber \\
&\left.\,\,\,\,\, +\,2
   v_{o \tau } u_{o \tau }+2 u_z v_{z}+8 g_{x\tau} v_{\tau  x}+8 u_z v_{\tau  x}+4 v_{o \tau }
   g_{x\tau}+2 u_z v_{o \tau }-4 N_f v_{\tau  x}^2\right)   \nonumber \\
&+ \dot{\mathscr{N}}_{pp}  \left( 2 g_o v_{z}-4 u_{x} v_{z}-4 g_o v_{\tau  x}-2 g_z v_{z}-2 g_o v_{o \tau }+8 u_{x}
   v_{\tau  x}+4 g_z v_{\tau  x}+4 u_{x} v_{o \tau }+2 g_z v_{o \tau }-2 u_{o \tau } v_{z}+4 v_{o\tau } g_{x\tau} \right. \nonumber \\
&\left. \,\,\,\,\, +\,4 u_{o \tau } v_{\tau  x}+2 v_{o \tau } u_{o \tau }+2 u_z v_{z}-8
   g_{x\tau} v_{\tau  x}-4 v_{o \tau } g_{x\tau}-4 u_z v_{\tau  x}-2 u_z v_{o \tau }\right)  \nonumber 
\\
&+  \dot{\mathscr{L}}_{ph} \left( -2 g_o v_{z}+8 g_o v_{\tau  x}+4 u_{x} v_{z}-2 g_z v_{z}-2 g_o v_{o \tau }-8 u_{x}
   v_{\tau  x}+4 u_{x} v_{o \tau }-2 g_z v_{o \tau }-2 u_{o \tau } v_{z}+4 v_{z} g_{x\tau} -2v_{o \tau } u_{o \tau }\right.\nonumber \\
&\left. \,\,\,\,\, -\,2 u_z v_{z} -8 g_{x\tau} v_{\tau  x}+8 u_z v_{\tau  x}+4 v_{o \tau }g_{x\tau}-2 u_z v_{o \tau }-4 N_f v_{\tau  x}^2 \right)  \nonumber 
\\
&+ \dot{\mathscr{L}}_{pp}  \left( -2 g_o v_{z}-4 u_{x} v_{z}-4 g_o v_{\tau  x}+2 g_z v_{z}+2 g_o v_{o \tau }-8 u_{x}
   v_{\tau  x}+4 g_z v_{\tau  x}+4 u_{x} v_{o \tau }-2 g_z v_{o \tau }+2 u_{o \tau } v_{z}+4 v_{o
   \tau } g_{x\tau}\right.\nonumber \\
&\left. \,\,\,\,\, +\,4 u_{o \tau } v_{\tau  x}-2 v_{o \tau } u_{o \tau }-2 u_z v_{z} +8g_{x\tau} v_{\tau  x}-4 v_{o \tau } g_{x\tau}-4 u_z v_{\tau  x}+2 u_z v_{o \tau } \right)   
\end{align}
\begin{align}
\label{vo}
\dot{v}_{o\tau} &=  \dot{\mathscr{N}}_{ph} \left( 4 u_{x} v_{z}+4 g_o v_{\tau  x}+4 g_z v_{z}+4 g_o v_{o \tau }+8 u_{x} v_{\tau  x}+4 g_z
   v_{\tau  x}+4 u_{x} v_{o \tau }+4 v_{z} g_{x\tau}+4 u_{o \tau } v_{\tau  x} + 4 v_{o \tau } u_{o
   \tau }+4 u_z v_{z}\right.\nonumber \\
&\left.\,\,\,\,\, + \,8 g_{x\tau} v_{\tau  x}+4 v_{o \tau } g_{x\tau}+4 u_z v_{\tau 
   x} \right)   \nonumber \\
&+\dot{\mathscr{N}}_{pp}  \left( -4 u_{x} v_{z}-4 g_o v_{\tau  x}-4 g_z v_{z}-4 g_o v_{o \tau }+8 u_{x} v_{\tau  x}+4 g_z
   v_{\tau  x}+4 u_{x} v_{o \tau }+4 v_{z} g_{x\tau}+4 u_{o \tau } v_{\tau  x} +4 v_{o \tau } u_{o
   \tau }+4 u_z v_{z}\right.\nonumber \\
&\left. \,\,\,\,\, - \,8 g_{x\tau} v_{\tau  x}-4 v_{o \tau } g_{x\tau}-4 u_z v_{\tau 
   x} \right)  \nonumber
\\
&+ \dot{\mathscr{L}}_{ph} \left( -4 u_{x} v_{z}-4 g_o v_{\tau  x}+4 g_z v_{z}+12 g_o v_{o \tau }+8 u_{x} v_{\tau  x}-4
   g_z v_{\tau  x}-20 u_{x} v_{o \tau }+8 g_z v_{o \tau }-4 v_{z} g_{x\tau}-4 u_{o \tau } v_{\tau 
   x} \right.\nonumber \\
&\left. \,\,\,\,\, -\, 4 v_{o \tau } u_{o \tau }+4 u_z v_{z}+8 g_{x\tau} v_{\tau  x}+12 v_{o \tau } g_{x\tau}-4
   u_z v_{\tau  x}-8 u_z v_{o \tau }-8 N_f v_{o \tau }^2 \right)  \nonumber
\\
&+ \dot{\mathscr{L}}_{pp}  \left( 4 u_{x} v_{z}+4 g_o v_{\tau  x}-4 g_z v_{z}-4 g_o v_{o \tau }+8 u_{x} v_{\tau  x}-4 g_z
   v_{\tau  x}-4 u_{x} v_{o \tau }-4 v_{z} g_{x\tau}-4 u_{o \tau } v_{\tau  x}+4 v_{o \tau } u_{o
   \tau }+4 u_z v_{z} \right.\nonumber \\
&\left. \,\,\,\,\,  -\, 8 g_{x\tau} v_{\tau  x}+4 v_{o \tau } g_{x\tau}+4 u_z v_{\tau 
   x} \right)  
\\
\label{uo}
\dot{u}_{o\tau} &=  \dot{\mathscr{N}}_{ph} \left(12 g_o u_{o \tau }+20 u_{x} u_{o \tau }+4 g_o g_{x\tau}+8 g_z u_{o \tau }+8 u_{x} u_{\tau 
   x}+4 g_z g_{x\tau}+4 u_{x} u_z+4 g_z u_z-8 N_f u_{o \tau }^2+8 u_{o \tau }^2-16 u_{o\tau } v_{z}\right.\nonumber \\
&\left. \,\,\,\,\, -\,32 u_{o \tau } v_{\tau  x}-16 v_{o \tau } u_{o \tau }+16 u_{o \tau }
   g_{x\tau}+8 u_z u_{o \tau }+4 v_{z}^2+8 v_{z} v_{\tau  x}+8 v_{\tau 
   x}^2+8 v_{o \tau } v_{\tau  x}+4 v_{o \tau }^2 \right)   \nonumber \\
&+ \dot{\mathscr{N}}_{pp}  \left( -4 g_o u_{o \tau }+4 u_{x} u_{o \tau }+4 g_o g_{x\tau}-8 u_{x} g_{x\tau}-4 g_z u_{\tau 
   x}-4 u_{x} u_z-4 g_z u_z  +4 v_{z}^2-8 v_{z} v_{\tau  x}+8 v_{\tau  x}^2+8
   v_{o \tau } v_{\tau  x}+4 v_{o \tau }^2 \right)  \nonumber 
\\
&+ \dot{\mathscr{L}}_{ph} \left( 4 g_o u_{o \tau }-4 u_{x} u_{o \tau }-4 g_o g_{x\tau}+8 u_{x} g_{x\tau}-4 g_z u_{\tau 
   x}-4 u_{x} u_z+4 g_z u_z+4 v_{z}^2-8 v_{z} v_{\tau  x}+8 v_{\tau  x}^2-8
   v_{o \tau } v_{\tau  x}+4 v_{o \tau }^2 \right)  \nonumber 
\\
&+\dot{\mathscr{L}}_{pp}  \left( -4 g_o u_{o \tau }-4 u_{x} u_{o \tau }-4 g_o g_{x\tau}-8 u_{x} g_{x\tau}+4 g_z u_{\tau 
   x}+4 u_{x} u_z-4 g_z u_z+4 v_{z}^2+8 v_{z} v_{\tau  x}+8 v_{\tau  x}^2 \right.\nonumber \\
&\left. \,\,\,\,\, -\,8v_{o \tau } v_{\tau  x}+4 v_{o \tau }^2 \right)  
\\
\label{gx}
 \dot{g}_{x\tau} &= \dot{\mathscr{N}}_{ph} \left( 2 g_o u_{o \tau }+8 g_o g_{x\tau}+4 u_{x} u_{o \tau }+2 g_z u_{o \tau }+2 g_o u_z+8 u_{x}
   g_{x\tau}+4 u_{x} u_z+2 g_z u_z-8 v_{z} g_{x\tau}+4 u_{o \tau } g_{x\tau}+2
   v_{z}^2 \right.\nonumber \\
&\left. \,\,\,\,\, +\,8 v_{z} v_{\tau  x}+4 v_{o \tau } v_{z}+8 v_{o \tau } g_{x\tau}-4 N_fg_{x\tau}^2-4 u_z g_{x\tau}+8 v_{\tau  x}^2+8 v_{o \tau } v_{\tau  x}+2 v_{o \tau }^2 \right)   \nonumber \\
&+ \dot{\mathscr{N}}_{pp}  \left( 2 g_o u_{o \tau }-4 u_{x} u_{o \tau }-4 g_o g_{x\tau}-2 g_z u_{o \tau }-2 g_o u_z+8 u_{x}
   g_{x\tau}+4 g_z g_{x\tau}+4 u_{x} u_z+2 g_z u_z-2 v_{z}^2 \right.\nonumber \\
&\left. \,\,\,\,\, +\,8 v_{z}
   v_{\tau  x}+4 v_{o \tau } v_{z}-8 v_{\tau  x}^2-8 v_{o \tau } v_{\tau  x}-2 v_{o \tau }^2 \right)  \nonumber 
\\
&+\dot{\mathscr{L}}_{ph} \left( -2 g_o u_{o \tau }+8 g_o g_{x\tau}+4 u_{x} u_{o \tau }-2 g_z u_{o \tau }-2 g_o u_z-8 u_{x}
   g_{x\tau}+4 u_{x} u_z-2 g_z u_z-8 v_{z} g_{x\tau}+4 u_{o \tau } g_{x\tau}-2
   v_{z}^2 \right.\nonumber \\
&\left. \,\,\,\,\, +\,8 v_{z} v_{\tau  x}-4 v_{o \tau } v_{z}+8 v_{o \tau } g_{x\tau}-4 N_f
   g_{x\tau}^2-4 u_z g_{x\tau}-8 v_{\tau  x}^2+8 v_{o \tau } v_{\tau  x}-2 v_{o \tau }^2 \right)  \nonumber 
\\
&+  \dot{\mathscr{L}}_{pp}  \left( -2 g_o u_{o \tau }-4 u_{x} u_{o \tau }-4 g_o g_{x\tau}+2 g_z u_{o \tau }+2 g_o u_z-8 u_{x}
   g_{x\tau}+4 g_z g_{x\tau}+4 u_{x} u_z-2 g_z u_z+2 v_{z}^2+8 v_{z}v_{\tau  x} \right.\nonumber \\
&\left. \,\,\,\,\, -\,4 v_{o \tau } v_{z}+8 v_{\tau  x}^2-8 v_{o \tau } v_{\tau  x}+2 v_{o \tau }^2 \right)  
\\
\label{uz}
\dot{u}_z &=  \dot{\mathscr{N}}_{ph} \left( 4 u_{x} u_{o \tau }+4 g_o g_{x\tau}+4 g_z u_{o \tau }+4 g_o u_z+8 u_{x} g_{x\tau}+4 g_z
   g_{x\tau}+4 u_{x} u_z  +8 v_{z} v_{\tau  x}+8 v_{o \tau } v_{z}+8v_{\tau  x}^2+8v_{o \tau } v_{\tau  x} \right)   \nonumber \\
&+ \dot{\mathscr{N}}_{pp}  \left( -4 u_{x} u_{o \tau }-4 g_o g_{x\tau}-4 g_z u_{o \tau }-4 g_o u_z+8 u_{x} g_{x\tau}+4 g_z
   g_{x\tau}+4 u_{x} u_z +8 v_{z} v_{\tau  x}+8 v_{o \tau } v_{z}-8 v_{\tau  x}^2-8
   v_{o \tau } v_{\tau  x} \right)  \nonumber 
\\
&+ \dot{\mathscr{L}}_{ph} \left( -4 u_{x} u_{o \tau }-4 g_o g_{x\tau}+4 g_z u_{o \tau }+12 g_o u_z+8 u_{x} g_{x\tau}-4
   g_z g_{x\tau}-20 u_{x} u_z+8 g_z u_z-16 u_z v_{z}+8 u_z u_{o \tau }-8 v_{o \tau
   } v_{\tau  x}\right.\nonumber \\
&\left. \,\,\,\,\, +\,8 v_{o \tau } v_{z} +32 u_z v_{\tau  x}-16 u_z v_{o \tau }-16 u_z g_{x\tau}-8 N_f
   u_z^2+8 u_z^2+8 v_{\tau  x}^2-8 v_{o \tau } v_{\tau  x} \right)  \nonumber 
\\
&+  \dot{\mathscr{L}}_{pp}  \left( 4 u_{x} u_{o \tau }+4 g_o g_{x\tau}-4 g_z u_{o \tau }-4 g_o u_z+8 u_{x} g_{x\tau}-4 g_zg_{x\tau}-4 u_{x} u_z-8 v_{z} v_{\tau  x}+8 v_{o \tau } v_{z}-8 v_{\tau  x}^2+8v_{o \tau } v_{\tau  x} \right)  
\end{align}

\newpage

\subsection{Flow equations for the order parameter vertices}
Through manipulations similar to those in the previous section, it is possible to derive non perturbative flow equations for the composite operators $\mathcal{O} \in  \psi^\dag \sigma^\mu \tau^\nu \eta^\lambda \psi$ and $\Delta \in  \psi^\dag \sigma^\mu \tau^\nu \eta^\lambda \psi^\dag$, corresponding to particle-hole and particle-particle (superconducting) instabilities respectively. The derivation proceeds by adding to the action a test vertex for the composite operator and then deriving a flow equation for the vertex. The manipulations are similar to those given in the previous section, so we state the resulting flow equations without an explicit proof.  The results are a set of gap equations for the associated order parameters, which generalise the BCS gap equation for superconductivity in that they include strong-coupling corrections. As with our expressions for the coupling constant flow equations, we neglect the contributions of the electron self energy. For the superconducting order parameters, we find 
  \begin{align}
 \partial_t\Delta = - \dot{\mathscr{L}}_{pp} V_{\mu\nu}\left(2\Omega^\mu \Delta \Omega^{\nu T}+\Omega^\mu \alpha_i \Delta  \alpha^i  \Omega^{\nu T}\right) - \dot{\mathscr{N}}_{pp}V_{\mu\nu}\left(2\Omega^\mu \Delta \Omega^{\nu T}-\Omega^\mu \alpha_i \Delta  \alpha^i  \Omega^{\nu T}\right)
 \end{align}
where as usual we employ Einstein summation notation, while for the particle-hole order parameters,
\begin{align}
\partial_t\mathcal{O} = \delta\mathcal{O}_1+\delta\mathcal{O}_2
\end{align}
with
\begin{align}
\delta\mathcal{O}_1 &=  \dot{\mathscr{L}}_{ph}V_{\mu\nu}\left(2\Omega^\mu \mathcal{O} \Omega^\nu-\Omega^\mu \alpha_i \mathcal{O}  \alpha^i  \Omega^\nu\right)+\dot{\mathscr{N}}_{ph}V_{\mu\nu}\left(2\Omega^\mu \mathcal{O} \Omega^\nu+ \Omega^\mu \alpha_i \mathcal{O}  \alpha^i  \Omega^\nu\right)\\
\delta\mathcal{O}_2 &= -\dot{\mathscr{L}}_{ph}V_{\mu\nu}\left(2\text{Tr}[\mathcal{O}\Omega^\mu]-\text{Tr}[\mathcal{O}\alpha_i\Omega^\mu\alpha^i] \right)\Omega^\nu-\dot{\mathscr{N}}_{ph}V_{\mu\nu}\left(2\text{Tr}[\mathcal{O}\Omega^\mu]+ \text{Tr}[\mathcal{O}\alpha_i\Omega^\mu\alpha^i] \right)\Omega^\nu
\end{align}
which now give us formally exact expressions for the order parameter vertices. These gap equations form a linear system; choosing a basis of operators which diagonalise this system gives us a set of equations of the form
\begin{align}
\label{OgapeqERG}
\partial_t \mathcal{O}= \lambda_\mathcal{O}(t) \, \mathcal{O} \\
\label{DgapeqERG}
\partial_t\Delta= \lambda_{\Delta}(t) \,\Delta 
\end{align}
where the eigenvalue functions take the form
\begin{gather}
\lambda_\mathcal{O}(t) \equiv \lambda^{\mathscr{L}}_\mathcal{O}(t)\dot{\mathscr{L}}_{ph}(t) + \lambda^{\mathscr{N}}_\mathcal{O}(t)\dot{\mathscr{N}}_{ph}(t)  \\
\lambda_\Delta(t) \equiv \lambda^{\mathscr{L}}_\Delta(t)\dot{\mathscr{L}}_{pp}(t) + \lambda^{\mathscr{N}}_\Delta(t)\dot{\mathscr{N}}_{pp}(t) 
\end{gather}
where $\lambda^{\mathscr{L},\mathscr{N}}$ are linear combinations of the couplings; $\lambda^{\mathscr{L}}$ correspond to the parquet couplings of Tables \ref{ph_order_params} and \ref{pp_order_params}. We find 18 distinct order parameter manifolds -- 12 particle-hole and 6 particle-particle. Solving for the flow of the order parameter vertex gives the formal expressions as a function of temperature
\begin{align}
\label{suscO}
\mathcal{O}(t)   = \mathcal{O}_0\, \exp \left( 2\int_1^{t'} \lambda_\mathcal{O}(t'') dt''  \right)  \\ 
\label{suscD}
\Delta(t)   = \Delta_0\, \exp \left( 2\int_1^{t'} \lambda_{\Delta}(t'') dt''  \right) 
\end{align}
where $\mathcal{O}_0 ,\Delta_0$ are nonuniversal constants. Our illustrations of the RG flow plot the exponent in this expression.

\begin{table*}[t]
\begin{center}
\caption{\textbf{Particle-hole order parameters and their eigenvalues.} Projection onto the flavour $f$ is realised via $\mathcal{O} \rightarrow \mathcal{P}_f \mathcal{O} \mathcal{P}_f$. Notation: indices can take any value from $a,b,c\in\{0,z\}$, $i\neq i',j,k\in \{x,y\}$, $\mu\in\{0,x,y,z\}$. In each order parameter, the factor $s_\mu\tau_a$ is chosen so that $\text{Tr}(\mathcal{P}_f s_\mu\tau_a)=N_f$; else the order parameter possesses the same eigenvalue as the order one row down. } 
\label{ph_order_params}
\vspace{0.1cm}
\begin{ruledtabular}
 \begin{tabular} {lllllllllllllllllllll} \\[-3.5mm]
Order parameter type &  Order parameter   & Parquet eigenvalue   \\[1mm] \hline \vspace{-0.2cm} \\ \vspace{0.2cm}
Graphene nematic  &  $(\sigma_x,  \sigma_y \tau_z)(s_\mu \tau_a)$ & \ $\lambda_1=2\left(g_o -2N_f  g_{x\tau} -g_z   +2 v_{o \tau } -2 v_z+u_{o\tau} -u_z\right)$  \\   \vspace{0.2cm}
(Un)polarised K-IVC  &  $\sigma _a\tau _i\eta _y$ & \  $\lambda_3=2\left (g_o -g_z  +2 v_{o \tau }-2 v_z+u_{o\tau} -u_z  \right)$  \\   \vspace{0.2cm}
Graphene-moir\'e nematic  &  $(\sigma_x,\tau_z\sigma_y)(\eta_x,\tau_z\eta_y)(s_\mu \tau_a)$ & \ $\lambda_5=2\left (g_o - g_z - 2N_f v_{x} - u_ {o\tau} + u_z \right)$   \\ \vspace{0.2cm}
(Un)polarised T-IVC    &  $\sigma _a\tau _i\eta _b$ & \  $\lambda_2= 2\left(g_o -g_z -u_{o\tau} +u_z\right)$  \\   \vspace{0.2cm}
Moir\'e polarised nematic  &  $(\sigma_x ,\sigma_y\tau_z) \eta _z (s_\mu \tau_a)$ & \ $\lambda_6=2\left (g_o-g_z-2 v_{o \tau }+2 v_z+u_{o\tau}-2N_f u_{x}-u_z\right)$   \\ \vspace{0.2cm}
(Un)polarised T-IVC   &  $\sigma _a\tau_i\eta _x$ & \  $\lambda_4=2\left(g_o -g_z   -2 v_{o \tau }+2 v_z +u_{o\tau} -u_z\right) $  \\   \vspace{0.2cm}
Sublattice polarised  &  $\sigma _z\tau _z\eta _0(s_\mu \tau_a)$ & \  $\lambda_{12}= 4\left (g_o-2 g_{x\tau}-2N_f g_z+g_z+2 v_{o \tau }-4v_{x}+2 v_z+u_{o\tau}-2 u_{x}+u_z\right) $  \\ \vspace{0.2cm}
K-IVC  &  $\sigma _x\tau _i\eta _y$ & \ $\lambda_7= 4\left (g_o-2 g_{x\tau}+g_z+2 v_{o \tau }-4 v_{x}+2v_z+u_{o\tau}-2 u_{x}+u_z\right) $   \\ \vspace{0.2cm}
Sublattice/moir\'e polarised  &   $\sigma_z\tau_ 0\eta_z (s_\mu \tau_a)$ & \ $\lambda_{11}= 4\left (g_o-2 g_{x\tau}+g_z-2 v_{o \tau }+4v_{x}-2 v_z+u_{o\tau}-2 u_{x}-2N_f u_z+u_z\right) $   \\ \vspace{0.2cm}
T-IVC  &  $\sigma _x\tau _i\eta _x$ & \ $\lambda_8= 4\left (g_o-2 g_{x\tau}+g_z-2 v_{o \tau }+4v_{x}-2v_z+u_{o\tau}-2 u_{x}+u_z\right) $   \\ \vspace{0.2cm}
Moir\'e density wave  &  $(\tau_z\eta _x,\eta _y)(s_\mu \tau_a)$ & \ $\lambda_9= 4\left (g_o-2 g_{x\tau}+g_z-2N_f v_{o\tau}-u_{o\tau}-2 u_{x}-u_z\right)$   \\ %\vspace{0.2cm}
(Un)polarised K-IVC &  $\sigma _y\tau _i\eta _a$ & \ $\lambda_{10}= 4\left(g_o-2 g_{x\tau}+g_z-u_{o\tau }-2 u_{x}-u_z\right) $   \\ 
 \end{tabular}
\end{ruledtabular}
\end{center}
\begin{center}
\caption{\textbf{Particle-particle order parameters (superconducting gap functions) and their eigenvalues.} Notation: indices can take any value from $a,b\in\{0,z\}$, $i,j,k\in \{x,y\}$, $\mu\in\{x,y,z\}$, such that the gap function is an antisymmetric matrix. Projection onto the flavour $f$ is realised via $\Delta \rightarrow \mathcal{P}_f \Delta \mathcal{P}_f$.} 
\label{pp_order_params}
\vspace{0.1cm}
\begin{ruledtabular}
 \begin{tabular} {lllllllllllllllllllll} \\[-3.5mm]
Order parameter type &    Superconducting gap   & Parquet eigenvalue   \\[1mm] \hline \vspace{-0.2cm} \\ \vspace{0.2cm}
$A_{1,2}$/$B_{1,2}$ sublattice intravalley singlet &  $\sigma _a\tau _b\eta _c is_y$ & \ $\lambda_1=-2(g_o+g_z + u_{o\tau}+u_z)$  \\   \vspace{0.2cm}
 $A_{1,2}$/$B_{1,2}$ intervalley singlet/triplet &  $\sigma _i\tau _j\eta _k (s_0,s_\mu)is_y$ & \ $\lambda_1=-2(g_o+g_z + u_{o\tau}+u_z)$  \\   \vspace{0.2cm}
$A_{1,2}$ sublattice intravalley singlet  &  $\sigma _a\tau _b\eta _x is_y$  & \  $\lambda_2=-2\left (g_o+g_z+2 v_{o \tau }+2 v_z-u_{o\tau}-u_z\right)$  \\   \vspace{0.2cm}
$A_{1,2}$/$B_{1,2}$ intervalley singlet/triplet  &  $\sigma _i\tau _j\eta _z (s_0,s_\mu)is_y$  & \  $\lambda_2=-2\left (g_o+g_z+2 v_{o \tau }+2 v_z-u_{o\tau}-u_z\right)$  \\   \vspace{0.2cm}
 $A_{1,2}$ intravalley triplet  &  $\sigma_a \tau_b\eta_y s_\mu is_y$ & \  $\lambda_3=-2\left (g_o+g_z-2 v_{o \tau }-2 v_z-u_{o\tau}-u_z\right)$  \\   \vspace{0.2cm}
 $A_{1,2}$/$B_{1,2}$ intervalley singlet/triplet  &  $\sigma_i\tau_j (s_0,s_\mu)is_y$ & \  $\lambda_3=-2\left (g_o+g_z-2 v_{o \tau }-2 v_z-u_{o\tau}-u_z\right)$  \\   \vspace{0.2cm}
$E_1$/$E_2$ intravalley singlet  &  $\sigma_x \tau_a\eta_b is_y$ & \  $\lambda_4=-4\left (g_o+2 g_{x\tau}-g_z+u_{o\tau}+2 u_{x}-u_z\right)$  \\   \vspace{0.2cm}
$E$ intervalley singlet/triplet  &  $\sigma_z \tau_i\eta_j(s_0,s_\mu)is_y$ & \  $\lambda_4=-4\left (g_o+2 g_{x\tau}-g_z+u_{o\tau}+2 u_{x}-u_z\right)$  \\   \vspace{0.2cm}
$A_2$/$B_2$ intravalley triplet  &   $\sigma_y \tau_a\eta_x s_\mu is_y$ & \ $\lambda_5=-4\left (g_o+2 g_{x\tau}-g_z+2 v_{o \tau }-4v_{x}-2 v_z-u_{o\tau}+2 u_{x}+u_z\right)$   \\ \vspace{0.2cm}
$A_2$/$B_2$ intervalley singlet/triplet  &   $\tau_i\eta_z (s_0,s_\mu)is_y$ & \ $\lambda_5=-4\left (g_o+2 g_{x\tau}-g_z+2 v_{o \tau }-4v_{x}-2 v_z-u_{o\tau}+2 u_{x}+u_z\right)$   \\ \vspace{0.2cm}
$A_1$/$B_1$ intravalley singlet   &  $\sigma _y\tau _a\eta _y  is_y$ & \  $\lambda_6=-4\left (g_o+2 g_{x\tau}-g_z-2 v_{o \tau }+4v_{x}+2 v_z-u_{o\tau}+2 u_{x}+u_z\right)$   \\ \vspace{0.2cm}
$A_1$/$B_1$ intervalley singlet/triplet  &  $\tau _i(s_0,s_\mu)is_y$  & \  $\lambda_6=-4\left (g_o+2 g_{x\tau}-g_z-2 v_{o \tau }+4v_{x}+2 v_z-u_{o\tau}+2 u_{x}+u_z\right)$ \\
 \end{tabular}
\end{ruledtabular}
\end{center}
\end{table*}

The effects of the logarithmically divergent contributions $\mathscr{L}_{ph}(t)$ shall be discussed further below, but first we call attention to the effect of the contributions to the susceptibilities arising from $\mathscr{N}_{ph}(t)$. One finds 
\begin{align}
\label{Stoner_factor}
\exp \left( 2\int_1^t \lambda^{\mathscr{N}}_\mathcal{O}(t')\dot{\mathscr{N}}_{ph}(t') dt'  \right) \approx \exp \left( 2\lambda^{\mathscr{N}}_\mathcal{O}(t\approx 1) \right)
\end{align}
a Stoner enhancement factor which depends exponentially on the values of the couplings  near the start of the RG flow $1\lesssim t \lesssim 2$, which we may roughly approximate with the bare values of the couplings. The approximate equality comes from the fact that $\dot{\mathscr{N}}_{ph}(t')$ decays exponentially at long RG times. This approximate equality breaks down at stronger bare couplings, where the initially zero bare couplings may obtain large renormalised values at early RG times, in which case these couplings contribute to the Stoner exponent as well.  

At stronger values of the bare couplings, the Stoner enhancement factors can significantly change the competition between the competing instabilities.  General comments aside, the precise interplay of the logarithmic and non-logarithmic contributions to the order parameter susceptibilities is encoded in the solutions we find by explicit integration of the equations \eqref{OgapeqERG} and \eqref{DgapeqERG}, which we shall present in a later section.

Several order parameter structures with nonzero parquet couplings $\lambda^{\mathscr{L}}$ have vanishing Stoner couplings $\lambda^{\mathscr{N}}$. Specifically, $\lambda^{\mathscr{N}}_7=\lambda^{\mathscr{N}}_8=\lambda^{\mathscr{N}}_9=\lambda^{\mathscr{N}}_{10}=\lambda^{\mathscr{N}}_{11}=\lambda^{\mathscr{N}}_{12}=0$ among the particle-hole order parameters and $\lambda^{\mathscr{N}}_1=\lambda^{\mathscr{N}}_2=\lambda^{\mathscr{N}}_3=0$ for the particle-particle orders. For the other order parameters, we find that $\lambda^{\mathscr{N}}=\lambda^{\mathscr{L}}$.

\subsection{Effect of the Dirac revivals on the interactions and flow equations}
Consider the flavour space $f\in\{\tau,s\}$ consisting of the four spin-valley states. Let $\mathcal{P}_{f}$ be the operator which projects on to the flavour subspace $f$. The electron operator $c_{a}$, where $a$ is the flavour index, transforms as
\begin{align}
c_{a} = [\mathcal{P}_f]_{ab} \, c_b
\end{align}
Conveniently, the projection operator is independent of momentum. When performing a projection, a particle-hole mass term in the Hamiltonian will transform as
\begin{align}
\mathcal{O}_{ab} c^\dag_a c_b \rightarrow [\mathcal{P}^\dag_f\mathcal{O} \mathcal{P}_f ]_{ab} c^\dag_a c_b
\end{align}
Meanwhile, a particle-particle mass term (superconducting gap) will transform as 
\begin{align}
\Delta_{ab} c_a c_b \rightarrow [\mathcal{P}_f\Delta \mathcal{P}^T_f ]_{ab} c_a c_b
\end{align}
If we choose the $z$ basis for spin and valley, then $\mathcal{P}$ is simply a real symmetric matrix. For instance, projecting onto a single valley $\tau$ and spin $s$,
\begin{align}
\mathcal{P}_{\tau s} = \sigma_0 \eta_0 \, ( \tau_0 +\tau \tau_z  ) ( s_0 +s s_z  )
\end{align}
Conveniently, one finds that the only effect this projection has on the flow equations for the couplings is the replacement of $N_f=8$ with $N_f \rightarrow \text{Tr}(\mathcal{P}_f s_\mu\tau_a)$ -- a result of the fact that the Green's function and interactions commute with the projection operator -- as can be seen by direct substitution of the projection operator into Eqs. \eqref{XIPPSUPP} -- \eqref{XIVERTSUPP}. Hence the RG analysis straightforwardly carries over to the flavour-polarised case: our flow equations and couplings are all identical, but with an appropriate choice of $N_f$ and $\mu$ to reflect the filling factor of interest.

\subsection{Parquet RG flow}
At long RG times $t\rightarrow \infty$, the $\mathscr{L}\gg\mathscr{N}$ and so the logarithmically divergent terms dominate the right hand side of the flow equation. In the limit where we only retain these terms, we arrive at the so-called parquet RG beta functions, defined so that for each coupling $V$ we have $t\tfrac{d}{dt}V=\beta_{V}$. To simplify our analysis, we encode the effect of finite chemical potential by taking $\mathscr{L}_{ph} \approx d \times \mathscr{L}_{pp}$ with $d=d(\mu)<1$ rather than using the full chemical potential dependent particle-hole susceptibility -- an approximation which makes analysis of the fixed trajectories more straightforward, and is standard in the parquet RG literature. With these simplifications, the beta functions  are given explicitly by
\begin{align}
\beta_{g_o} &=  -4 g_o u_{x}-2g_o^2+4 u_{x} g_z-4 u_{x}^2-2 g_z^2-2 u_{o \tau }^2-4 u_{o \tau }g_x-4 v_{o\tau }^2-8 v_{o \tau } v_{x \tau}-4 g_x^2+4 u_z g_x-2 u_z^2-8v_{x \tau}^2\nonumber\\
&+8 v_{o\tau} v_{x \tau}-4 v_{o\tau}^2 + d(\mu) \left(-4 g_o u_{x}+2 g_o^2-4 u_{x} g_z+4 u_{x}^2+2 g_z^2+2 u_{o \tau }^2-4 u_{o\tau } g_x+4 v_{o \tau }^2-8 v_{o \tau } v_{x \tau}+4 g_x^2 \right. \nonumber \\
&\left.-4 u_zg_x+2 u_z^2+8 v_{x \tau}^2-8 v_{o\tau} v_{x \tau}+4v_{o\tau}^2\right) \\
\beta_{u_{x}} &=  -4 g_ou_{x}+2 g_o g_z-g_o^2+4 u_{x} g_z-4 u_{x}^2-g_z^2-u_{o \tau }^2-4 u_{o \tau } g_x+2 u_z u_{o \tau }-2 v_{o \tau }^2-8 v_{o \tau } v_{x \tau}+4 v_{o\tau} v_{o \tau }-4g_x^2 +4 u_z g_x\nonumber\\
&-u_z^2-8 v_{x \tau}^2+8 v_{o\tau} v_{x \tau}-2 v_{o\tau}^2+d(\mu )\left(4 u_{x} u_{o \tau }+8 u_{x} v_{o \tau }+8 g_o u_{x}-2 g_o g_z-g_o^2-4 u_{x}u_z-8 u_{x} v_{o\tau} -4 N u_{x}^2-4 u_{x}^2-g_z^2\right. \nonumber \\
&\left.-u_{o \tau }^2+4 u_{o \tau } g_x-2 u_zu_{o \tau }-2 v_{o \tau }^2+8 v_{o \tau } v_{x \tau}-4 v_{o\tau} v_{o\tau }-4 g_x^2+4 u_z g_x-u_z^2-8 v_{x \tau}^2+8 v_{o\tau} v_{x \tau}-2 v_{o\tau}^2\right) \\
\beta_{g_z} &= 4 g_ou_{x}-4 g_o g_z-4 u_{x} g_z+4 u_{x}^2+4 u_{o \tau } g_x-4 u_z u_{o \tau }+8 v_{o\tau } v_{x \tau}-8 v_{o\tau} v_{o \tau }+4 g_x^2-4 u_z g_x+8 v_{x \tau }^2-8 v_{o\tau} v_{x \tau} \nonumber \\
&+d(\mu ) \left(8 g_z u_{o \tau }+16 g_z v_{o \tau }-4 g_o u_{x}+12 g_o g_z-16 g_z g_x+8 g_z u_z-32 g_z v_{x \tau}+16 g_z v_{o\tau}-20 u_{x} g_z+4 u_{x}^2-8 N g_z^2 \right. \nonumber \\
&\left.+8 g_z^2-4u_{o \tau } g_x+4 u_z u_{o \tau }-8 v_{o \tau } v_{x \tau}+8 v_{o\tau} v_{o \tau}+4 g_x^2-4 u_z g_x+8 v_{x \tau}^2-8 v_{o\tau} v_{x \tau}\right) \\
\beta_{v_{o\tau}} &= 4 u_{x} v_{o \tau }+4 g_o v_{x \tau}-4 g_z v_{o \tau }-4 g_o v_{o\tau}+8 u_{x} v_{x \tau}-4 g_z v_{x \tau}-4 u_{x} v_{o\tau}-4 v_{o \tau } g_x-4 u_{o \tau } v_{x \tau}+4 v_{o\tau} u_{o\tau }+4 u_z v_{o\tau } \nonumber \\
&-8 g_x v_{x \tau}+4 v_{o\tau} g_x+4 u_z v_{x \tau}+d(\mu ) \left(-4 u_{x} v_{o \tau }-4 g_o v_{x \tau}+4 g_z v_{o \tau }+12 g_o v_{o\tau}+8 u_{x}v_{x \tau}-4 g_z v_{x \tau} -20 u_{x} v_{o\tau}\right. \nonumber \\
&\left. +8 g_z v_{o\tau}-4 v_{o \tau } g_x-4 u_{o
   \tau } v_{x \tau}-4 v_{o\tau} u_{o \tau }+4 u_z v_{o \tau }+8 g_x v_{x \tau}+12v_{o\tau} g_x-4 u_z v_{x \tau}-8 u_z v_{o\tau}-8 N v_{o\tau}^2\right) \\
\beta_{u_z} &=  4 u_{x}u_{o \tau }+4 g_o g_x-4 g_z u_{o \tau }-4 g_o u_z+8 u_{x} g_x-4 g_zg_x-4 u_{x} u_z-8 v_{o \tau } v_{x \tau}+8 v_{o\tau} v_{o \tau }-8 v_{x \tau}^2+8v_{o\tau} v_{x \tau}\nonumber \\
&d(\mu ) \left(-4 u_{x} u_{o \tau }-4 g_o g_x+4 g_z u_{o \tau }+12 g_o u_z+8 u_{x}g_x-4 g_z g_x-20 u_{x} u_z+8 g_z u_z-16 u_z v_{o \tau }+8 u_z u_{o\tau }-8 v_{o \tau } v_{x \tau}\right.\nonumber \\
&\left.+8 v_{o\tau} v_{o \tau }+32 u_z v_{x \tau}-16 u_z v_{o\tau}-16u_z g_x-8 N u_z^2+8 u_z^2+8 v_{x \tau}^2-8 v_{o\tau} v_{x \tau}\right) \\
\beta_{v_{z}} &=  -4 g_o v_{o \tau }-4 u_{x} v_{o \tau }-4 g_o v_{x \tau }-8 u_{x} v_{x \tau}+4 g_z v_{x \tau}+4 u_{x} v_{o\tau}-4 g_z v_{o\tau}+4 u_{o \tau } v_{o \tau }+4 v_{o \tau } g_x+4 u_{o \tau } v_{x \tau}+8 g_x v_{x \tau}\nonumber\\
&-4 v_{o\tau} g_x-4 u_z v_{x \tau}+4 u_z v_{o\tau} + d(\mu ) \left(4 g_o v_{o \tau }-4 u_{x} v_{o \tau }-4 g_o v_{x \tau}+8 u_{x} v_{x \tau }-4 g_z v_{x \tau}-4 u_{x} v_{o\tau}+4 g_z v_{o\tau}+4 u_{o \tau } v_{o \tau }\right. \nonumber \\
&\left.-4 v_{o \tau }g_x-4 u_{o \tau } v_{x \tau}+8 g_x v_{x \tau}-4 v_{o\tau} g_x-4u_z v_{x \tau}+4 u_z v_{o\tau}\right)
\end{align}
\begin{align}
\beta_{v_{x\tau}} &=  -2 g_o v_{o \tau}-4 u_{x} v_{o \tau }-4 g_o v_{x \tau}+2 g_z v_{o \tau }+2 g_o v_{o\tau}-8 u_{x} v_{x \tau }+4 g_z v_{x \tau}+4 u_{x} v_{o\tau}-2 g_z v_{o\tau}+2 u_{o \tau } v_{o \tau }+4 v_{o \tau }g_x+4 u_{o \tau } v_{x \tau}\nonumber\\
&-2 v_{o\tau} u_{o \tau }-2 u_z v_{o \tau }+8 g_x v_{x \tau}-4 v_{o\tau} g_x-4 u_z v_{x \tau}+2 u_z v_{o\tau} +d(\mu ) \left(-2 g_o v_{o \tau }+8 g_o v_{x \tau}+4 u_{x} v_{o \tau }-2 g_z v_{o \tau}-2 g_o v_{o\tau}-8 u_{x} v_{x \tau}\right.\nonumber\\
&\left.+4 u_{x} v_{o\tau}-2 g_z v_{o\tau}-2 u_{o \tau } v_{o \tau }+4 v_{o\tau } g_x-2 v_{o\tau} u_{o \tau }-2 u_z v_{o \tau }-8 g_x v_{x \tau}+8u_z v_{x \tau}+4 v_{o\tau} g_x-2 u_z v_{o\tau}-4 N v_{x \tau}^2\right) \\
\beta_{u_{o\tau}} &=  -4 g_o u_{o \tau }-4 u_{x} u_{o \tau }-4g_o g_x-8 u_{x} g_x+4 g_z g_x+4 u_{x} u_z-4 g_z u_z+4 v_{o \tau }^2+8 v_{o \tau } v_{x \tau}+8 v_{x \tau}^2-8 v_{o\tau} v_{x \tau}+4 v_{o\tau}^2 \nonumber \\
&\!\!\!\!\!\!\!\!\!\!\!\!\!\!\!\! +d(\mu ) \left(4 g_o u_{o \tau }-4 u_{x} u_{o \tau }-4 g_o g_x+8 u_{x} g_x-4 g_z g_x-4 u_{x} u_z+4 g_z u_z+4 v_{o \tau }^2-8 v_{o \tau } v_{x \tau}+8v_{x \tau}^2-8 v_{o\tau} v_{x \tau}+4 v_{o\tau}^2\right)\\
\beta_{g_{x\tau}} &=  -2 g_o u_{o \tau }-4 u_{x} u_{o \tau }-4 g_o g_x+2 g_z u_{o \tau }+2 g_o u_z-8 u_{x} g_x+4 g_z g_x+4 u_{x} u_z-2 g_z u_z+2 v_{o \tau}^2+8 v_{o \tau } v_{x \tau}\nonumber\\
&-4 v_{o\tau} v_{o \tau }+8 v_{x \tau}^2-8 v_{o\tau} v_{x \tau}+2v_{o\tau}^2 + d(\mu ) \left(-2 g_o u_{o \tau }+8 g_o g_x+4 u_{x} u_{o \tau}-2 g_z u_{o \tau }-2 g_o u_z-8 u_{x} g_x+4 u_{x} u_z-2 g_z u_z\right. \nonumber \\
&\left. -8 v_{o \tau } g_x+4 u_{o \tau } g_x-2 v_{o \tau }^2+8 v_{o \tau } v_{x \tau}-4 v_{o\tau} v_{o \tau }+8 v_{o\tau} g_x-4 N g_x^2-4 u_z g_x-8 v_{x \tau}^2+8 v_{o\tau} v_{x \tau}-2 v_{o\tau}^2\right)
\end{align}
In the simple limit of $d(\mu)=1$, which applies when the chemical potential is tuned right to the quadratic band-touching point (i.e. close to integer filling of the moir\'e unit cell), these beta functions take an even more simple form,
\begin{align}
\beta_{g_o} &= -8 g_o u_{x}-8 u_{o\tau } g_x-16 v_{o\tau }v_{x\tau }\\
\beta_{u_{x}} &= 4 g_o u_{x}-2 g_o^2+4 u_{x} u_{o\tau}+8 u_{x} v_{z}-4 u_{x} u_z-8 u_{x} v_{o\tau}-4 N u_{x}^2+4 u_{x} g_z-8 u_{x}^2-2 g_z^2-2 u_{o\tau}^2-4 v_{z}^2\nonumber \\
&\ \ +8 u_z g_{x\tau}-2 u_z^2-8 g_{x\tau}^2+16 v_{o\tau} v_{x\tau}-4 v_{o\tau}^2-16 v_{x\tau}^2\\
\beta_{g_z} &= 8 g_o g_z+8 g_z u_{o\tau}+16 g_z v_{z}+8 g_z u_z-16 g_zg_{x\tau}+16 g_z v_{o\tau}-32 g_z v_{x\tau}-24 u_{x} g_z+8 u_{x}^2-8 Ng_z^2\nonumber \\
   & \ \ +8 g_z^2-8 u_z g_{x\tau}+8 g_{x\tau}^2-16 v_{o\tau} v_{x\tau}+16 v_{x\tau}^2 \nonumber \\
\beta_{v_{o\tau}} &= 8 g_o v_{o\tau}-24 u_{x} v_{o\tau}+8 g_z v_{o\tau}+16 u_{x} v_{x\tau}-8 g_z v_{x\tau}+8 v_{z} u_z-8 v_{z} g_{x\tau}-8 u_{o\tau} v_{x\tau}-8 u_z v_{o\tau} \nonumber\\
&\ \ +16 g_{x\tau} v_{o\tau}-8 N v_{o\tau}^2 \\
\beta_{u_z} &= 8 g_o u_z-24 u_{x} u_z+8 g_z u_z+16 u_{x} g_{x\tau}-8 g_z g_{x\tau}+8 u_{o\tau} u_z-16 v_{z} u_z+16 v_{z} v_{o\tau}-16 v_{z} v_{x\tau}-16 u_z g_{x\tau}\nonumber \\
&\ \ -16 u_z v_{o\tau}+32 u_z v_{x\tau}-8 N u_z^2+8 u_z^2\\
\beta_{v_{z}} &= -8 g_o v_{x\tau}-8 u_{x} v_{z}+8 u_{o\tau} v_{z}+8 u_z v_{o\tau}-8 u_z v_{x\tau}-8 g_{x\tau} v_{o\tau}+16 g_{x\tau} v_{x\tau}\\
\beta_{v_{x\tau}} &= -4 g_o v_{z}+4 g_o v_{x\tau}+8 u_{x} v_{o\tau}-4 g_z v_{o\tau}-16 u_{x}
   v_{x\tau}+4 g_z v_{x\tau}-4 v_{z} u_z-4
   u_{o\tau} v_{o\tau}+4 u_{o\tau} v_{x\tau} \nonumber \\
   &\ \ +8 v_{z} g_{x\tau}+4 u_z v_{x\tau}-4 N v_{x\tau}^2\\
\beta_{u_{o\tau}} &= -8 g_o g_{x\tau}-8 u_{x} u_{o\tau}+8 v_{z}^2-16 v_{o\tau}
   v_{x\tau}+8 v_{o\tau}^2+16 v_{x\tau}^2\\
\beta_{g_{x\tau}} &= -4 g_o u_{o\tau}+4 g_o g_{x\tau}+8 u_{x} u_z-4 g_z u_z-16 u_{x} g_{x\tau}+4 g_z g_{x\tau}-8 v_{z} g_{x\tau}+4 u_{o\tau} g_{x\tau}-8 v_{z} v_{o\tau} \nonumber \\
&\ \ +16 v_{z} v_{x\tau}-4 u_z g_{x\tau}+8 g_{x\tau} v_{o\tau}-4 N g_{x\tau}^2
   \end{align}
Analogous equations have been derived in the context of untwisted Bernal-stacked bilayer graphene, in the work of Yang and Vafek \cite{Yang2010}, where similar divergent corrections arise due to a quadratic band-touching. However, even upon setting the moir\'e-valley-dependent couplings to zero, our equations do not agree -- this results from the fact our single-particle Hamiltonian possesses the anticommuting chiral symmetry $\mathcal{S}=\sigma_x\tau_x\eta_y$ absent in Bernal-stacked bilayer graphene. If one were to remove the factor of $\eta_z$ from our single-particle Hamiltonian and then set the moir\'e-valley-dependent couplings to zero, a different closed set of nine couplings appears, as discussed in the previous sections -- defined as those possessing vertices which commute with $\mathcal{S}$. The resulting parquet equations are given by
\begin{align}
\beta_{g_o} &= -8 g_o g_x \nonumber\\
\beta_{g_x} &= -4 N_f g_x^2+4 g_x g_z+4 g_0 g_x-8 g_x^2-2 g_z^2-2 g_0^2\nonumber\\
\beta_{g_z} &= -24 g_x g_z+8 g_x^2-8 N_f g_z^2+8 g_z^2+8 g_0 g_z\nonumber
\end{align}
which agree with Yang and Vafek (up to a prefactor of 2 arising from the definition of the couplings). In addition to the $\eta$-dependent chiral symmetry, our present analysis differs from that in Bernal BLG due to the $\eta$-dependent couplings and the fact that the degeneracy $N_f$ varies as a function of density in TBG, due to the Dirac revival resetting the dispersion at each integer filling. Moreover, our field theory treatment is not strictly limited to the weak coupling regime -- the results of the previous sections allow us to compute corrections to the beta functions at early RG times encoded by the $\mathscr{N}$ functions.

\subsection{Fixed rays at long RG times}
At long RG times, the solution of the parquet equations results in a diverging subset of the couplings. In this limit, the diverging couplings tend towards fixed constant ratios of each other referred to as {\it fixed rays}. The relative magnitudes of the couplings determine which ground state dominates. All possible choices of bare initial coupling values flow to one of these possible sets of ratios in the deep infrared. In this section we derive the fixed rays of the parquet equations as a function of $N_f$, which corresponds to the filling factor via $N_f=2(4-\lfloor \nu \rfloor)$, and as a function of  $d(\mu)$, which corresponds to the deviation from the filling factor -- starting from $d(\mu)=1$ at the Dirac points located at integer filling and decreasing $d(\mu)<1$ upon doping above the Dirac point.

To obtain the RG fixed rays, we insert the scaling form $\{g_{i}, v_i, u_i\} = \{G_{i}, V_i, U_i\} \mathtt{S}$, with $\mathtt{S} =1/(t_c-t)$, into the parquet RG equations, which at long times $t \rightarrow t_c$ allows the differential flow equations to be reduced to nonlinear algebraic equations for the $t$-independent coefficients $\{G_{i}, V_i, U_i\}$.  We sumarise the procedure, which has been discussed in detail elsewhere, c.f Ref. \cite{Scammell2023}: 
\begin{enumerate}
    \item Denote the set of running couplings as $\gamma_i=\{g_{i}, v_i, u_i\}$ and the set of scaling coefficients $\Gamma_i= \{G_{i}, V_i, U_i\}$. 

\item The fixed rays are found via $\dot{\Gamma}_i = \left[\dot{\gamma}_i - \Gamma_i \dot{\mathtt{S}}\right]/\mathtt{S}=\left[\beta_i[\{\Gamma\}] - \Gamma_i\right] \mathtt{S}=0$.  

\item To analyse the stability of the fixed points, we examine the matrix
\begin{align}
T_{ij} = \frac{\partial}{\partial \gamma_j} \left(\beta_i[\{\Gamma\}] - \Gamma_i\right).
\end{align}
Evaluating $T_{ij}$ at the fixed points (i.e. at the solutions to $\Gamma_i = \beta_i[\Gamma_j]$), we discard those fixed points with greater than one positive eigenvalues. The stable fixed rays satisfy this condition.
\end{enumerate}

Below we compute the stable fixed rays; we make use of the result that weak-coupling instabilities require the associated order parameter eigenvalue coefficient $\Lambda_i\geq1/2$ \cite{Maiti2013}, where the eigenvalue is $\lambda_i  = \Lambda_i {\mathtt S}$. This follows from requiring that the susceptibility diverge as $t\rightarrow t_c$.

\section{Fixed Rays}
Some general trends emerge. Firstly, increasing the degree of particle-hole asymmetry $d$ by doping above the Dirac point enhances superconductivity and eventually reduces the tendency towards particle-hole orderings. Secondly, decreasing $N_f$ lowers the requisite value of $d$ needed for superconductivity to emerge -- e.g. when $N_f=8$ superconductivity only appears as a possible fixed point when $d<0.75$ whereas for $N=2,4$ it is already a fixed point in the particle-hole symmetric limit $d=1$. Third, the most common particle-hole orders are $\Omega_9, \Omega_{11}, \Omega_{12}$ while the most common superconducting orders are $\Delta_4, \Delta_5$, though several other orderings are possible.

%%%%%%%%%%%%%%%%%%%%%%%%%%%%%
\begin{figure*}[t!]
\includegraphics[width=0.88\textwidth]{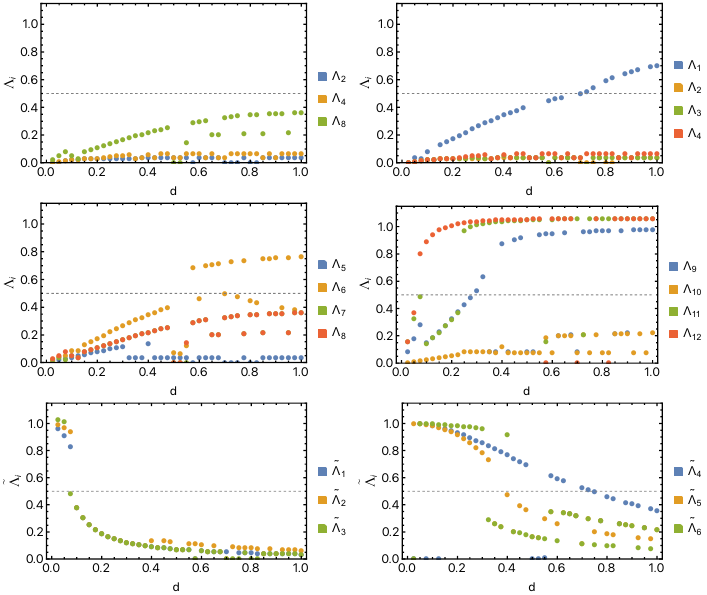}
    \caption{ $\bm{0\leq\nu<1}$ ($\bm{N_f=8}$). The reduced order parameter eigenvalues, $\Lambda_i = \lambda_i/\mathtt{S}$, evaluated at the stable fixed rays at each value of $d$, and with only the maximum $\Lambda_i$ at each $d$ presented. For this reason the curves are discontinuous. Top left panel focuses on the three distinct T-IVC orders.} \label{phase}
\end{figure*}
%%%%%%%%%%%%%%%%%%%%%%%%%%%%%

%%%%%%%%%%%%%%%%%%%%%%%%%%%%%
\begin{figure*}[t!]
\includegraphics[width=0.88\textwidth]{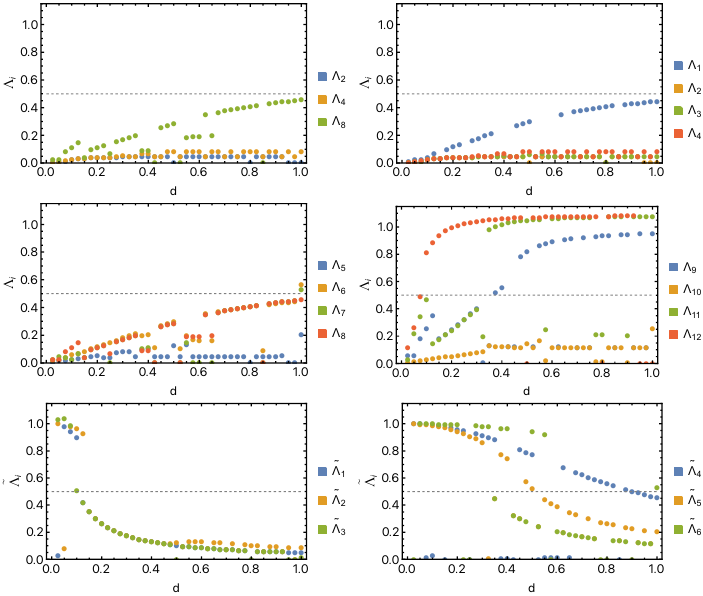}
    \caption{ $\bm{1\leq\nu<2}$ ($\bm{N_f=6}$). The reduced order parameter eigenvalues, $\Lambda_i = \lambda_i/\mathtt{S}$, evaluated at the stable fixed rays at each value of $d$, and with only the maximum $\Lambda_i$ at each $d$ presented.  }\label{phase}
\end{figure*}
%%%%%%%%%%%%%%%%%%%%%%%%%%%%%

%%%%%%%%%%%%%%%%%%%%%%%%%%%%%
\begin{figure*}[t!]
\includegraphics[width=0.88\textwidth]{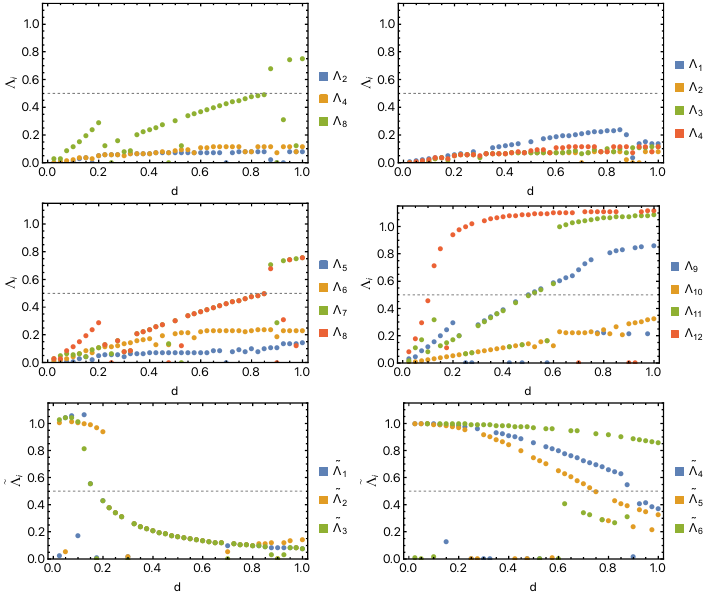}
    \caption{ $\bm{2\leq\nu<3}$ ($\bm{N_f=4}$). The reduced order parameter eigenvalues, $\Lambda_i = \lambda_i/\mathtt{S}$, evaluated at the stable fixed rays at each value of $d$, and with only the maximum $\Lambda_i$ at each $d$ presented.    }\label{phase}
\end{figure*}
%%%%%%%%%%%%%%%%%%%%%%%%%%%%%

%%%%%%%%%%%%%%%%%%%%%%%%%%%%%
\begin{figure*}[t!]
\includegraphics[width=0.88\textwidth]{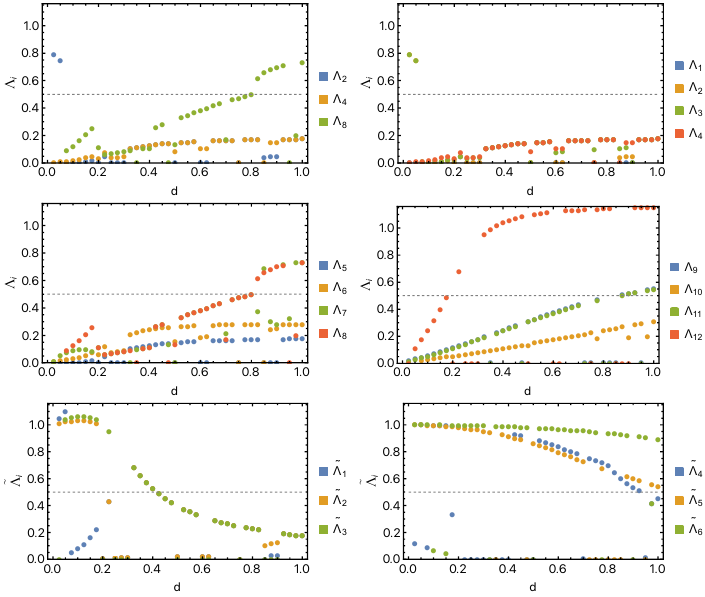}
    \caption{ $\bm{3\leq\nu<4}$ ($\bm{N_f=2}$). The reduced order parameter eigenvalues, $\Lambda_i = \lambda_i/\mathtt{S}$, evaluated at the stable fixed rays at each value of $d$, and with only the maximum $\Lambda_i$ at each $d$ presented.    }\label{phase}
\end{figure*}
%%%%%%%%%%%%%%%%%%%%%%%%%%%%%

\newpage
{\color{white} ... }
\newpage
\newpage
{\color{white} ... }
\newpage
\newpage
{\color{white} ... }
\newpage
\newpage
{\color{white} ... }
\newpage

\section{Wess-Zumino-Witten terms}
For completeness and to be self-contained, we here detail the criteria for determining the possible WZW terms discussed in the main text. We focus on the scenario with a single superconducting order parameter (two real degrees of freedom), described by the matrix $\Delta=-\Delta^T$ in the vicinity of the Dirac cones, and three particle-hole order parameters $m_j=m_j^\dagger$, $j=1,2,3$, coupling to the electrons as
\begin{equation}
    \mathcal{H}_C = \sum_{\vec{k}} \left[\psi^\dagger_{\vec{k}} \Delta \psi^\dagger_{-\vec{k}} + \text{h.c.} \right] + \sum_{\vec{k},j} \psi^\dagger_{\vec{k}} m_j \psi^\pdagger_{\vec{k}}.
\end{equation}
At $\nu=0$, the operators $\psi^\dagger_{\vec{k}}$ are the same as those in Eq.~(\ref{V1234}) of the main text, i.e., 16-component spinors in sublattice, mini-valley, valley, and spin space, and, hence, $\Delta$ and $m_j$ are $16 \times 16$ matrices. Due to the Dirac revival, these are only 8-component field operators and, thus, $8\times 8$ order parameters at $\nu=2$. Focusing on spin-valley locking, these remaining eight flavor degrees of freedom arise from sublattice, mini-valley, and the combined spin-valley index (with Pauli matrices $\gamma_j$).

As shown in Ref.~\cite{Christos2020}, a WZW term arises if all of the following properties hold
\begin{subequations}\begin{align}
    \Gamma_i \Delta &= - \Delta \Gamma_i^T \neq 0, \quad\quad i=1,2, \\
    m_j \Delta &= \Delta m_j^T \neq 0, \quad\quad \, \, j =1,2,3, \\
    \text{tr}\left[ \Gamma_{i_1} \Gamma_{i_2} m_{j_1} m_{j_2} m_{j_3} \right] &= k \, \epsilon_{i_1,i_2,j_1,j_2,j_3}, \quad k\neq 0.
\end{align}\label{WZWCondition}\end{subequations}
Here, $\Gamma_{i=1,2}$ are the two Dirac matrices parametrising the non-interacting Hamiltonian, i.e., $\Gamma_1 = \tau_z\alpha_x = \tau_z \sigma_x$, $\Gamma_2 = \tau_z\alpha_y = \sigma_y$ for $\nu=0$ [as directly follows from Eq.~(\ref{singleHeq})] and $\Gamma_1 = \gamma_z \sigma_x$, $\Gamma_2 = \sigma_y$ for spin-valley locking at $\nu=2$.

By systematically checking which combinations $\{\Delta,m_{1,2,3}\}$ of the order parameters in Table~\ref{symmetries} and \ref{nu2} satisfy Eq.~(\ref{WZWCondition}), associated with a fixed ray in Table~\ref{f:rays} at the respective filling, we find all possible WZW scenarios.


\begin{thebibliography}{34}
\bibitem{endnote4} In the Bistritzer-MacDonald continuum model, this requires setting $w_0=0$. Here we do not require $w_0=0$, and so ${\cal C}$ is an approximate low-energy symmetry of the Dirac theory, rather than an exact symmetry of the full theory away from the Dirac points.
\bibitem{endnote5} In the presence of lattice relaxation, there is no angle where the velocity exactly vanishes in the full Bistritzer-MacDonald model, where $\mathcal{P}$ is only an approximate symmetry.
\bibitem{endnote7} Note that the inter-moir\'e-valley states break moir\'e translational symmetry, and so are properly classified by irreducible representations of the moir\'e space group. We refer to this state as $E$ to denote its irrep under $D_3$, as per Table \ref{symmetries}.
\end{thebibliography}
\end{document}